\documentclass[useAMS,usenatbib]{mn2e}
\usepackage[dvips]{graphicx}
\usepackage{amsmath}
\usepackage[german]{babel}

\title[Chemical element ratios of SDSS early-type galaxies]{Chemical element ratios of SDSS early-type galaxies}
\author[J. Johansson, D. Thomas and C. Maraston]{Jonas Johansson$^{1,2}$, Daniel Thomas$^{1,3}$ and Claudia Maraston$^{1,3}$\\
$^1$Institute of Cosmology and Gravitation, Dennis Sciama Building, Burnaby Road, Portsmouth PO1 3FX\\
$^2$Max-Planck Institut f{\"u}r Astrophysik, D-85741 Garching, Germany\\
$^3$SEPnet, South East Physics Network}

\begin{document}

\pagerange{\pageref{firstpage}--\pageref{lastpage}} \pubyear{2009}
\maketitle
\label{firstpage}
\begin{abstract} 

We discuss chemical enrichments of $\sim$4000 SDSS early-type galaxies using as tracers a large variety of element abundance ratios, namely [C/Fe], [N/Fe], [O/Fe], [Mg/Fe], [Ca/Fe] and [Ti/Fe]. 
We utilise the stellar population models of absorption line indices from \citet*{TMJ10} which are based on the MILES stellar library.  We confirm previous results of increasing age, [Z/H] and [O/Fe] ratios (most often represented by [$\alpha$/Fe] in the literature) with velocity dispersion. We further derive identical correlations with velocity dispersion for the abundance ratios [O/Fe], [Mg/Fe] and [C/Fe], implying that C/Mg and C/O are close to solar values. This sets a lower limit on the formation time-scales and star-burst components of early-type galaxies to $\sim$0.4 Gyr, which is the lifetime of a 3M$\odot$ star, since the full C enrichment must be reached.  
[N/Fe] correlates with velocity dispersion, but offset to lower values and with a steeper slope compared to the other element ratios.
We do not find any environmental dependencies for the abundances of C and N, contrary to previous reports in the literature.
[Fe/H] does not correlate with velocity dispersion over the entire parameter range covered, but for fixed age we find a steep trend for the [Fe/H]-$\sigma$ relation. This trend is weaker than the analogous for total metallicity (which also shows steeper trends at fixed age) owing to the lower Fe contribution from SN Ia for more massive early-type galaxies.
We find [Ca/Fe] ratios that are close to solar values over the entire velocity dispersion range covered. Tentative, due to large scatter, the results for [Ti/Fe] indicate that Ti follows the trends of Ca. 
This implies a significant contribution from SN Ia to the enrichment of heavy $\alpha$-elements and puts strong constraints on supernova nucleosynthesis and models of galactic chemical evolution. 
\end{abstract}


\section{introduction}
\label{intro}

The chemical compositions of stellar atmospheres are tracers of the element
abundances of the parent gas clouds forming the stars throughout the formation history of a galaxy. Some elements are also affected by dredge-up during stellar evolution \citep*[e.g.][]{sweigart89}. 
Stellar populations 
are therefore a powerful tool to extract information on chemical evolution in the Universe.
Element abundances can be directly determined for individual stars of resolved stellar populations in the Milky Way 
or in nearby dwarf galaxies, using absorption lines measured in high resolution stellar spectra 
\citep*[e.g.][]{edvardsson93,fuhrmann98,bensby04,feltzing09,bensby10}. 
Light averaged spectra must instead be used for determining element abundances of distant unresolved stellar 
populations. The absorption features of such spectra are sensitive to multiple elements due to velocity 
dispersion broadening. The Lick system of absorption line indices \citep*[e.g.][]{worthey94,trager98} have been frequently 
used for measuring 25 prominent absorption features in galaxy spectra.

Elements are produced in stellar nucleosynthesis besides the primordial nucleosynthesis of H and He. The chemical enrichment of stellar populations depends on the star formation history, initial mass function, fraction of exploding supernovae etc. The chemical pattern of the parent gas clouds will be carried on to new stellar generations. Thus the chemical enrichment of stellar populations is also affected by mechanisms affecting the interstellar medium (ISM) such as the efficiency of stellar winds to mix newly synthesised elements with the ISM, efficiency of outflow from galactic winds to remove enriched gas, inflow of less enriched gas from gas reservoirs etc. \citep[e.g.][]{matteucci89,matteucci94}. Chemical enrichment sets stringent constraints on galaxy formation and evolution. 


Studies beginning in the late 1970's have revealed non-solar abundance ratios for the stellar populations of 
early-type galaxies \citep*[e.g.][]{oconnell76,peterson76,burstein84,worthey92,davies93,surma95}, indicating
different chemical enrichment histories. 
This triggered more detailed investigations showing that the ratio between $\alpha$-elements and Fe-peak elements increases with
increasing galaxy mass for early-type galaxies \citep[e.g.][]{trager00b,thomas05,bernardi06,clemens06,thomas10}.
Most interestingly, the [$\alpha$/Fe] ratio participates in the E-E dichotomy, i.e. elliptical galaxies with low [$\alpha$/Fe] ratios have core-less central profiles, while [$\alpha$/Fe]-enhanced galaxies with short formation time-scales have cores \citep{Kormendy09}.

Individual element abundance ratios, in addition to $\alpha$/Fe, can further disentangle the formation of different 
stellar populations. Since the individual elements are produced in different stellar evolutionary phases they trace varying formation histories. 
\citet{edvardsson93} and \cite{bensby10} derive as many as 12 different element abundance ratios to distinguish between the different 
formation histories of the stellar populations in the Milky~Way.


A number of studies have derived individual element abundance ratios for early-type galaxies \citep{sanchez03,sanchez06,clemens06,kelson06,graves07,graves08,smith09,price11}, in several cases for fairly small samples. 
Different methods are applied in these studies, but they are all based on absorption line indices. The results are dependent on the method applied and the sample used. The aim of this work is to simultaneously derive all element abundance ratios allowed by the sensitivity of the adopted absorption line indices. The maximum amount of information is extracted from the indices to reliably derive the abundance ratios and state of the art models of stellar populations of absorption indices are utilised to obtain as accurate results as possible.

To fully interpret observed element abundance ratio trends, 
stellar nucleosynthesis needs to be understood. \citet{pipino10} find up-to-date models of chemical evolution to struggle in simultaneously reproducing observed abundance ratios for Carbon and Nitrogen from the unresolved stellar populations of early-type galaxies. 
The $\alpha$-element Ca is a puzzle as it is has been found to trace Fe instead of other $\alpha$-elements for early-type galaxies 
\citep{cenarro03,thomas03b,smith09,Saglia02}. This has been interpreted as Ca being contributed by SNIa as well as SNII \citep*{TJM10}. Thus element abundance ratios of unresolved stellar populations are also useful for constraining stellar nucleosynthesis. 
 

We present 
a technique for deriving a wide range of element abundance ratios 
for unresolved stellar populations, 
including 
[O/Fe] (representing [$\alpha$/Fe]), [C/Fe], [N/Fe], [Mg/Fe], [Ca/Fe] and [Ti/Fe]. The method is based on
new flux-calibrated stellar population models of absorption line indices presented in \citet{TMJ10}. 
We analyse a 
sample of 3802 SDSS early-type galaxies for which we investigate element ratio scaling relations with velocity dispersion.

The paper is organised as follows. The SDSS early-type galaxy sample used is presented in Section~\ref{data} and the technique for deriving the 
element abundance ratios is described in Section~\ref{spp}. The results of derived element abundance ratios for the data sample 
are presented in Section~\ref{results} and further discussed and compared with the literature in Section~\ref{disc}. Conclusions are given in Section~\ref{conc}.

\section{the data sample}
\label{data}

The selected sample is part of the MOSES catalogue (MOrphologically Selected Early-types in SDSS). This is
described in detail in \citet{schawinski07} and \citet{thomas10} (T10) and only a brief description is 
given here. The MOSES sample consists of 48 023 galaxies from the SDSS Data Release 4, selected 
to have a magnitude $r<16.8$ in the redshift range $0.05<z<0.10$. The selection criteria ensured
a reliable visual inspection of galaxy morphology and the full sample was divided into late-types
(31 521) and early-types (16 502) through a purely visual classification scheme.

As comprehensively described in T10 the 25 standard Lick absorption line indices were 
measured on the galaxy spectra downgraded from the SDSS spectral resolution to the Lick/IDS resolution and corrected for emission line fill in using GANDALF/PPXF \citep{ppxf,sarzi06}. The stellar kinematics and best-fit stellar templates from PPXF were used in T10 for correcting the
measured indices for stellar velocity dispersion broadening effects. The velocity dispersion measurements derived and published in T10 are used in Section~\ref{results} as proxies for galaxy mass. 

The visual classification does not bias against star forming galaxies and the sub-sample of early-type galaxies 
will therefore include galaxies with blue colours, having possible on-going star formation or recent star 
formation. These galaxies can have possible emission line contamination in the absorption features, but the use of GANDALF gives reliable Lick index measurements for the full sample. 

The final sample (3802 objects) used in this work was selected according to T10.
This is a sub-sample of early-type classified galaxies in the narrow redshift range $0.05<z<0.06$, chosen 
to minimise 
evolutionary effects present when using
the full range of redshifts, affecting the derivation of stellar population parameters.


\section{The TMJ models}
\label{models}

In \citet*{TMJ10} (TMJ) we present new stellar population models of Lick absorption-line indices with variable element abundance ratios. The model is an extension of the \citet{TMB03,thomas04} (TMB/K) model, which is based on the evolutionary stellar population synthesis code of \citet{maraston98,maraston05}. A calibration on galactic globular clusters was performed in \citet{TJM10} for the range of element abundance ratios covered. For basic information on the model we refer the reader to \citet{TMB03,thomas04} and TMJ. Here we provide a brief summary of the main features of our new models.

\subsection{New features}
\label{features}

The key novelty compared to the TMB model is that the TMJ model is flux-calibrated, hence not tied anymore to the Lick/IDS system. This is because the new models are based on our calibrations of absorption-line indices with stellar parameters
\citep{johansson10} derived from the flux-calibrated stellar library MILES \citep{miles}. The MILES library consists of spectra covering the wavelength range 3500-7400 \AA~and with a spectral resolution of 2.54 \AA~\citep[as revised by][]{Beifiori11} of 985 stars selected to produce a sample with extensive stellar parameter coverage. Most importantly it has been carefully flux-calibrated, making standard star-derived offsets unnecessary.
The data release now provides two model versions partially based on different stellar evolutionary tracks, \citet{cassisi97} and Padova \citep{girardi00} at high metallicities. 
In the present study we use the former version, since these tracks were used in the TMB/K model adopted in T10, it was calibrated on galactic globular clusters in \citet{TJM10} and our Padova model version only cover ages down to 2.8 Gyr. 

Finally, we calculate models in which we selectively enhance, by 0.3 dex, each of the elements C, N, Na, Mg, Si, Ca, Ti, and Cr in turn, released with TMJ. Further extended model calculations with arbitrary enhancement factors, hence a finer grid in element ratios, are used in the present work.

\begin{figure*}
\centering
\includegraphics[scale=0.45,angle=-90]{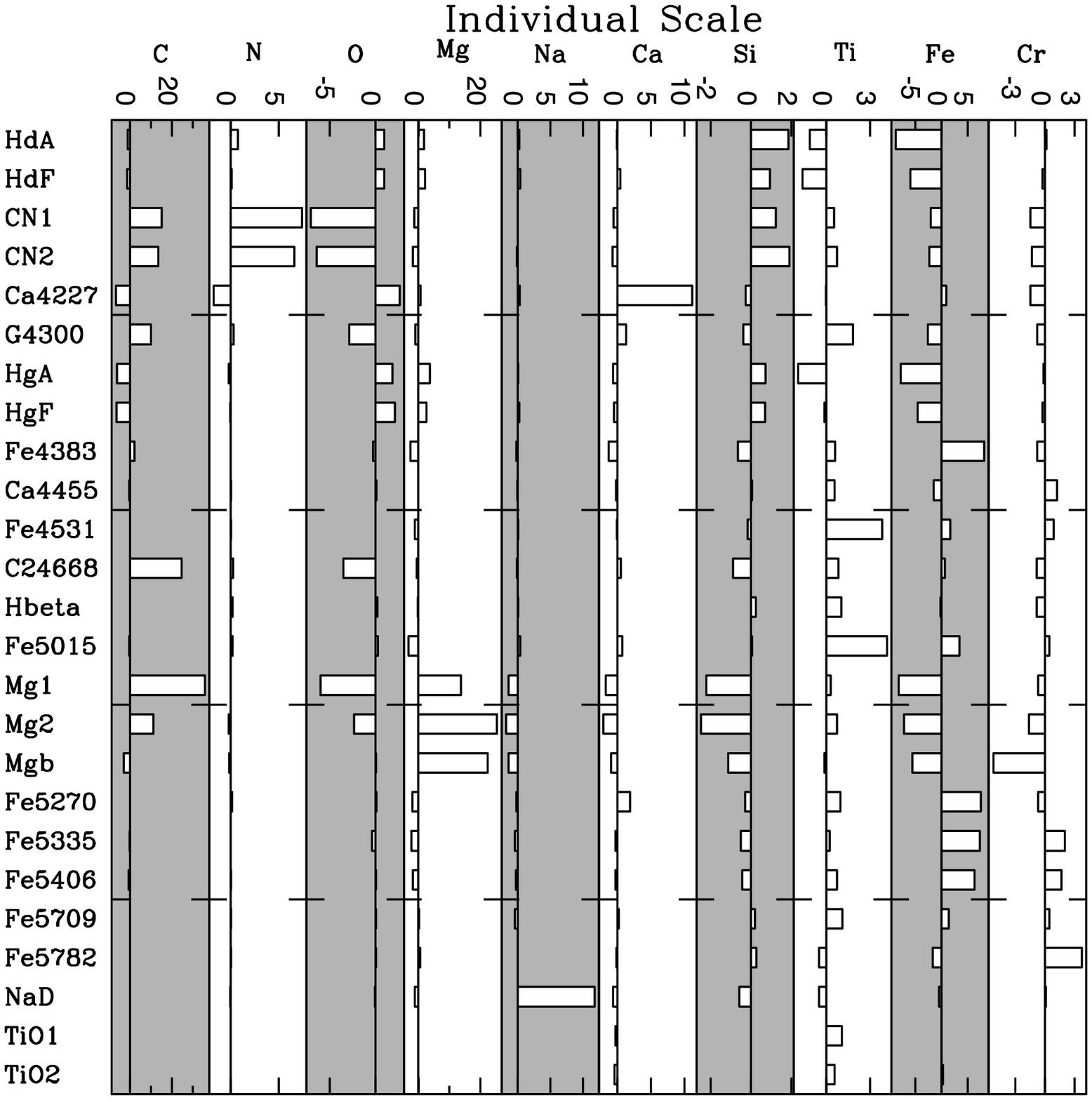}\hspace{0.4cm}\includegraphics[scale=0.45,angle=-90]{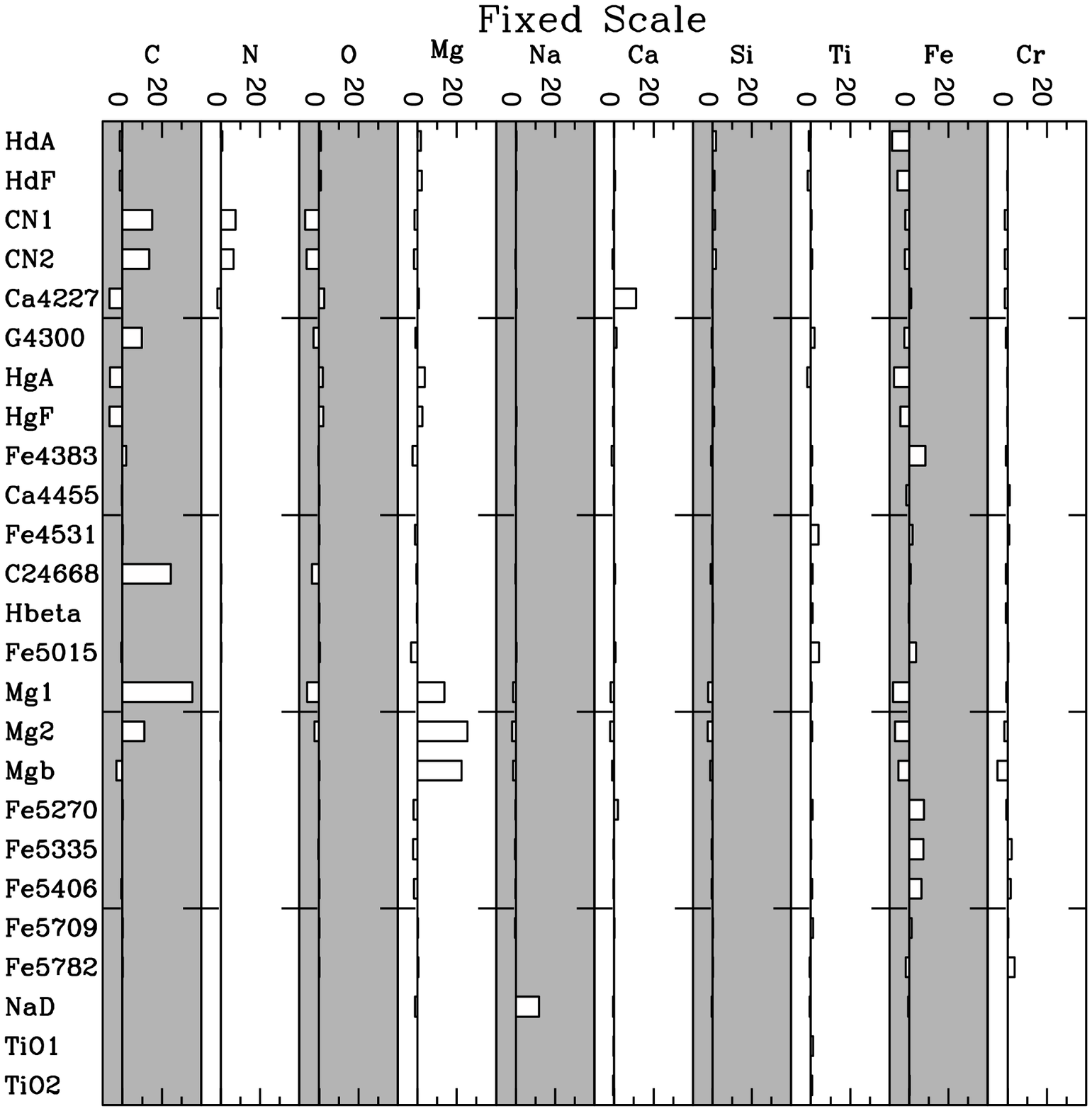}
\caption{The response of the 25 Lick indices to changes in individual element abundances for a 12 Gyr solar metallicity modelled stellar population. The response is defined as the fractional index changes for enhancements of a specific element abundance by a factor two and normalised to the typical observational index errors presented in \citet{johansson10}. For each element we display the index responses on an individual scale (left-hand panel) as well as those referred to a fixed scale (right-hand panel, see text for details). 
}
\label{response}
\end{figure*}

\subsection{Element abundance ratios}
\label{Eratios}


As in \citet{TMB03} and \citet{TMJ10} we keep the total metallicity
fixed while varying the element abundance ratio [$\alpha$/Fe]. Briefly, 
Fe-peak elements (Fe and Cr) are locked together (depressed group) and 
the rest of the considered elements (C, N, O, Na, Mg, Si, Ca and Ti) are locked to the $\alpha$-elements (enhanced group). 
Solar values of element ratios are known from measurements of the individual abundances \citep*{grevesse96}.
A non-solar [$\alpha$/Fe] ratio is computed by simultaneously changing the abundances of elements in the enhanced and depressed group 
to counter-balance a change of total metallicity. A non-solar [$\alpha$/Fe] ratio mainly means a change in the
abundances for the Fe-peak elements, since $\alpha$-elements and in particular Oxygen dominate total metallicity in the Sun, as discussed in
\citet{trager00a} and \citet{TMB03}. 

We keep the total metallicity fixed also when we produce models with enhancements of the individual 
elements C, N, Mg,
Ca and Ti. 
Starting from an [$\alpha$/Fe] ratio the models are perturbed by 
enhancing/depressing the individual element E. 
Hence we actually consider elements ratios of given element over $\alpha$, e.g. a N-enhanced model have varying [N/$\alpha$] ratios. The [E/Fe] ratio is then calculated with [E/Fe]=[E/$\alpha$]+[$\alpha$/Fe], where [$\alpha$/Fe] is the starting element ratio.
The element E is detached from the rest of the elements in the enhanced group, while 
the ratio between elements in the enhanced and depressed group is locked. 
Thus when enhancing/depressing the individual element E the total metallicity is conserved and the abundance ratios between the rest of the elements remain unchanged. 
Varying [E/Fe] mainly means a change in the abundance of the element E, since the rest
of the elements, locked together, dominate total metallicity. 

\subsection{Re-calculation of [Fe/H]}
\label{feh}


In addition to the parameters discussed above we derive [Fe/H]. Following \citet*{TCB98}, \citet{trager00a} and \citet{TMB03} we have the relationship between iron abundance [Fe/H] and total metallicity [Z/H] when only considering the element ratio [$\alpha$/Fe]

\begin{equation}\label{alpha}
[Fe/H]=[Z/H]+A[\alpha/Fe]=[Z/H]+A[O/Fe]
\end{equation}
if we assume that $\alpha$/Fe reflect O/Fe (see Section~\ref{signals}). Hence Eq.~\ref{alpha} remains valid if the analysis is restricted to [$\alpha$/Fe] ([O/Fe]). Eq.~\ref{alpha} instead needs to be revised when considering more element ratios besides [$\alpha$/Fe] ([O/Fe]). 
Modifying the individual element ratios with [O/Fe] as a starting point as in this work (see Section~\ref{Eratios}), perturbations to Eq.~\ref{alpha} arise in the form [E/O] for element E. We rewrite the general relationship of Eq.~\ref{alpha} to

\begin{equation}\label{general}
[Fe/H]=[Z/H]+A[O/Fe]+\sum_xB_x[E_x/O]
\end{equation}
for x number of individual elements E$_x$ with corresponding coefficients B$_x$. Changing an individual element abundance while keeping  total metallicity and
the ratio between the rest of the elements fixed, Eq.~\ref{general} becomes

\begin{equation}\label{individual}
\Delta[Fe/H]=B_x\Delta[E_x/O]=B_x(\Delta[E_x/H]-\Delta[O/H])
\end{equation}
leading to
\begin{equation}\label{individual2}
B_x=\frac{\Delta[Fe/H]}{\Delta[E_x/H]-\Delta[O/H]}
\end{equation}
Following \citet{TMB03}, when varying the [$\alpha$/Fe] ratio only, total metallicity is kept fixed while enhancing the $\alpha$-elements through
\begin{equation}\label{metfix}
f_{\alpha}X^++f_{Fe}X^-=X^++X^-
\end{equation}
where X$^+$ and X$^-$ are the mass fractions of the enhanced and depressed groups (see Section~\ref{Eratios}), respectively, changed by the factors f$_{\alpha}$ and f$_{Fe}$. Hence for an increase in the [$\alpha$/Fe]-ratio a higher abundance of the $\alpha$-elements is counter-balanced by a decrease in the abundance of Fe-like elements to keep total metallicity fixed. 
Further, changing an individual element abundance keeping total metallicity and the ratio between the rest of the elements fixed, Eq.~\ref{metfix} is extended to
\begin{multline}\label{metfix1}
f_E(f_{\alpha}X^E)+f_O(f_{Fe}X^-+f_{\alpha}(X^+-X^E))=\\
f_{\alpha}X^++f_{Fe}X^-=X^++X^-
\end{multline}
where the individual element abundance is now multiplied by the factor f$_E$ and the rest of the elements by f$_O$. Hence if the abundance of element E is changed by the factor f$_E$, the abundances of the rest of the elements are changed by the factor f$_O$ to counter-balance a change in total metallicity. Also, since all other elements besides E are changed by the same factor the ratios between these elements remain constant. 
This implies secondary abundance changes due to f$_E$ and f$_O$ besides that of f$_{\alpha}$ and f$_{Fe}$. 
The logarithmically solar scaled iron abundance is defined as
\begin{equation}\label{FeHold}
[Fe/H]_{old}=\log\bigg(\frac{X^{Fe}}{X^H}\bigg)-\log\bigg(\frac{X^{Fe}_{\odot}}{X^H_{\odot}}\bigg)
\end{equation}
where X$^{Fe}$ and X$^H$ are the mass fractions of Fe and H, while X$^{Fe}_{\odot}$ and X$^H_{\odot}$ are the corresponding solar values. If the mass fraction is changed by the factor f$_O$ the iron abundance becomes
\begin{equation}\label{FeHnew}
[Fe/H]_{new}=\log\bigg(\frac{f_O}{1}\bigg)+\log\bigg(\frac{X^{Fe}}{X^H}\bigg)-\log\bigg(\frac{X^{Fe}_{\odot}}{X^H_{\odot}}\bigg)
\end{equation}
Eq.~\ref{FeHold}-\ref{FeHnew} give $\Delta$[Fe/H]=[Fe/H]$_{new}$-[Fe/H]$_{old}$=$\log(f_O)$. In the same way we get $\Delta$[O/H]=$\log(f_O)$ and $\Delta$[E$_x$/H]=$\log(f_E)$ such that Eq.~\ref{individual2} becomes
\begin{equation}\label{individual3}
B_x=\frac{\log(f_O)}{\log(f_E)-\log(f_O)}
\end{equation}
Similarly to Eq.~\ref{FeHold}-\ref{FeHnew} we can enhance the element ratio [E/O] from 
\begin{equation}\label{EOold}
[E/O]_{old}=\log\bigg(\frac{f_{\alpha}}{f_{\alpha}}\bigg)+\log\bigg(\frac{X^E}{X^O}\bigg)-\log\bigg(\frac{X^E_{\odot}}{X^O_{\odot}}\bigg)
\end{equation}
to
\begin{multline}\label{EOnew}
[E/O]_{new}=\log\bigg(\frac{f_{E}}{f_{O}}\bigg)+\log\bigg(\frac{f_{\alpha}}{f_{\alpha}}\bigg)+\\
\log\bigg(\frac{X^E}{X^O}\bigg)-\log\bigg(\frac{X^E_{\odot}}{X^O_{\odot}}\bigg)
\end{multline}
For a new abundance ratio [E/O]$_{new}$=0.3 starting from solar values [E/O]$_{old}$=0.0 Eq.~\ref{EOold} and Eq.~\ref{EOnew} give
\begin{equation}\label{EOfinal}
\log\bigg(\frac{f_E}{f_O}\bigg)=0.3
\end{equation}
\begin{table}
\center
\caption{Solar abundance fractions (X$_{\odot}^E$) derived from \citet{grevesse96}, abundance enhancement factors (f$_E$ and f$_O$) and final element ratio coefficients (B$_x$) for the various elements considered.}
\label{abund}
\begin{tabular}{lllll}
\hline
\bf E$_x$ & \bf X$_{\odot}^E$ & \bf f$_E$ & \bf f$_O$ & \bf  B$_x$ \\
\hline
C   & 0.172  & 1.693 & 0.847 &   -0.24 \\
N   & 0.053  & 1.894 & 0.947 &   -0.079 \\
Mg & 0.038  & 1.923 & 0.962 &   -0.056 \\
Ca & 0.003  & 1.994 & 0.997 &   -0.0043 \\
Ti   & 0.0002 & 1.9996 & 0.9998 & -0.00029\\
\hline
\end{tabular}
\end{table}Following \citet{TMB03} with the difference of having C locked to the enhanced group we re-derive f$_{Fe}$ and f$_{\alpha}$. Adopting the solar abundance fractions X$^+$=0.91 and X$^-$=0.079 derived from \citet{grevesse96} we find f$_{Fe}$=0.521 and f$_{\alpha}$=1.042, resulting in A=0.94. Individual abundance fractions derived from \citet{grevesse96} are presented in Table~\ref{abund} along with the corresponding values for f$_E$ and f$_O$ derived using Eq.~\ref{metfix1} and Eq.~\ref{EOfinal}. Eq.~\ref{individual3} then gives the final coefficients B$_x$, also presented in Table~\ref{abund}.
The final relationship becomes
\begin{multline}\label{final}
[Fe/H]=[Z/H]-0.94[O/Fe]-0.24[C/O]-0.079[N/O]-\\
0.056[Mg/O]-0.0043[Ca/O]-0.00029[Ti/O]
\end{multline}

\subsection{Index Responses}
\label{signals}

Fig.~\ref{response} shows the response of the 25 Lick indices to individual element abundance changes for a 12 Gyr, solar metallicity stellar population. The fractional index change is calculated for an enhancement of the respective element by a factor two normalised to the typical observational measurement error for MILES stars from \citet{johansson10}. The scale on the x-axis in the left hand panel of Fig.~\ref{response} is different for each element. Hence the left-hand panel should be read vertically demonstrating the most sensitive indices for the individual elements. 
 The scale on the x-axis is instead kept fixed for all elements in the right hand panel. This panel should then instead be read horizontally to identify easily those elements that are best traced by a specific index for the current set of models. It can be seen that the elements C, N, Na, Mg, Ca, Ti, and Fe are best accessible.

The abundance of nitrogen is obtained from the CN indices that are also highly sensitive to C abundance. However, this degeneracy can be easily broken through other C sensitive indices such as C$_2$4668 and Mg$_1$. The Mg indices Mg$_1$, Mg$_2$, and Mgb are very sensitive to Mg abundance. Note, however, that all three additionally anti-correlate with Fe abundance \citep{trager00a,TMB03}. Ca can be measured well from Ca4227, except that this particular index is quite weak and requires good data quality. Na abundance can be derived quite easily from NaD in principle. However, in practise this is problematic as the stellar component of this absorption feature is highly contaminated by interstellar absorption \citep{TMB03}. Iron is well sampled through the Fe indices.

There are two among the Fe indices, however, that are also sensitive to Ti abundance besides Fe. These are Fe4531 and Fe5015. They offer the opportunity to estimate also Ti abundance. We will only use Fe4531, as Fe5015 is contaminated by a non-negligible Mg sensitivity besides Fe, which weakens its usefulness for Ti abundance determinations.

The remaining three elements O, Si and Cr cannot easily be measured through the available indices. As discussed extensively in \citet{TMB03}, however, oxygen has a special role. O is by far the most abundant metal and clearly dominates the mass budget of 'total metallicity'. Moreover, the $\alpha$/Fe ratio is actually characterised by a depression in Fe abundance relative to all light elements (not only the $\alpha$ elements), hence $\alpha$/Fe reflects the ratio between total metallicity to iron ratio rather than $\alpha$ element abundance to iron. As total metallicity is driven by oxygen abundance, the $\alpha$/Fe can be most adequately interpreted as O/Fe ratio. We therefore re-name the parameter $\alpha$/Fe to O/Fe under the assumption that this ratio provides an indirect measurement of oxygen abundance.

\begin{figure*}
\centering
\includegraphics[scale=0.5]{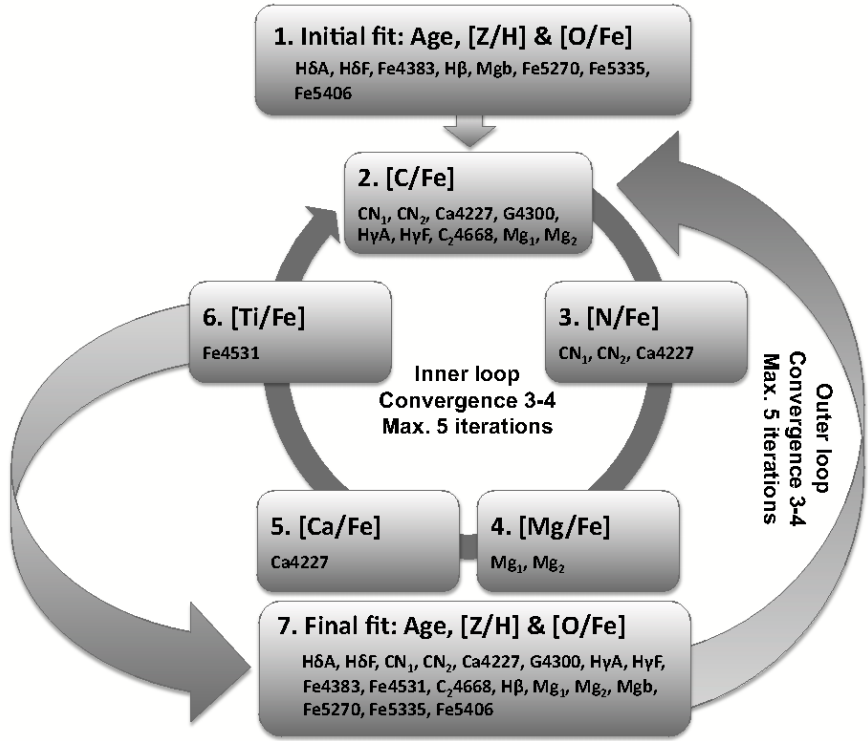}
\caption{Method to derive element abundance ratios. A $\chi^2$-minimisation routine is used at all steps to find the best fit model. Step 1 computes an initial fit for the base parameters age, [Z/H] and [O/Fe] using a base set of indices sensitive to these parameters only. The base parameters are fixed when deriving the individual element abundance ratios (inner loop, step 2-6) by adding indices sensitive to the element considered in each step to the base set of indices. At convergence, i.e. the element abundance ratios remain unchanged, the routine exit the inner loop and the base parameters are re-derived for the computed set of abundance ratios using all indices.  This outer loop is iterated until the $\chi^2$ at step 7 stops improving by less than 1$\%$. Models with varying abundance ratios of the element considered are produced in step 2-6, while models perturbed around the previously derived base parameters are produced in step 7. The indices used at each step are presented in each box and the typical number of iterations are indicated as convergence along with the maximum number iterations allowed for both the inner and outer loop.}
\label{iter_fig}
\end{figure*}

\begin{figure*}
\Large \bf $\quad$ Column 1 $\quad$$\quad$$\quad$$\quad$ Column 2 $\quad$$\quad$$\quad$$\quad$ Column 3 $\quad$$\quad$$\quad$$\quad$$\quad$ Column 4\\
\centering
\includegraphics[scale=0.2]{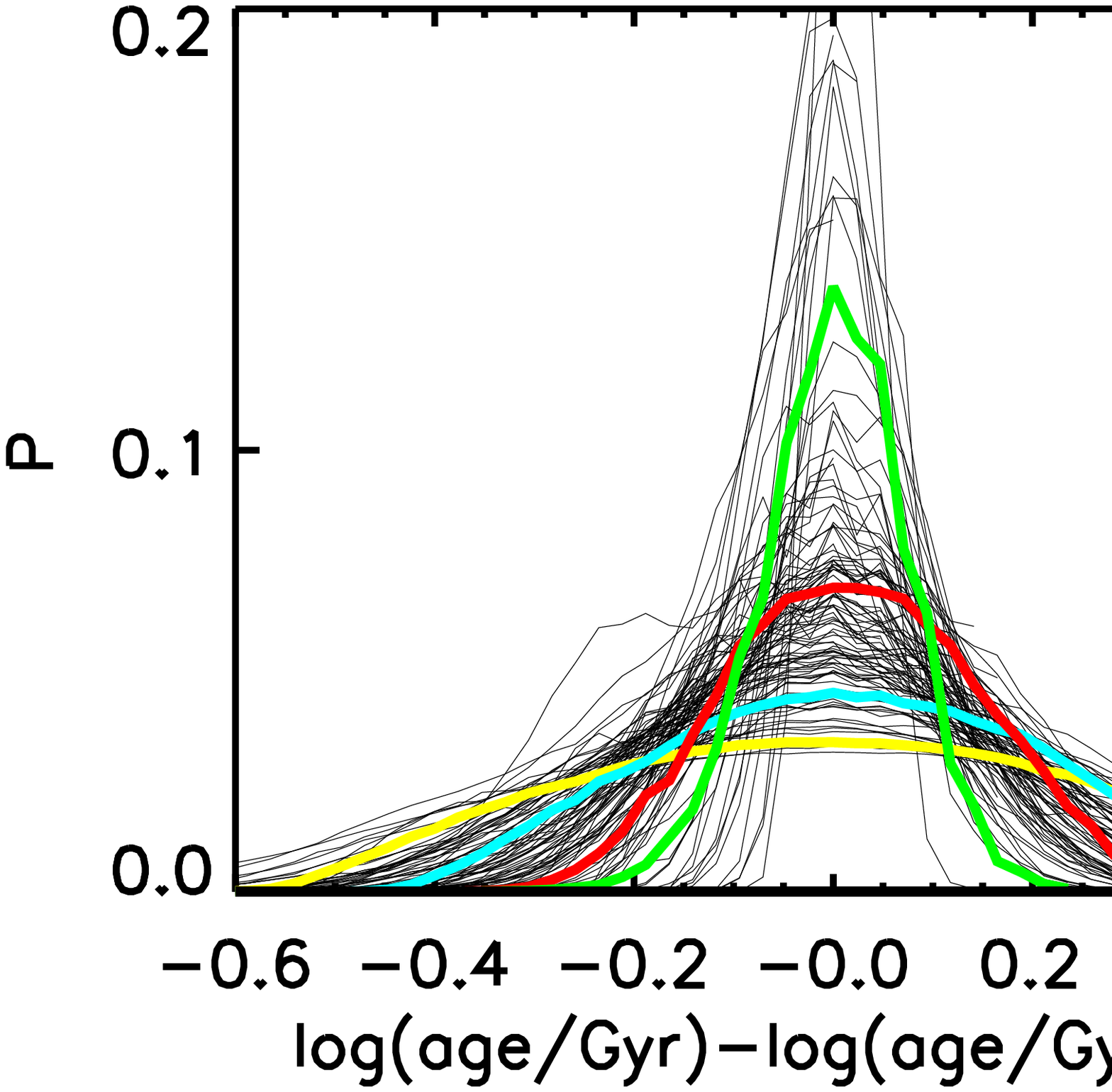}\includegraphics[scale=0.17]{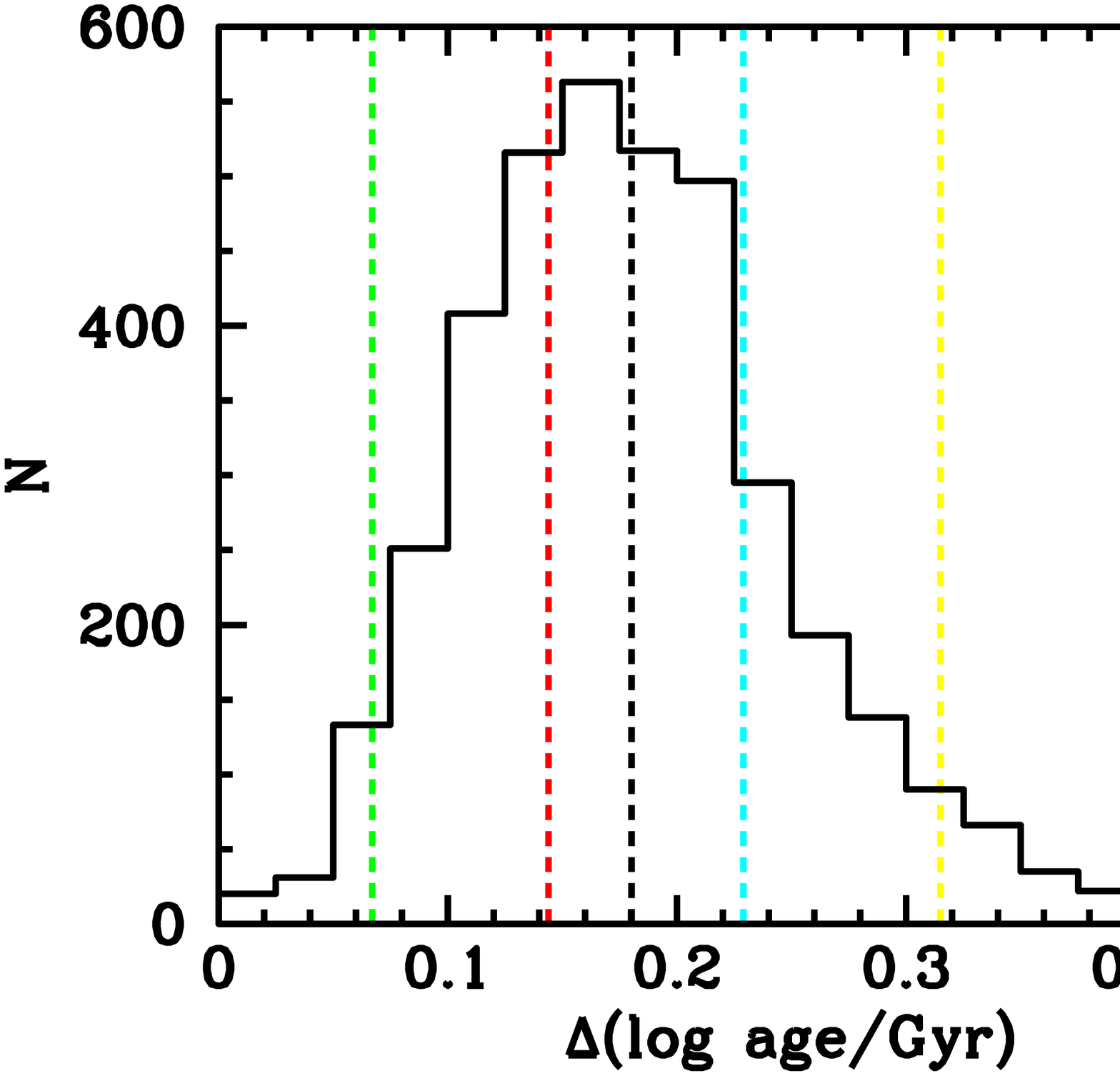}\includegraphics[scale=0.2]{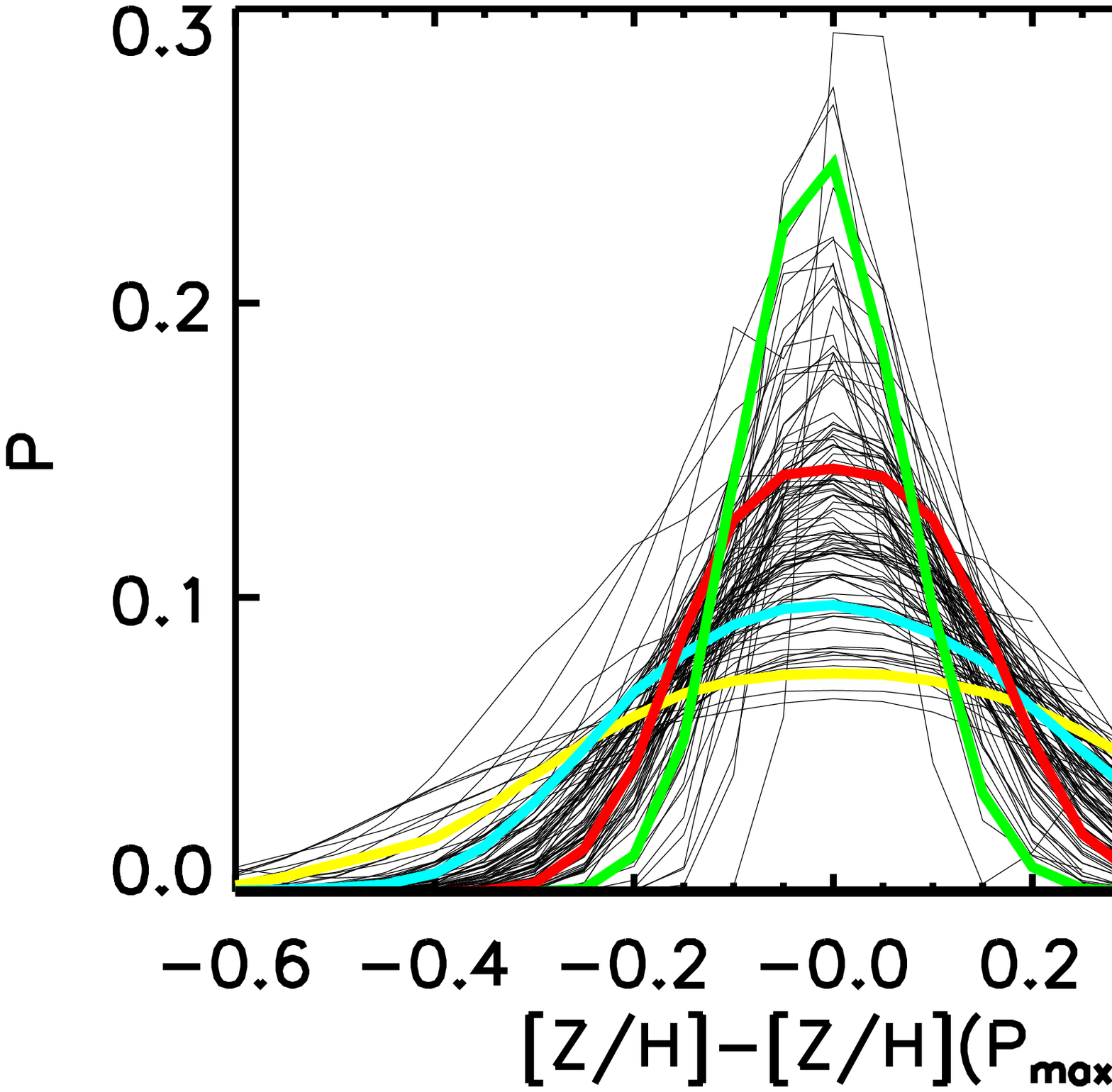}\includegraphics[scale=0.17]{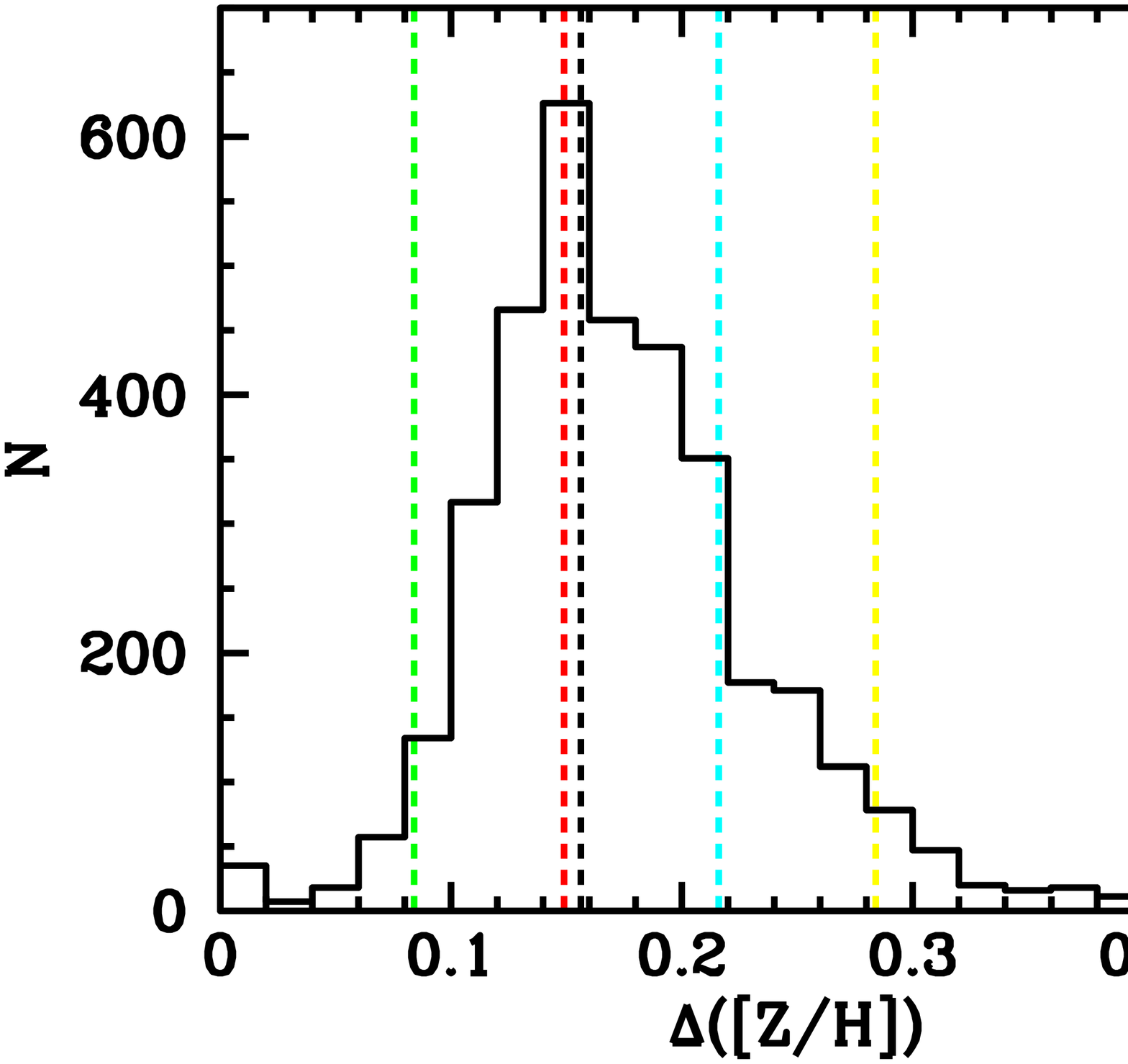}
\includegraphics[scale=0.2]{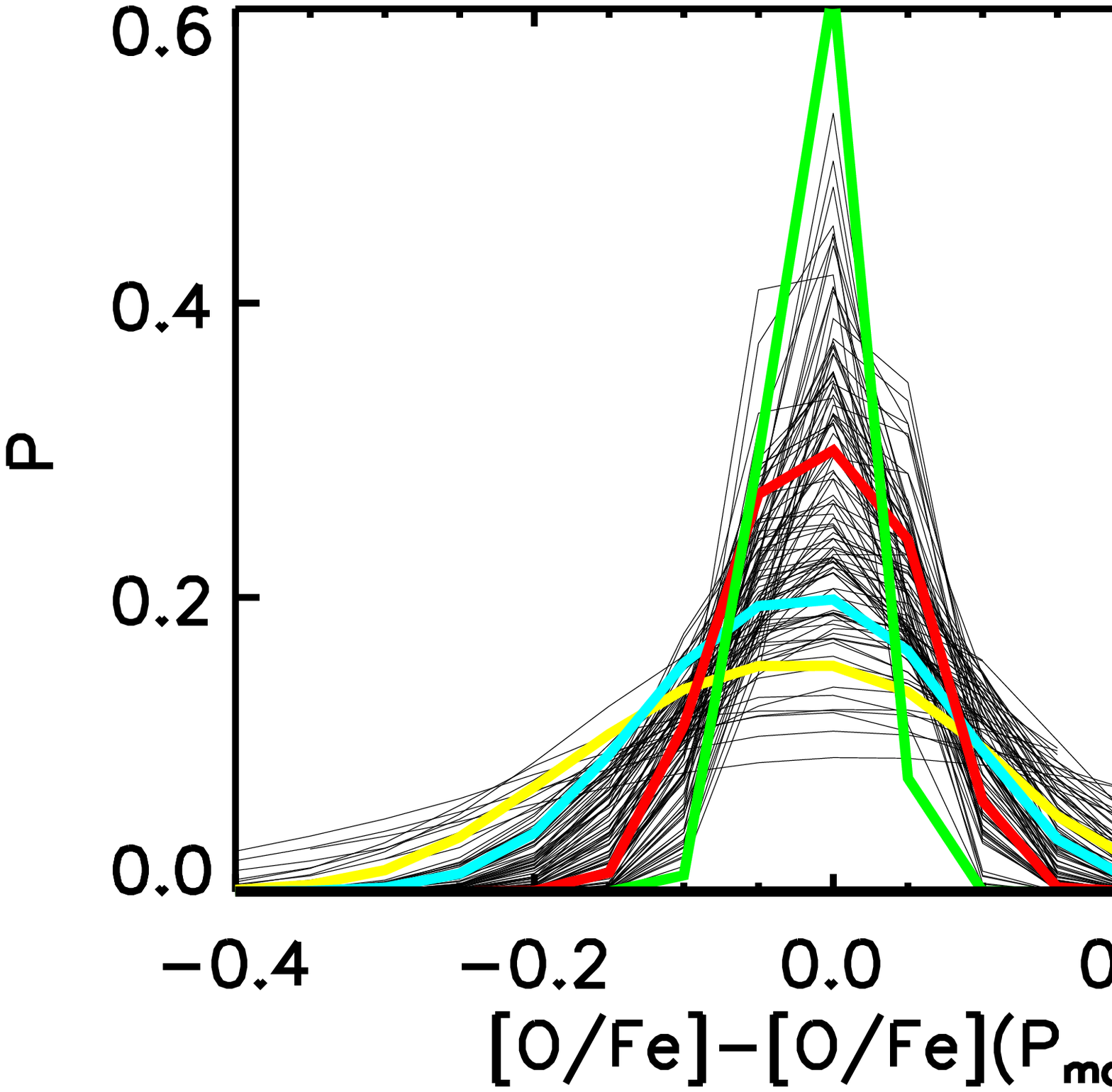}\includegraphics[scale=0.17]{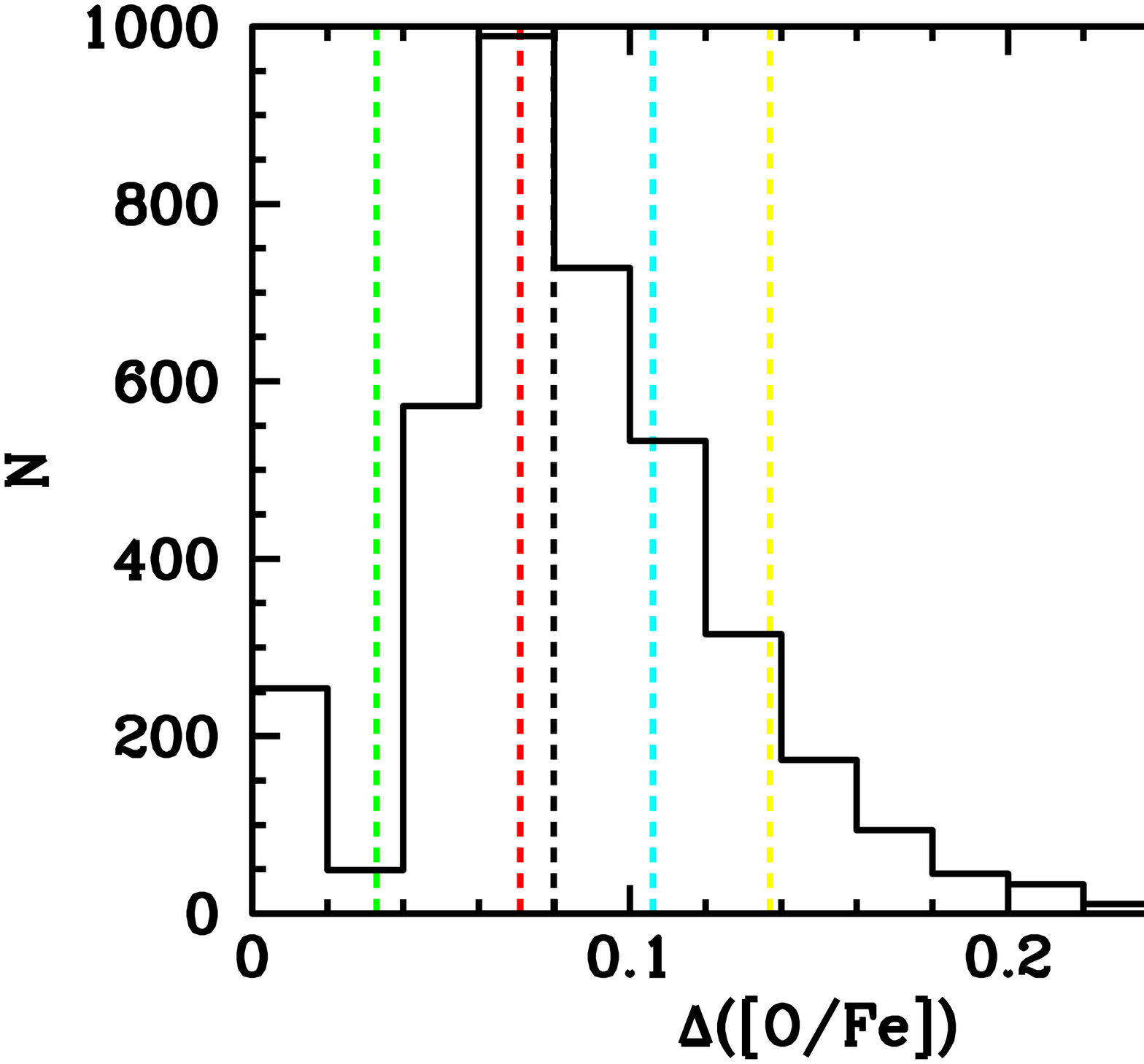}\includegraphics[scale=0.2]{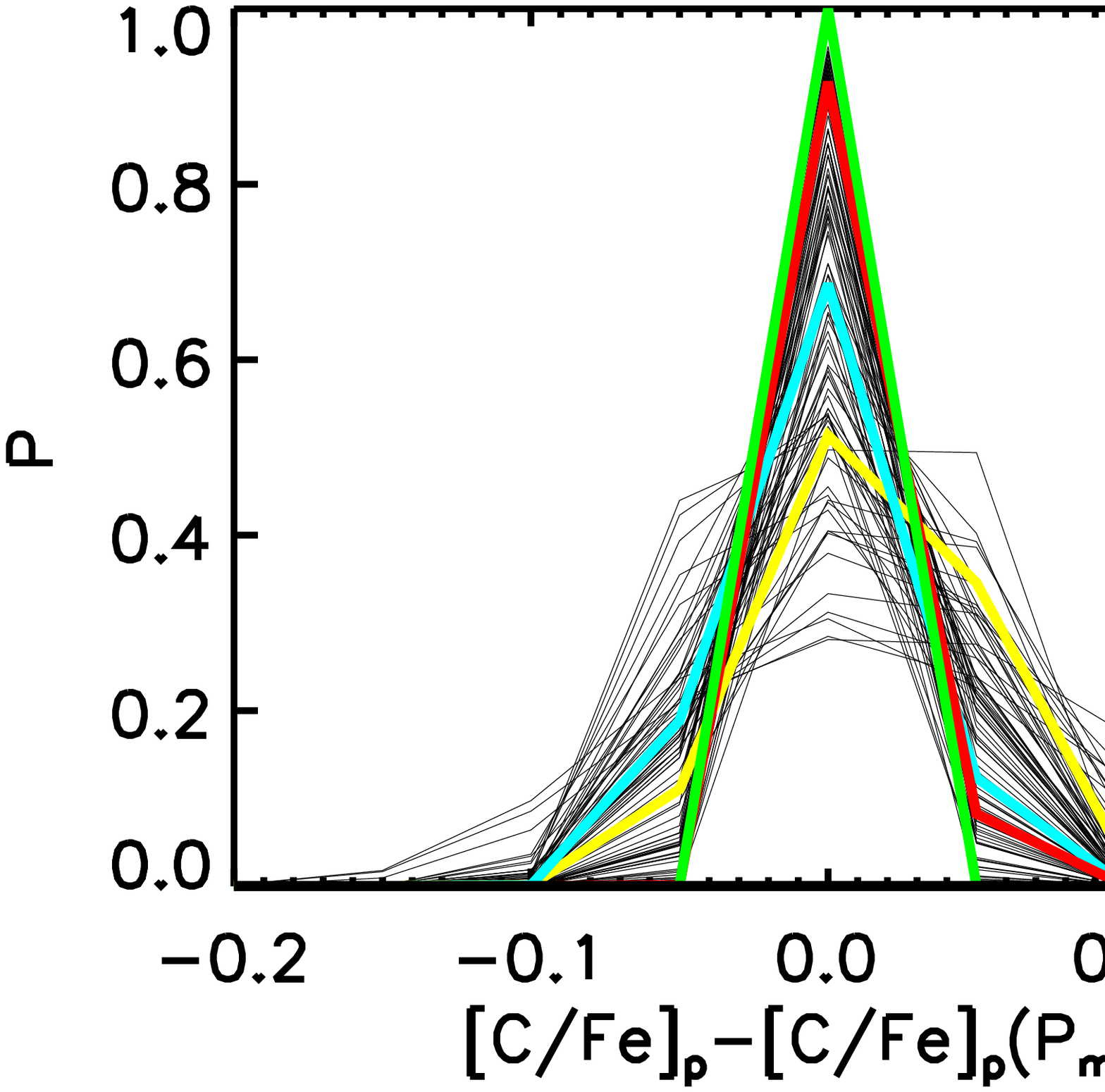}\includegraphics[scale=0.17]{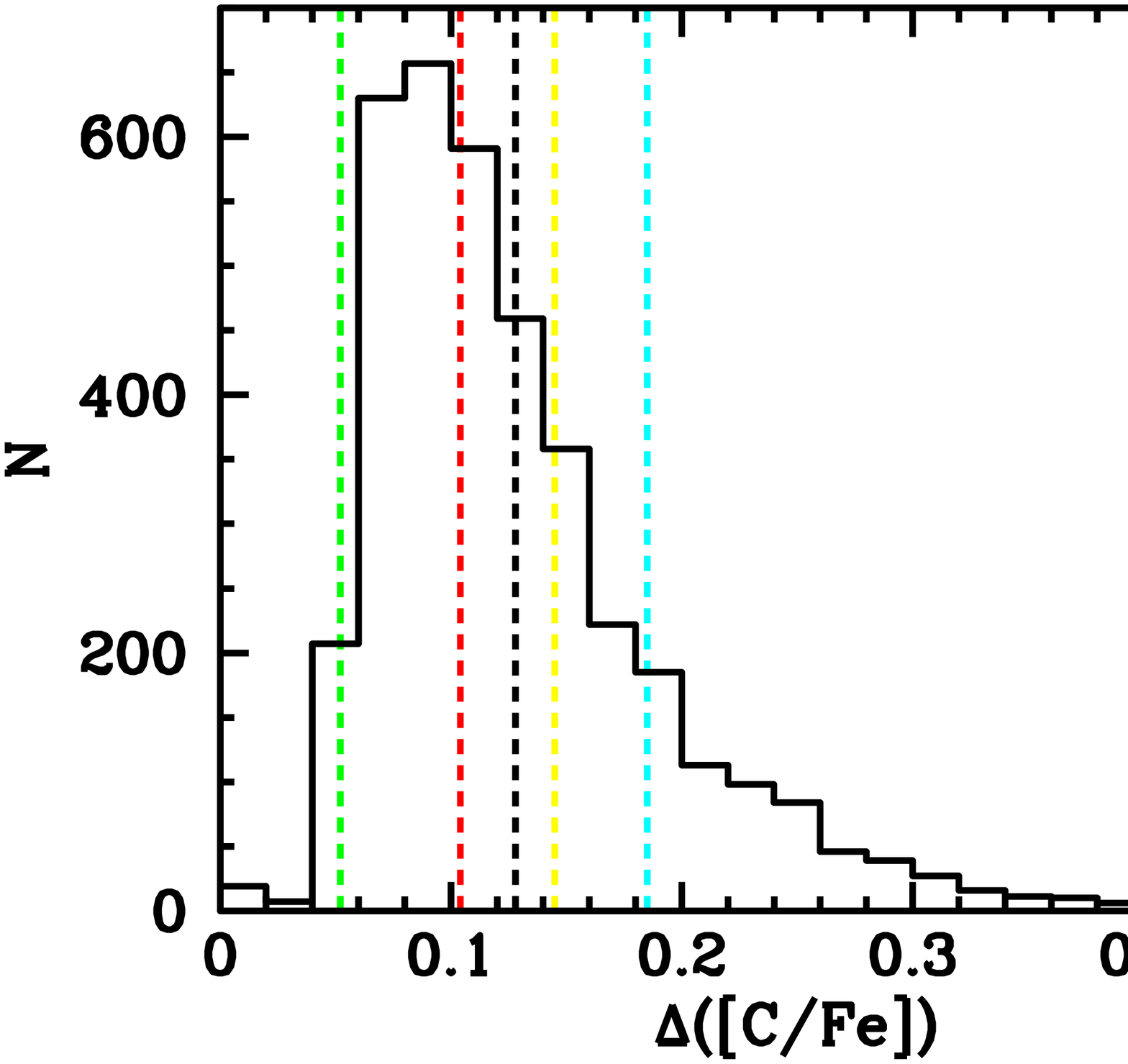}
\includegraphics[scale=0.2]{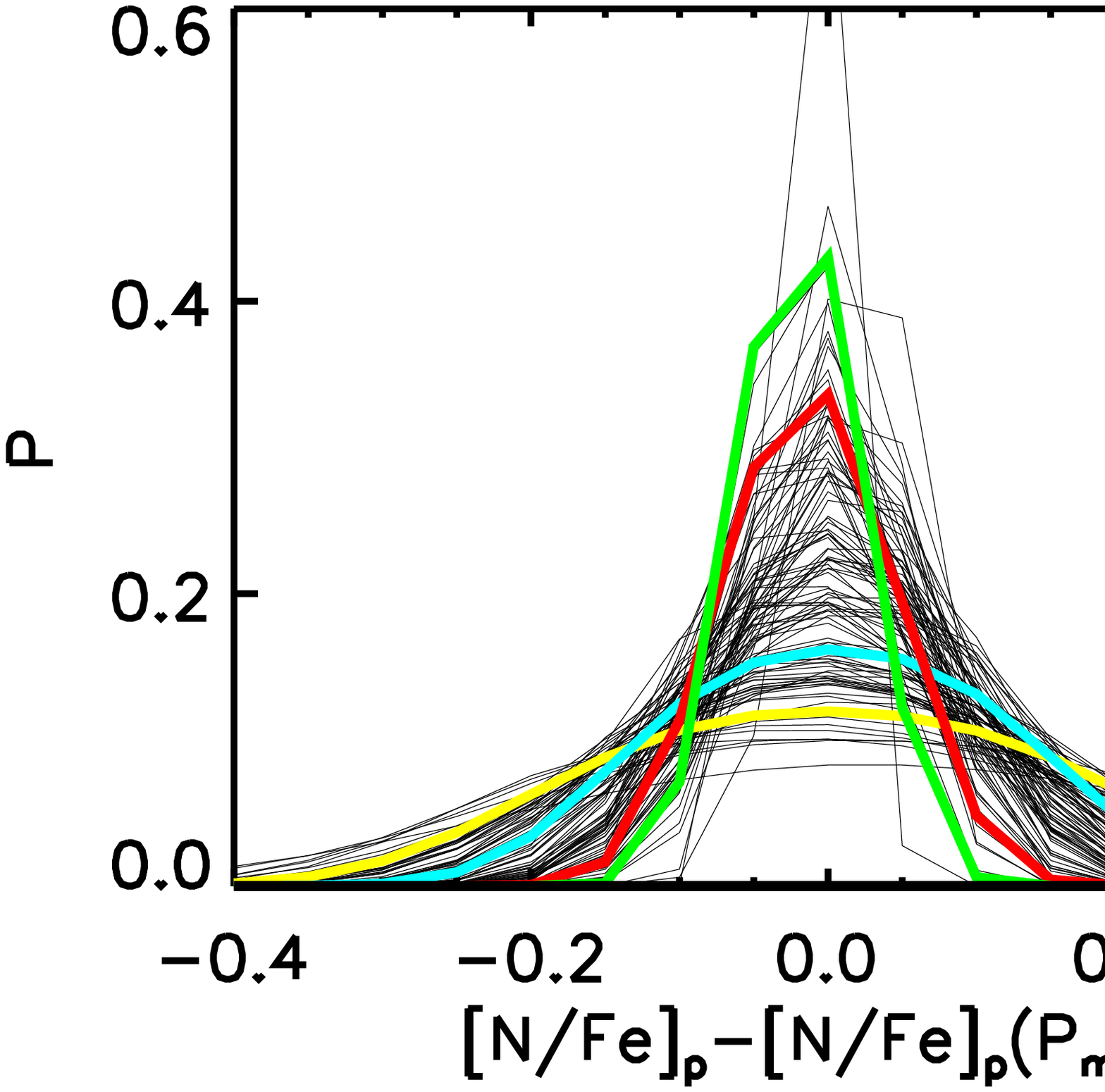}\includegraphics[scale=0.17]{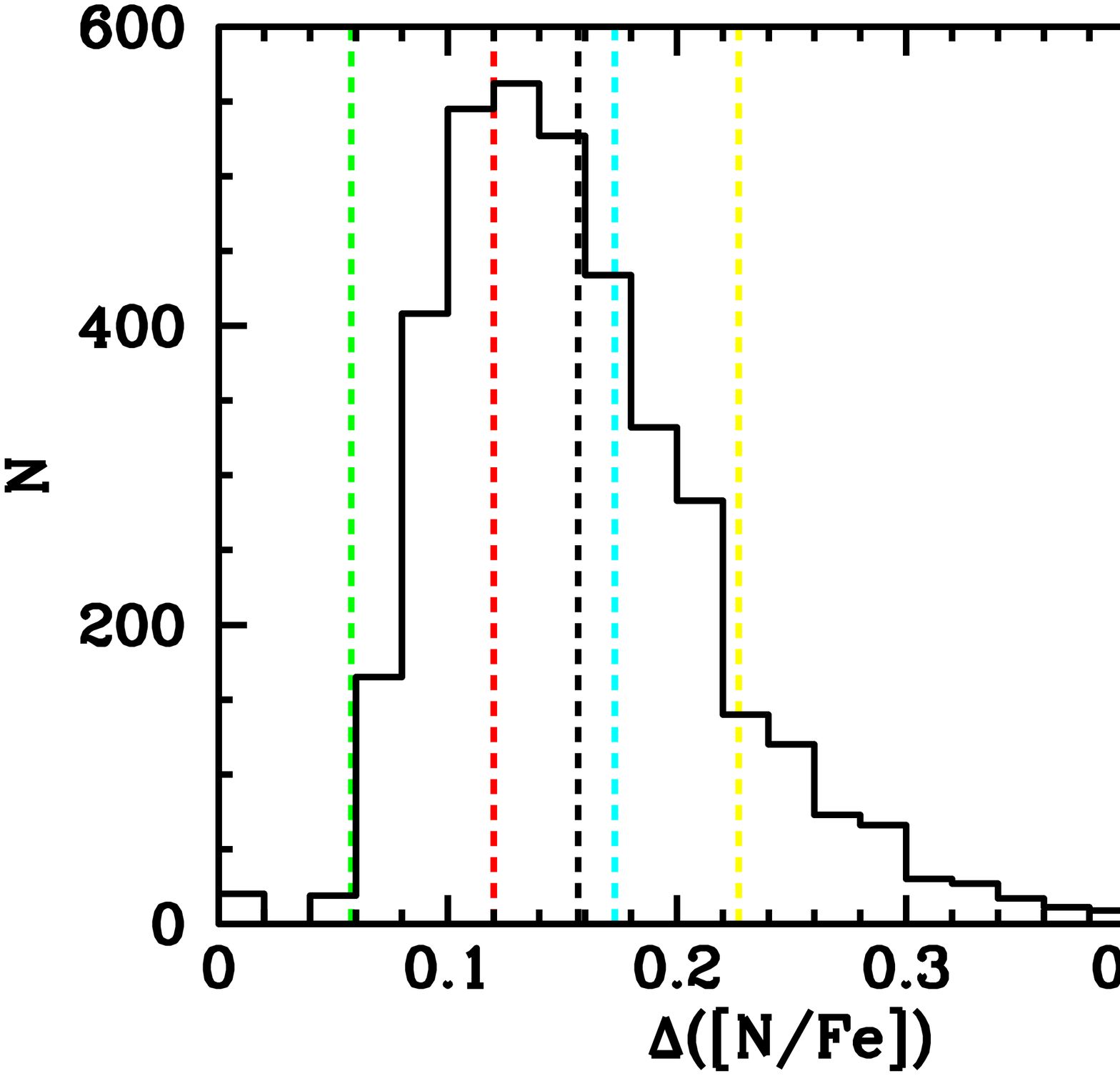}\includegraphics[scale=0.2]{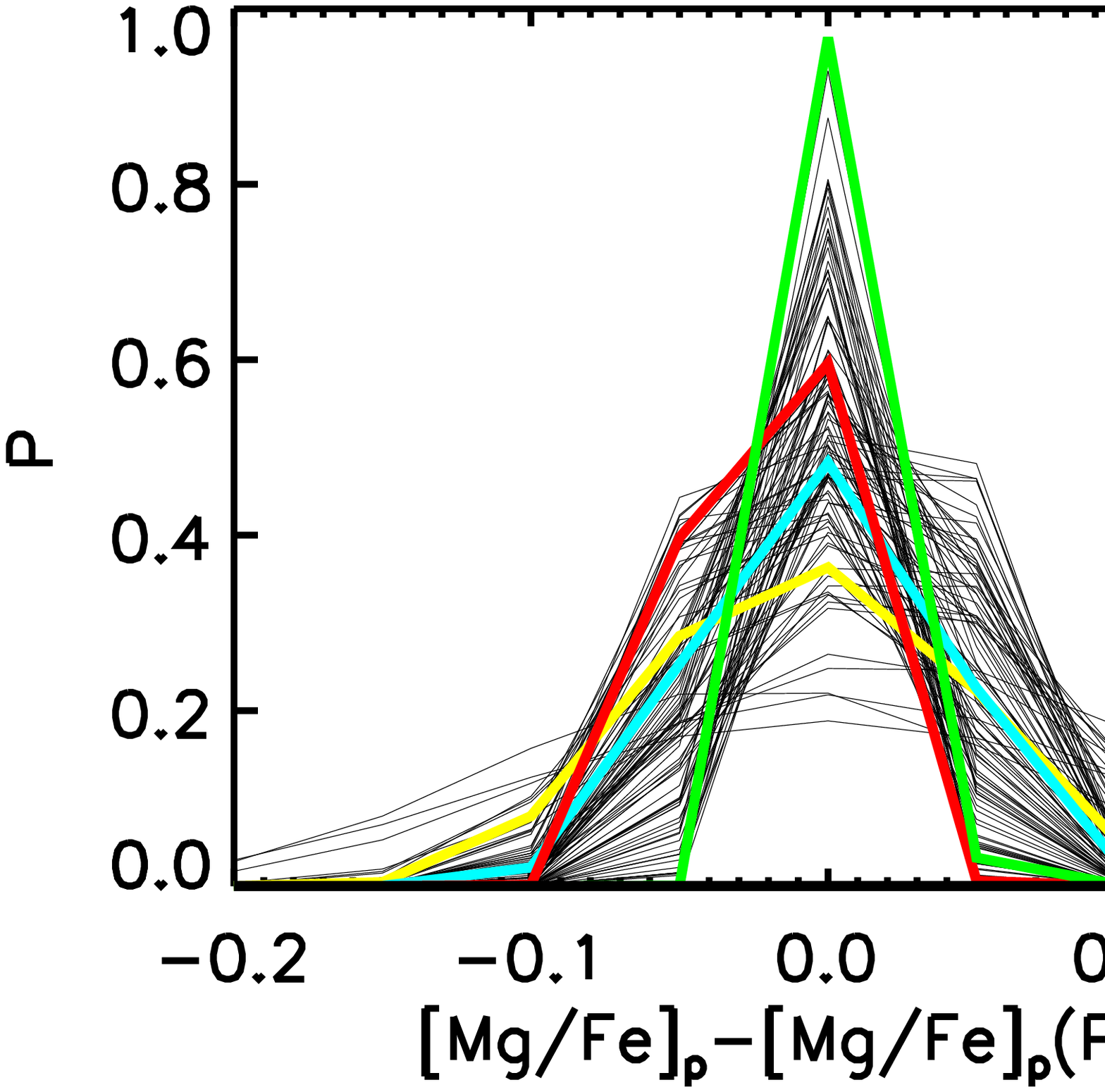}\includegraphics[scale=0.17]{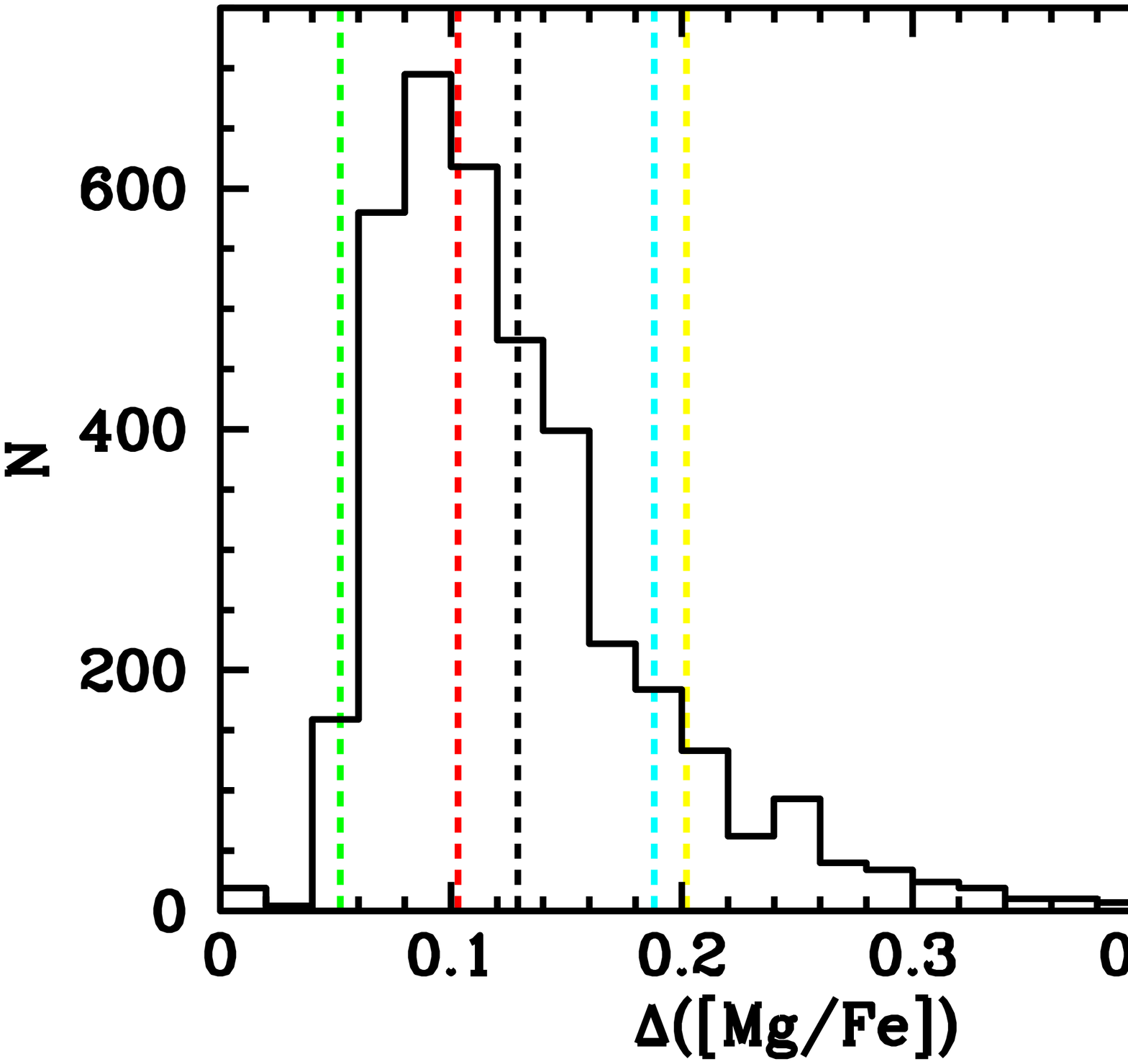}
\includegraphics[scale=0.2]{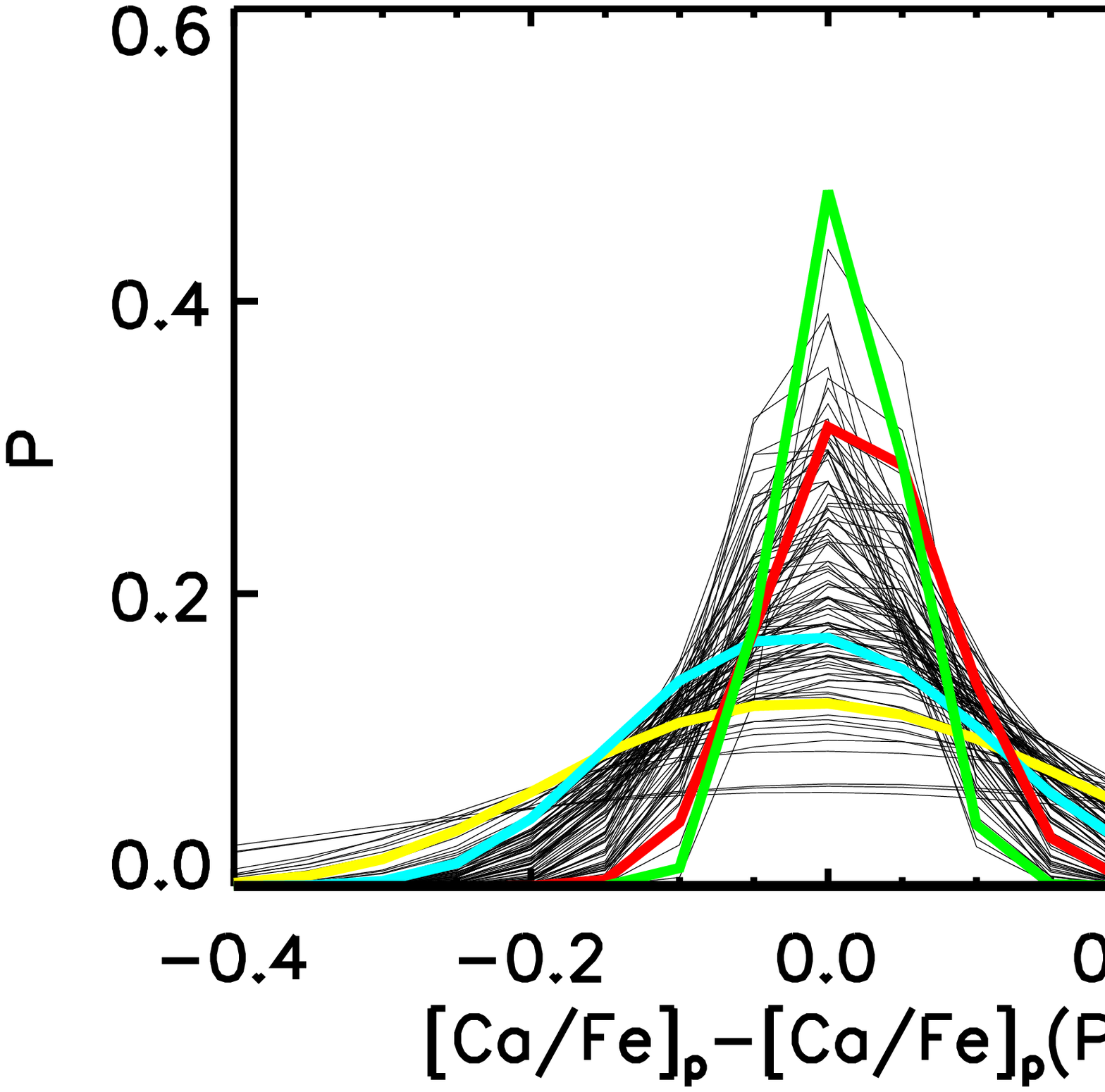}\includegraphics[scale=0.17]{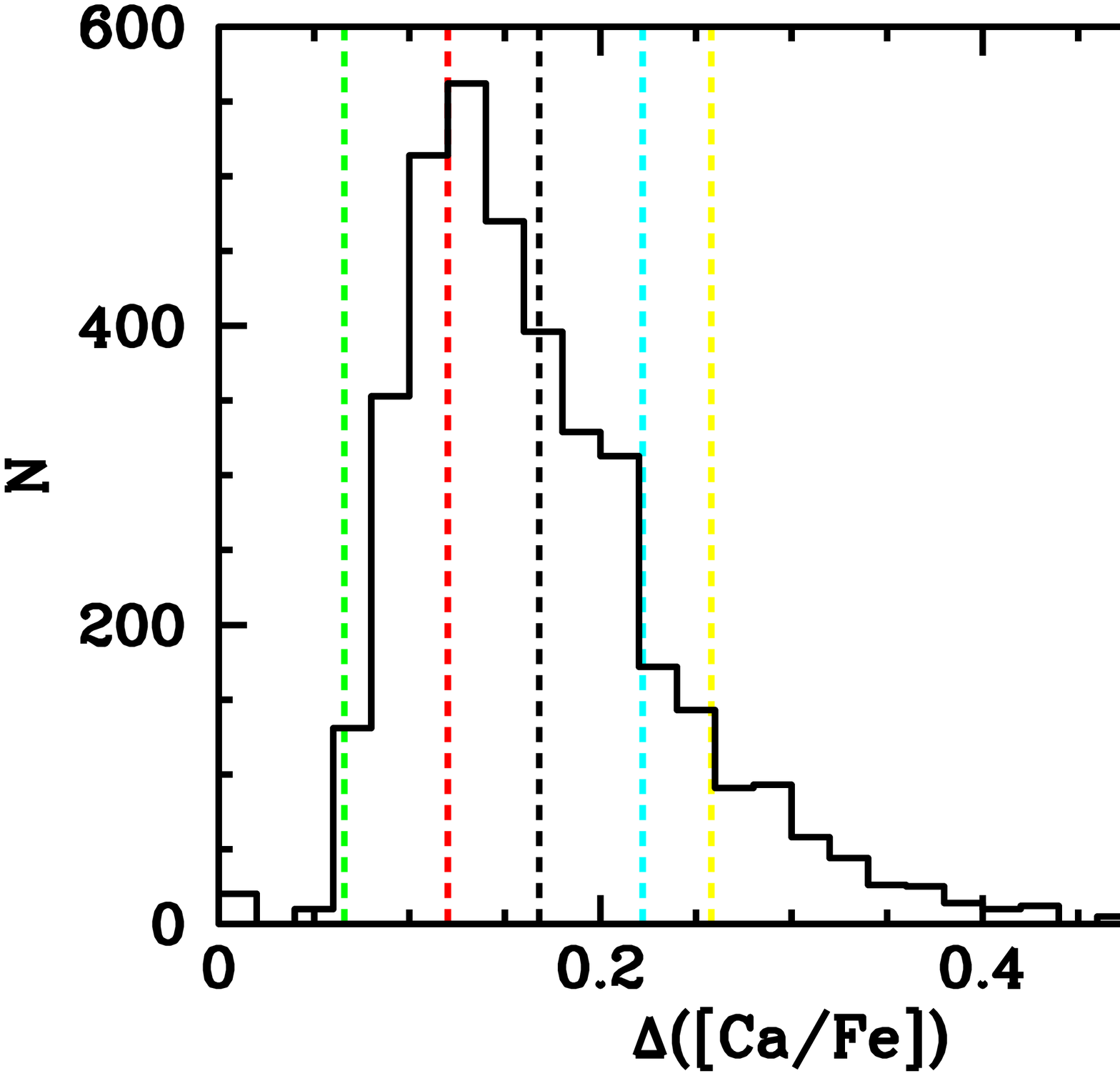}\includegraphics[scale=0.2]{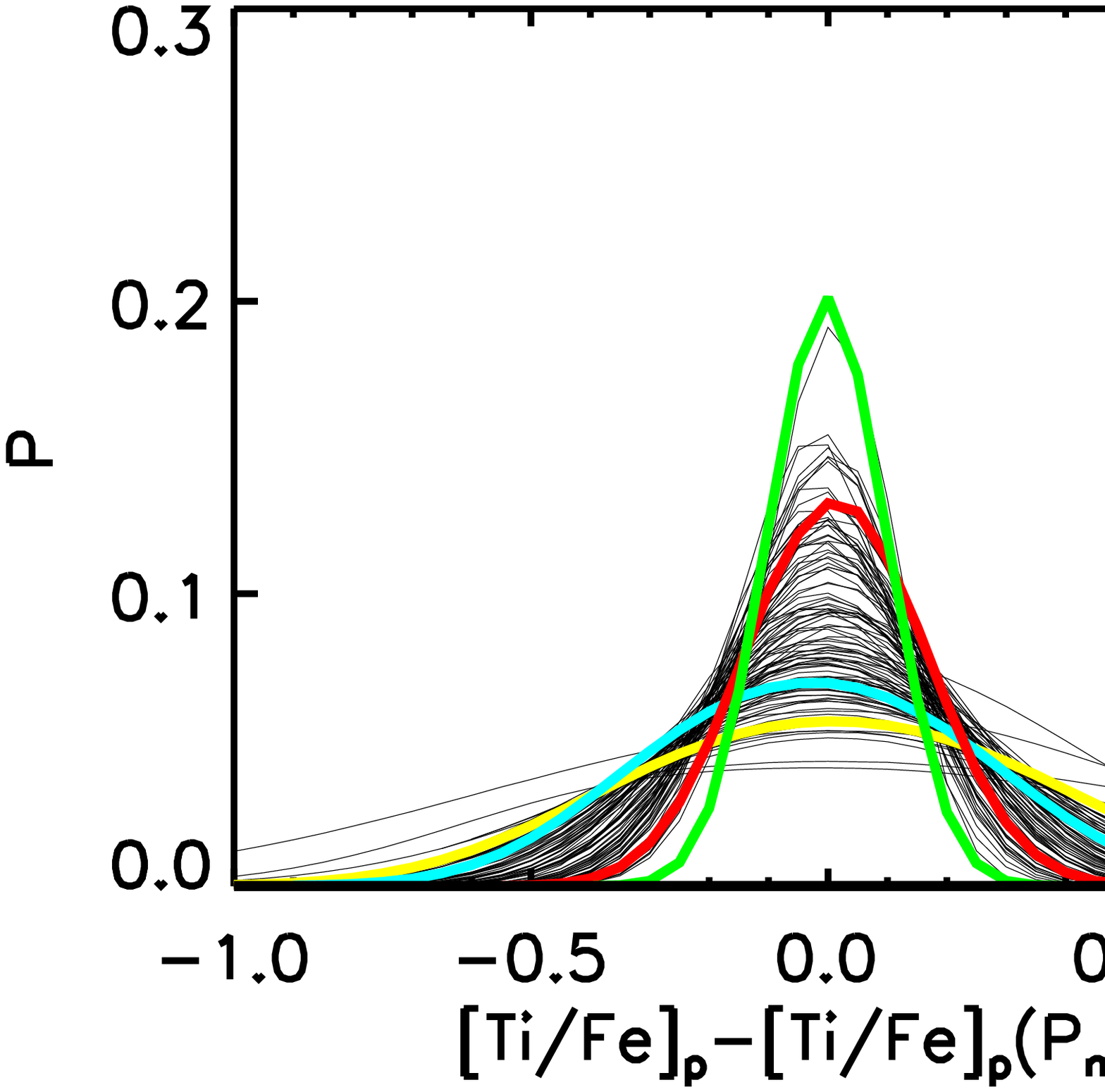}\includegraphics[scale=0.17]{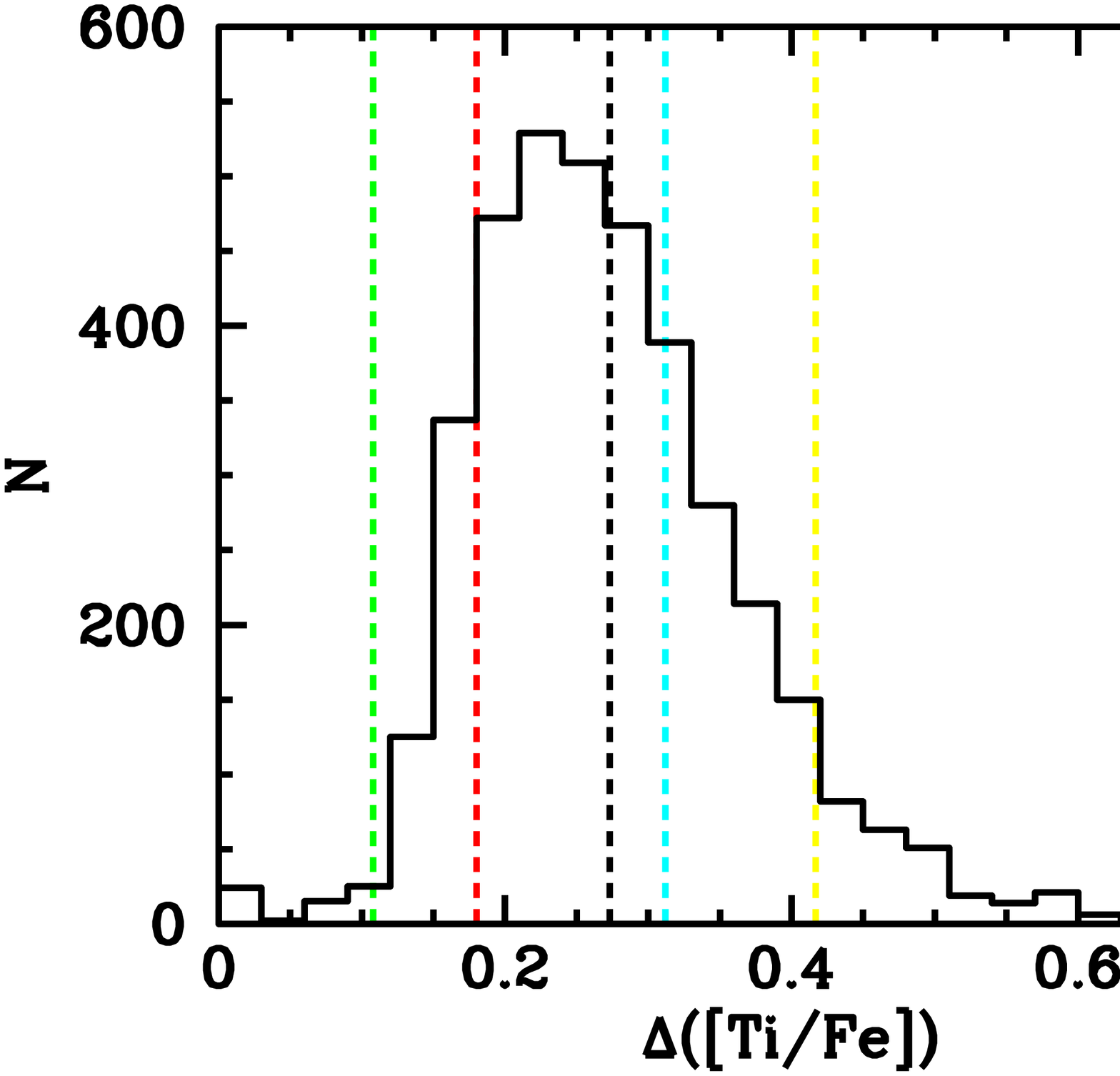}
\caption{Column 1 and 3: Probability distributions for 100 randomly selected objects (black gaussians) and 4 highlighted objects with varying parameter errors (coloured gaussians). Column 2 and 4: Distributions of parameter errors, which are estimated from the probability distributions (see text for more details), for the the full sample. Black dashed line shows the average errors, while colour dashed lines shows the errors for the highlighted probability distributions in column 1 and 3 with corresponding colours.}
\label{errdistr}
\end{figure*}

\section{Derivation of Stellar population parameters}
\label{spp}

The aim of this work is to develop a method for deriving a comprehensive set of element ratios for unresolved 
stellar populations to assess the early-type galaxy sample
described in Section~\ref{data}. This continues the work of T10 who determined 
the
stellar population parameters Age, [Z/H]
and [$\alpha$/Fe]. Including these parameters we extend the work by simultaneously deriving the
individual element abundance ratios [C/Fe], [N/Fe], [Mg/Fe], [Ca/Fe] and [Ti/Fe] using the stellar population models described in Section~\ref{models}. 
We
include the element Ti, 
one of the heaviest $\alpha$-elements before the Fe-group. 
In the following sections we introduce the method. 

\subsection{Method}
\label{method}

First we determine the traditional light-averaged stellar population parameters age, total metallicity, and [O/Fe] (representing [$\alpha$/Fe], see Section~\ref{Eratios}) from a base set of indices, using a $\chi^2$ minimization routine (described in T10 and \citet{TJM10}).
This base set of indices includes Mg~b, the Balmer indices H$_{\delta A}$, H$_{\delta F}$ and H$\beta$ and the iron indices Fe4383, Fe5270, Fe5335, Fe5406, chosen as they are well calibrated with galactic globular clusters without individual element abundance variations \citep{TJM10,TMJ10}.  
The models used at this point have step sizes of 0.024 dex for log(age) and  
0.05 dex for both [Z/H] and [O/Fe], over the ranges 0$<$age$<$20 Gyr, -2.25$<$[Z/H]$<$0.7 and -0.3$<$[O/Fe]$<$0.5. 
Only indices that are sensitive to these three parameters are included in the base set. 
Mgb is included in the base set of indices since it is 
useful in combination with Iron indices to constrain [O/Fe] \citep[e.g.][]{TMB03}.

In the subsequent steps we add in turn particular sets of indices that are sensitive to the element the abundance of which we want to determine (see step 2-6, Fig.~\ref{iter_fig}). In each step we re-run the $\chi^2$ fitting code with a new set of models to derive the abundance of this element. This new set of models is a perturbation to the solution found for the base set of indices. It is constructed by keeping the stellar population parameters age, metallicity, and O/Fe fixed and by modifying the element abundance of the element under consideration by $\pm$1 dex in steps of 0.05 dex around the base value.

A new best fit model is obtained from the resulting $\chi^2$ distribution. Then we move on to the next element. Due to the fact that several indices respond to the same elements (see Fig.~\ref{response}) an 
iterative method is needed to simultaneously derive the individual element abundance ratios, illustrated in Fig.~\ref{iter_fig}. The derivation of individual abundance ratios (Step 2-6) is iterated until the abundance ratios remain unchanged within the model grid step of 0.05 dex. A fast convergence of 3-4 iterations is generally found for this inner loop and a maximum of 5 iterations is set as an upper limit.

At the end of the sequence we re-determine the overall $\chi^2$ and re-derive the base parameters age, metallicity, and O/Fe for the new set of element ratios. At this final step we use models with the base parameters age, total metallicity, and O/Fe ratio perturbed around the previously derived values by $\Delta log(age)/\Delta [Z/H]/\Delta [O/Fe]=\pm0.1$ dex
and with step sizes of 0.02 dex for log(age) and [Z/H] and 0.05 dex for [O/Fe]. A bigger step size for [O/Fe] improves upon speed for the routine and it was found that it did not affect the final results.
All indices considered in the inner loop (step 2-6) are used at the final step together with the base set of indices. Then we go back to the second step and use the new base parameters to derive individual element abundances. This outer loop is iterated until the final $\chi^2$ stops improving by more than 1 per cent. The method converges relatively fast, again generally requiring 3-4 iterations and 5 iterations is set as an upper limit.

In more detail, the sequence of elements is as follows. The first element in the loop is carbon, for which we use the indices CN1, CN2, Ca4227, G4300, H$_{\gamma A}$, H$_{\gamma F}$, C24668, Mg$_1$, and Mg$_2$ on top of the base set. Next we drop these C-sensitive indices and proceed deriving N abundance, for which we use the N-sensitive indices CN$_1$, CN$_2$, and Ca4227. Then we move on to Mg$_1$ and Mg$_2$ for Mg abundance, Ca4227 for Ca, and finally Fe4531 for the element Ti. 



\begin{table}
\center
\caption{Discard percentage for all Lick indices}
\label{discard}
\begin{tabular}{lccc}
\hline
\bf Index & & & \bf Discard \% \\
\hline
H$_{\delta A}$  & & &  $<$ 1.0\\
H$_{\delta F}$  & & &  $<$ 1.0\\
CN$_1$  & & & $<$ 1.0 \\
CN$_2$  & & & $<$ 1.0 \\
Ca4277  & & & $<$ 1.0 \\
G4300  & & & 3.8 \\
H$_{\gamma A}$  & & & $<$ 1.0 \\
H$_{\gamma F}$  & & & $<$ 1.0 \\
Fe4383  & & & $<$ 1.0 \\
Ca4455  & & & not used \\
Fe4531  & & & $<$ 1.0 \\
C$_2$4668  & & & $<$ 1.0 \\
H$\beta$  & & & $<$ 1.0 \\
Fe5015 & & & not used \\
Mg$_1$  & & & $<$ 1.0 \\
Mg$_2$  & & & $<$ 1.0 \\
Mgb  & & & $<$ 1.0 \\
Fe5270  & & & 1.1 \\
Fe5335  & & & $<$ 1.0 \\
Fe5406  & & & $<$ 1.0 \\
Fe5709  & & & not used \\
Fe5782  & & & not used \\
NaD  & & &  not used\\
TiO$_1$  & & & not used \\
TiO$_2$  & & & not used \\
\hline
\end{tabular}
\end{table}

At step 1 and 7 we allow the procedure to discard indices with a bad $\chi^2$, explained in detailed in T10. Briefly, the probability distribution around the $\chi^2$-minimum is computed with the incomplete $\Gamma$-function for the model grid used at each step. This distribution gives the probability P that the true model has a $\chi^2$ lower than the value  obtained in the fit. If P $\le$ 0.999 the fit is considered to be unacceptable and the index with the largest $\chi^2$ is discarded. This procedure is repeated until P $>$ 0.999. In 5.2\% of the cases one index was discarded, while more than one index was discarded for 1.6\% of the 3802 objects. Compared to T10 the percentage where at least one index is discarded has been reduced from $>$30\% to 5.8\%. This is partly due to the fact that some indices used in T10 have not been considered here, e.g. Ca4455 and NaD that were the most frequently discarded indices in T10. The inclusion of individual element abundance ratios improve the fit for several indices and is thus also responsible for the lower number of discarded indices. Table~\ref{discard} gives the discard percentage for all Lick indices. G4300 is the index most frequently removed, as it was discarded in 3.8\% of the cases, while Fe5270 reached 1.1\% and the rest of the indices had a discard percentage less than 1.0\%.

\subsection{Errors}
\label{method:error}

Errors on the parameters are estimated by taking the FWHM of the probability distributions (see previous section) and converting to 1$\sigma$ errors using $\sigma$=FWHM/2.355. $\Delta\log$(age), $\Delta[Z/H]$ and $\Delta[O/Fe]$ are derived at step 1 (see previous section) since the models grids are too narrow at step 7 to reliably estimate the errors. Since the individual abundance ratios are derived as perturbations to the [O/Fe] ratio, errors on these parameters are estimated by in quadrature adding the errors on the derived perturbations $\Delta[E/Fe]_p$ to $\Delta[O/Fe]$, i.e. $\Delta[E/Fe]^2$=$\Delta[O/Fe]^2$+$\Delta[E/Fe]^2_p$. Fig.~\ref{errdistr} shows the normalised probability distributions for 100 randomly selected objects (Eiffel tower plots in 1st and 3rd column, black gaussians) of our sample (see Section~\ref{data}) together with the  distributions of the estimated errors (2nd and 4th column). Four objects are highlighted demonstrating the relationship between the sizes of the errors and the widths of the probability functions. The dashed vertical lines in column 2 and 4 show the error estimates of these objects and are coloured according to the highlighted probability distributions in column 1 and 2. Black dashed vertical lines in column 2 and 4 are average errors. The probability distributions are moved to a common position by shifting them with the parameter value at peak probability (P$_{max}$). In general we find very well defined one-peaked gaussians. For increasing errors the probability distributions clearly increase in width.

\begin{table}
\center
\caption{Linear error-S/N relationships below S/N$\sim$30 and S/N$\sim$45 for the g-band and r-band, respectively, on the form error=a$\times$S/N+b. }
\label{SNfits}
\begin{tabular}{clcc}
\hline
\bf Band & \bf Parameter & \bf a & \bf b \\
\hline
g  & $\Delta\log$(age) & -0.0077 & 0.38 \\
g  & $\Delta$[Z/H] & -0.0075 & 0.35 \\
g  & $\Delta$[O/Fe] & -0.0043 & 0.18 \\
r  & $\Delta\log$(age) & -0.0062 & 0.43 \\
r  & $\Delta$[Z/H] & -0.0047 & 0.34 \\
r  & $\Delta$[O/Fe] & -0.0038 & 0.22 \\
\hline
\end{tabular}
\end{table}

\begin{figure*}
\centering
\includegraphics[scale=0.525]{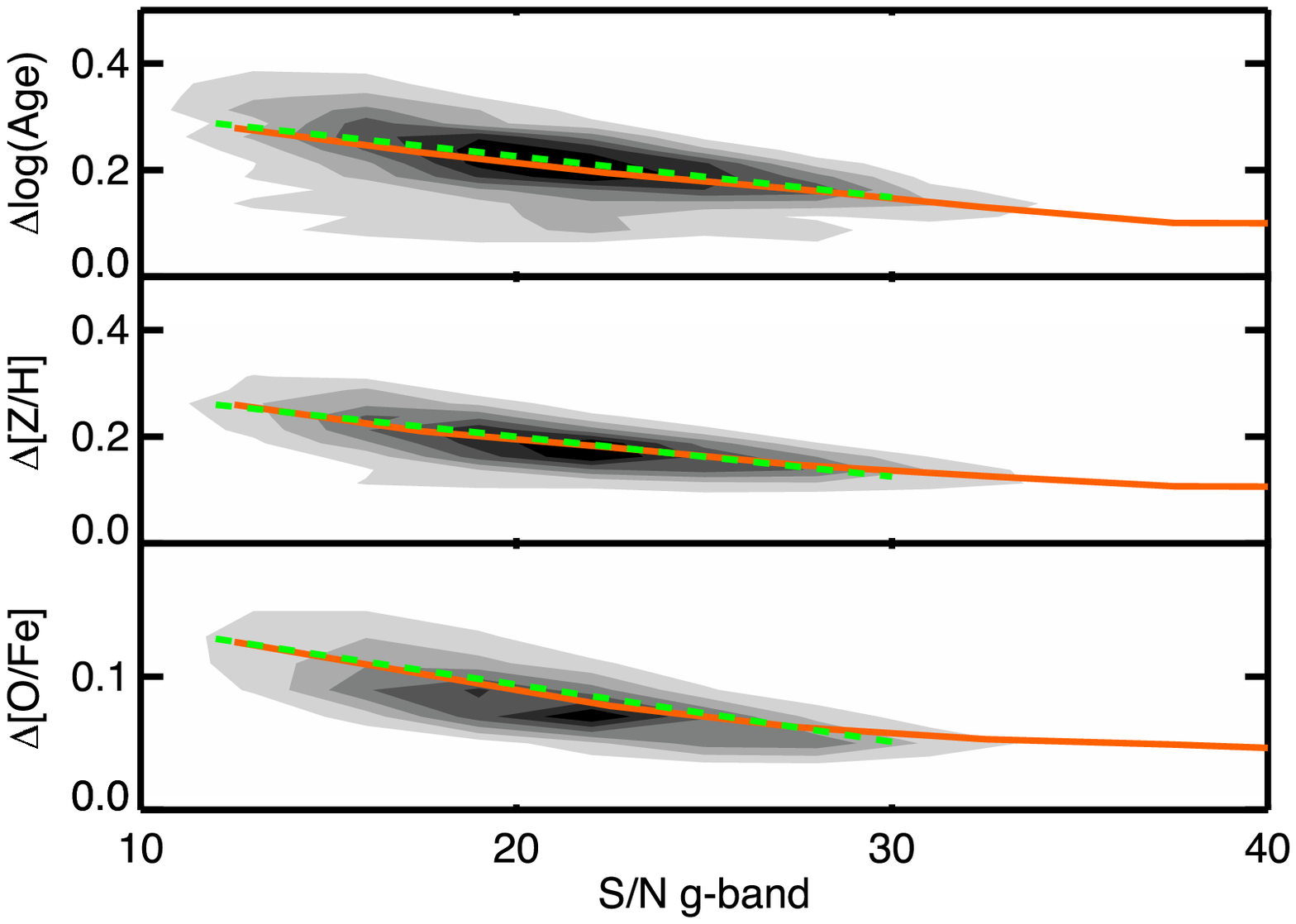}\hspace{0.2cm}\includegraphics[scale=0.525]{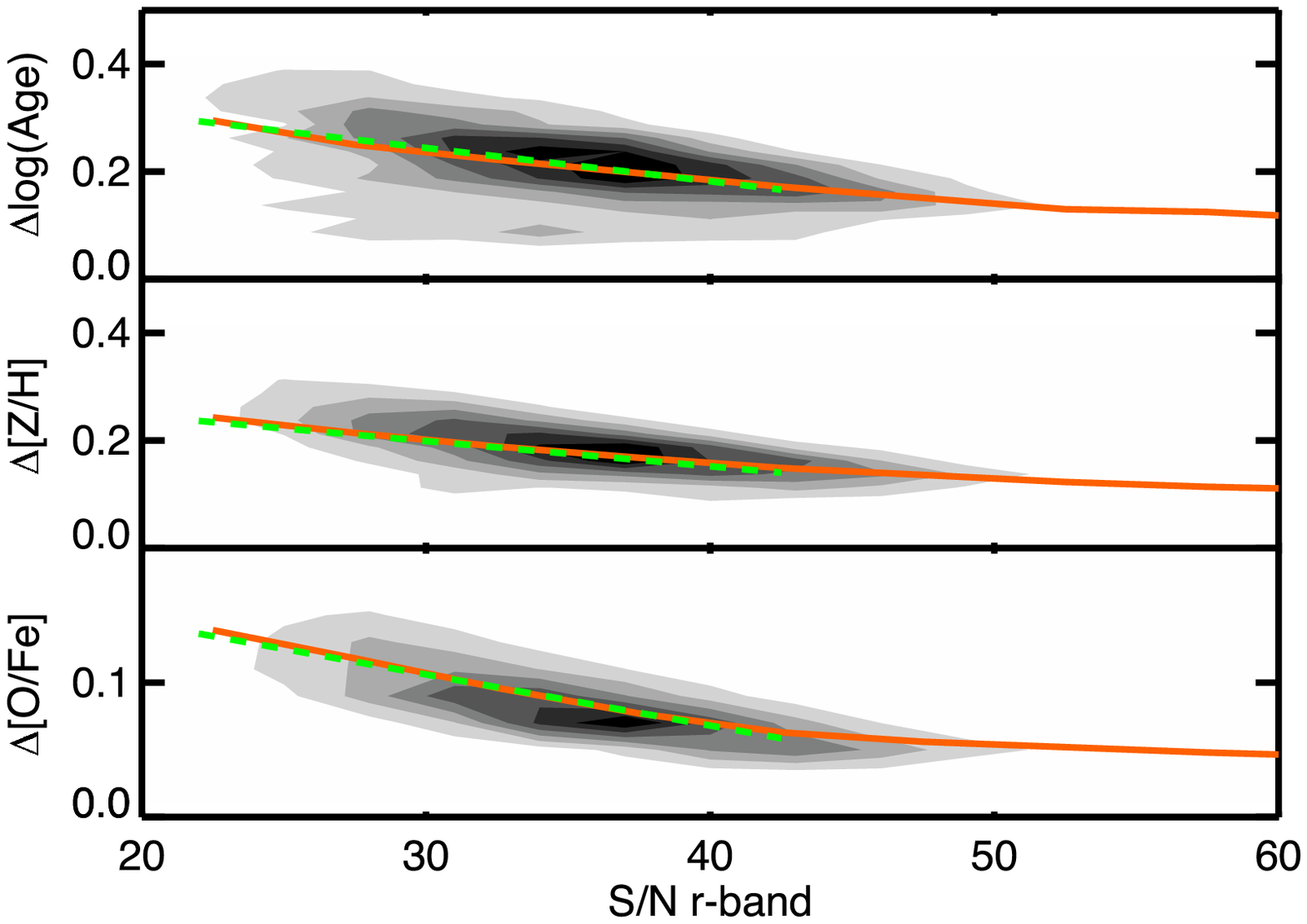}
\caption{Relationship between the estimated errors of the base parameters age, [Z/H] and [O/Fe] with S/N ratio in the g-band (left hand panel) and r-band (right hand panel) for the 3802 early-type galaxies studied (contours). Orange solid lines are mean errors in bins of S/N with a width of 5 dex 
and green dashed lines show linear relationships below a S/N of $\sim$30 and $\sim$45 for the g-band and r-band, respectively, on the form error=a$\times$S/N+b. The fit parameters a and b are given in Table~\ref{SNfits}.}
\label{errvssn}
\end{figure*}

Fig`~\ref{errvssn} shows the relationship between the estimated errors for the base parameters (age, [Z/H] and [O/Fe]) and the signal-to-noise (S/N) ratio of the galaxy spectra for the 3802 early-type galaxies studied in this work (contours). The left hand and right hand panels show the relationship for S/N in the g-band and r-band, respectively. The S/N ratios are the median values of S/N per pixel in each band as given by the SDSS spectroscopic fits-headers. 
Orange solid lines are mean errors in bins of S/N with a width of 5 dex. Below a S/N of $\sim$30 and $\sim$45 for the g-band and r-band, respectively, the error-S/N relationships are close to linear. Thus the green dashed lines show linear relationships in these regimes on the form error=a$\times$S/N+b with the values of the fit parameters a and b given in Table~\ref{SNfits}.

\subsection{Globular cluster calibration}
\label{GCcal}

The method described in the previous section is used in \citet{TJM10} to obtain estimates of stellar population parameters and element ratios from integrated light spectroscopy of galactic globular clusters. We show that the model fits to a number of indices improve considerably when various variable element ratios are considered. 
Our derived ages, metallicities and abundance ratios agree generally very well with the literature values from photometry and high-resolution spectroscopy of individual stars. 

\subsection{Comparison with methods in the literature}
\label{method_lit}

\citet{trager00a} introduced models with varying element abundance ratios by using the single stellar population models (SSPs) of \citet{wortheysingle94} together with the Lick index response functions of \citet{TB95}. They investigate the nature of the elements O and C by using different model treatments of these elements, i.e. locked to the enhanced group (see Section~\ref{Eratios}), locked the depressed group or kept fixed at solar values. 
\citet{sanchez06} follow the technique of \citet{trager00a}, but use the SSPs of \citet{vazdekis10} and modelled indices with different treatments for C, N and Mg to find the description that in general best resembles the overall behaviour of the data considered.

\citet{sanchez03} indirectly predict differences in the element abundances of C and N. They use absorption line indices known to be sensitive to variations of the abundances of these elements. If these indices show differences, while indices insensitive to abundance variations of the elements considered show no difference the stellar populations are believed to have different element abundance ratios. They also consider the SSPs of \citet{vazdekis96}, but without element abundance ratios.

\citet{clemens06} treat [C/H] as a separate variable and consequently fit modelled indices \citep[SSPs from][]{annibali07} with the free parameters age, [Z/H], [$\alpha$/Fe] and [C/H] to data. \citet{kelson06} consider modelled indices (TMB/K) with the free parameters age, [Z/H], [$\alpha$/Fe] and [$\alpha$/C], [$\alpha$/N] and [$\alpha$/Ca], but not all parameters simultaneously.

\begin{figure*}
\centering
\includegraphics[angle=90,scale=0.26]{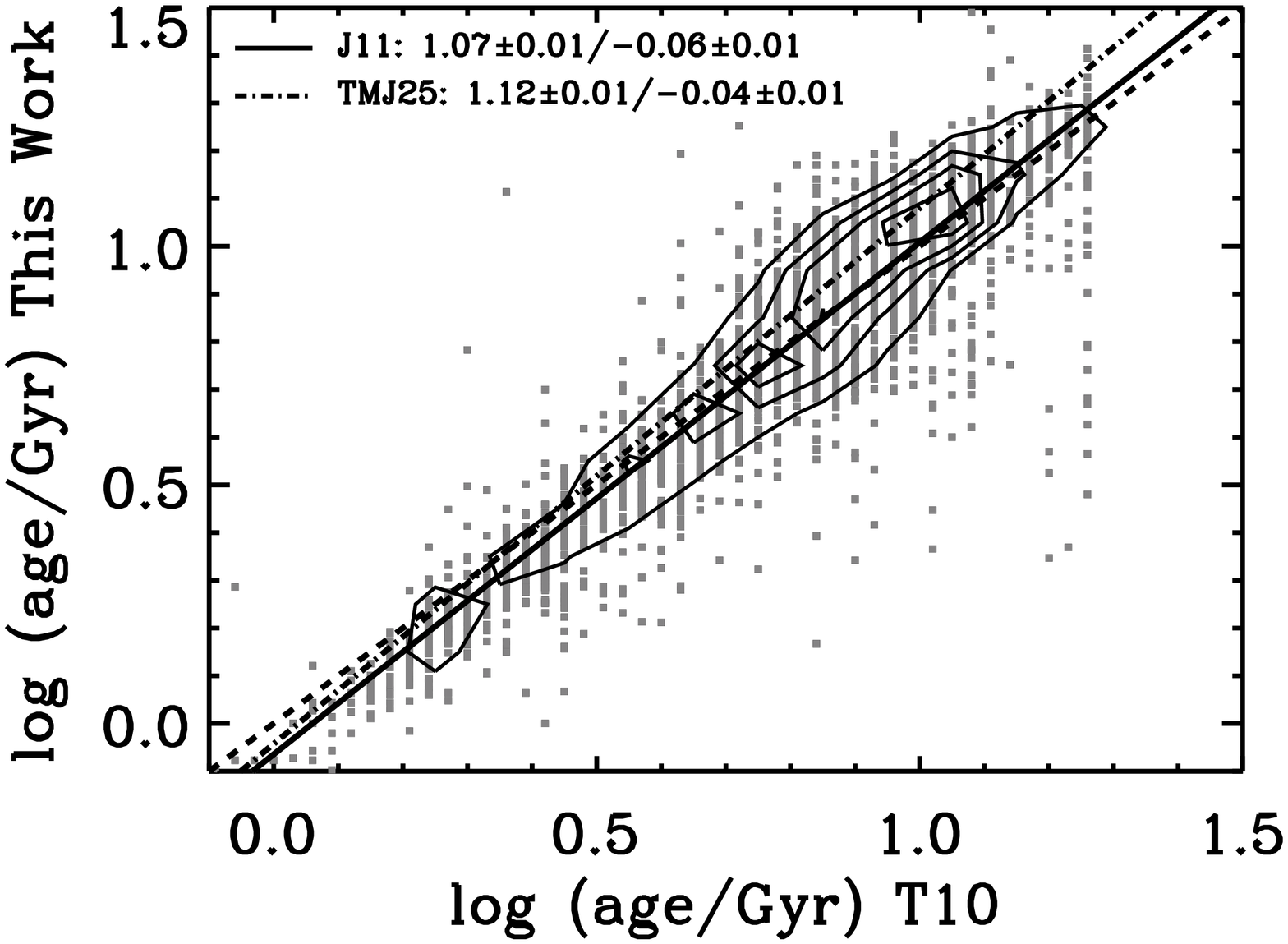}\includegraphics[angle=90,scale=0.26]{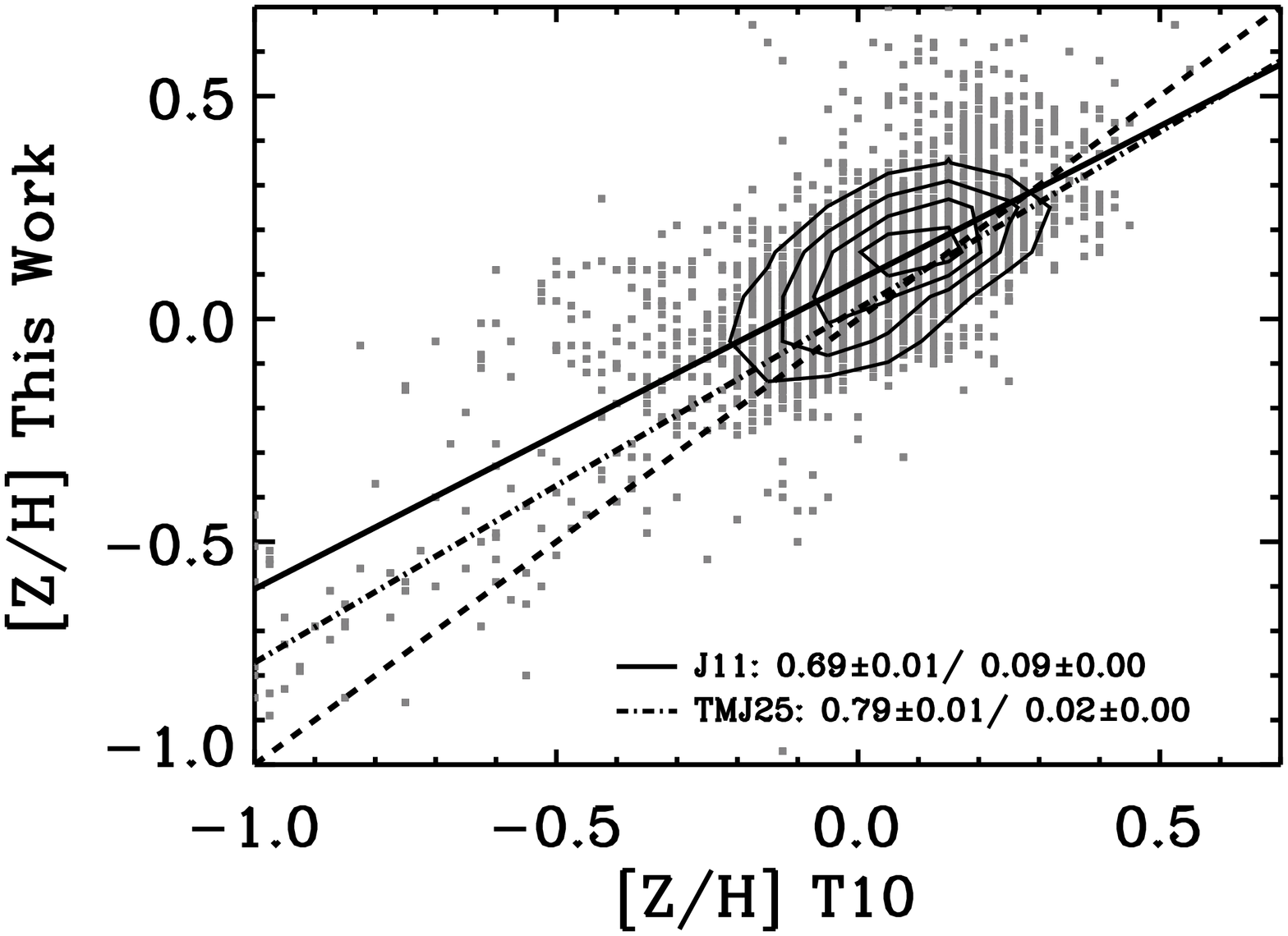}\hspace{0.12cm}\includegraphics[angle=90,scale=0.26]{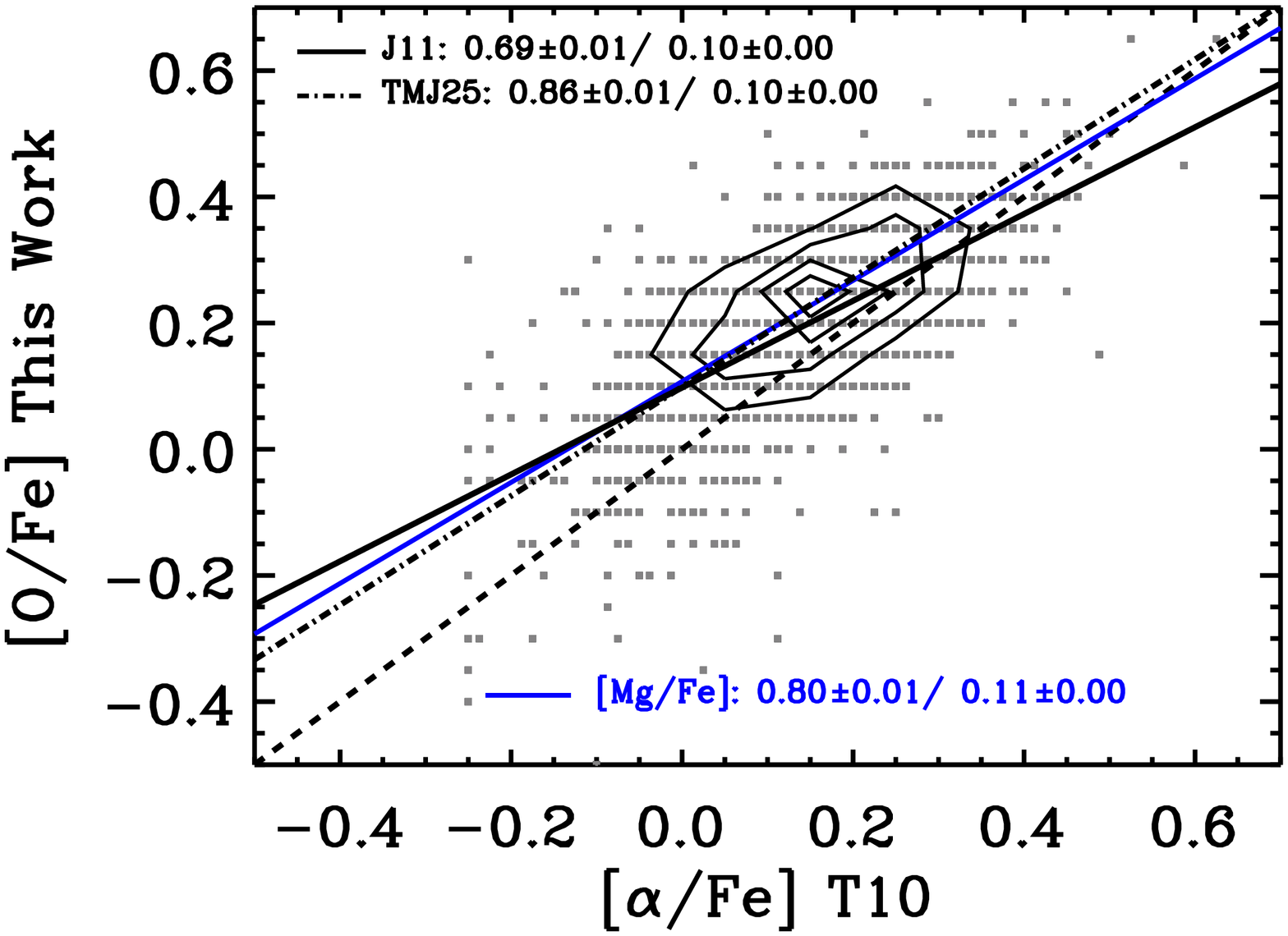}
\caption{Comparison between the derived parameters of this work (y-axes) and T10 (x-axes) for age (left hand panel), total metallicity (middle panel) and [O/Fe] (right hand panel). For the latter we compare to the [$\alpha$/Fe] ratios from T10 as it corresponds to [O/Fe] of this work (see Section~\ref{signals}). Contours show the density of data points and solid lines are least-square fits. These results are derived using the method presented in this work (J11 setup, see Section~\ref{spp}). For the TMJ models  (see Section~\ref{models}) we have also derived the stellar population parameters with the same setup as used in T10. Least-square fit representing the relationship between these results and T10 are shown as dot-dashed lines (T10 setup). Fit parameters are given by the labels and dashed lines are 1-to-1 relationships
}
\label{T10_fig}
\end{figure*}

The first authors to develop a method that simultaneously consider a full grid of element abundance ratios were \citet{graves08}, based on the models of \citet{schiavon07}. 
They start by deriving a fiducial Age and [Fe/H] 
using the index H$\beta$ in combination with $<$Fe$>$. H$\beta$ is then exchanged with an index sensitive to a 
specific element. This element is then either enhanced or depressed until the fiducial Age and [Fe/H] is matched. 

The method developed in this work is based on a different philosophy. 
In our method a $\chi^2$ routine finds the model with an enhancement of a 
specific element that best fits the data keeping a set of base (fiducial) parameters (Age, [Z/H] and [O/Fe]) fixed. Initially deriving [O/Fe] together with age and [Z/H] gives more freedom to the choice of age-sensitive
indices. Thus we can use the Fe-sensitive higher order Balmer indices (H$\delta_A$, H$\delta_F$, H$\gamma_A$ and H$\gamma_F$) besides 
H$\beta$ for initially constraining age.
A major difference is that we allow for iteration between the base model and the solution with varying element ratios. For the sample described in Section~\ref{data} the average difference between the initially derived base parameters and the final values are 0.11 dex for log(age), 0.09 dex for [Z/H] and 0.04 dex for [O/Fe]. 

For each element we use all indices that are sensitive to its abundance (see Fig.~\ref{response}, Section~\ref{signals}), 
while \citet{graves08} use the most sensitive index only for each element. Using all indices extracts all information available
and protects against anomalies in individual indices, i.e. noise and emission line fill affecting the absorption features. 
Still, more sensitive indices will have a greater weight than less sensitive indices.
\citet{graves08} set a fixed value of [O/Fe] and simultaneously derive [C/Fe], [N/Fe], [Mg/Fe] and [Ca/Fe]. We extend this and further include [Ti/Fe], while we also trace [O/Fe] inferred from [$\alpha$/Fe] (see Section~\ref{signals}).






\section{results}
\label{results}

In the following we present stellar population parameters and element abundance ratios as functions of the stellar velocity dispersion measurements derived in T10 (see Section~\ref{data}). 
Since this is a continuation of the work of T10 we compare the results of the base parameters (age, Z/H and O/Fe) with their results, while the new element ratios considered are compared with the full literature. 


\subsection{A direct comparison with T10}
\label{T10Sec}

In Fig.~\ref{T10_fig} we compare the ages, [O/Fe] ratios and total metallicities derived using the method presented in Section~\ref{spp} (J11 setup)
to the corresponding parameters from T10 (where [$\alpha$/Fe] corresponds to [O/Fe] of this work, see Section~\ref{signals}). Contours show the density of data points and dashed lines are 1-to-1 relationships. Solid lines are least-square fits with fit parameters given by the labels. We find a very good agreement for the derived ages and generally good agreements for total metallicity and [O/Fe]. 
For the latter two this work shows overall higher values and the differences increase towards lower parameter values. Still, the average differences are small within the parameter range containing the majority of data points ($<$0.2 dex for [Z/H] and $<$0.1 dex for [O/Fe]) . 

Using the TMJ models (see Section~\ref{models}) we have also derived the stellar population parameters age, [$\alpha$/Fe] and total metallicity using the same setup as used in T10 (TMJ25 setup), i.e. using all 25 Lick indices and not considering element ratios beyond [$\alpha$/Fe]. With these results we can disentangle the effect on the derived parameters from using both new models and accounting for individual element ratios. 
A direct comparison between the TMJ25 and T10 is also included in Fig.~\ref{T10_fig} (least-square fit dash-dotted lines). Again we find a good agreement between the derived ages. In the old age regime the TMJ25 ages are somewhat higher ($\sim$0.1 dex). Hence in this regime the TMJ models seem to result in slightly older ages compared to the TMB/K models, while accounting for element ratios results in younger ages. The TMJ25 and T10 metallicities are very similar within the parameter range containing the majority of data points (contours). Hence, the metallicities do not change considerably when using the new models. Instead, the metallicities are slightly lower if individual element ratios are not taken into account.

\begin{figure*}
\centering
\includegraphics[angle=90,scale=0.35]{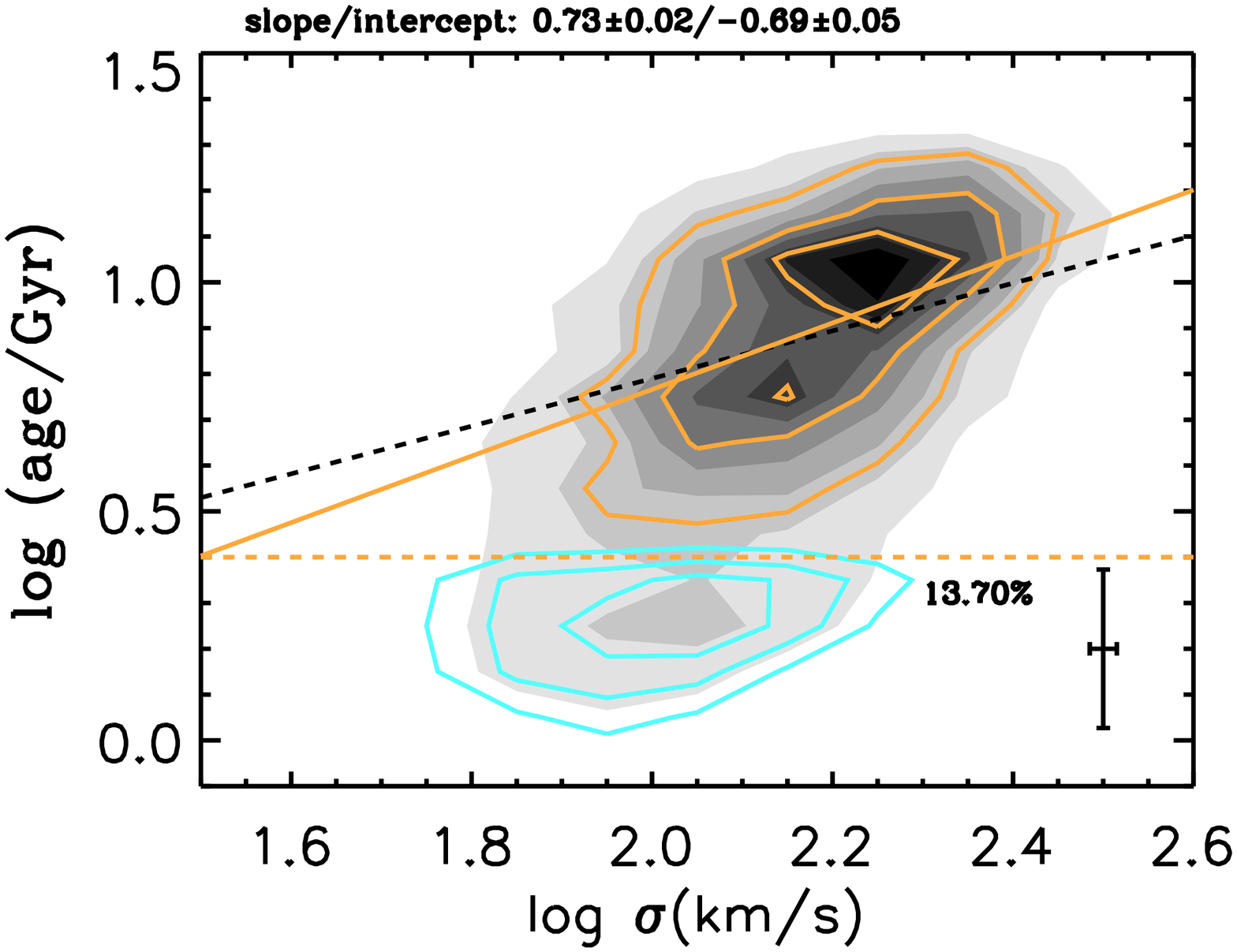}\includegraphics[scale=0.3]{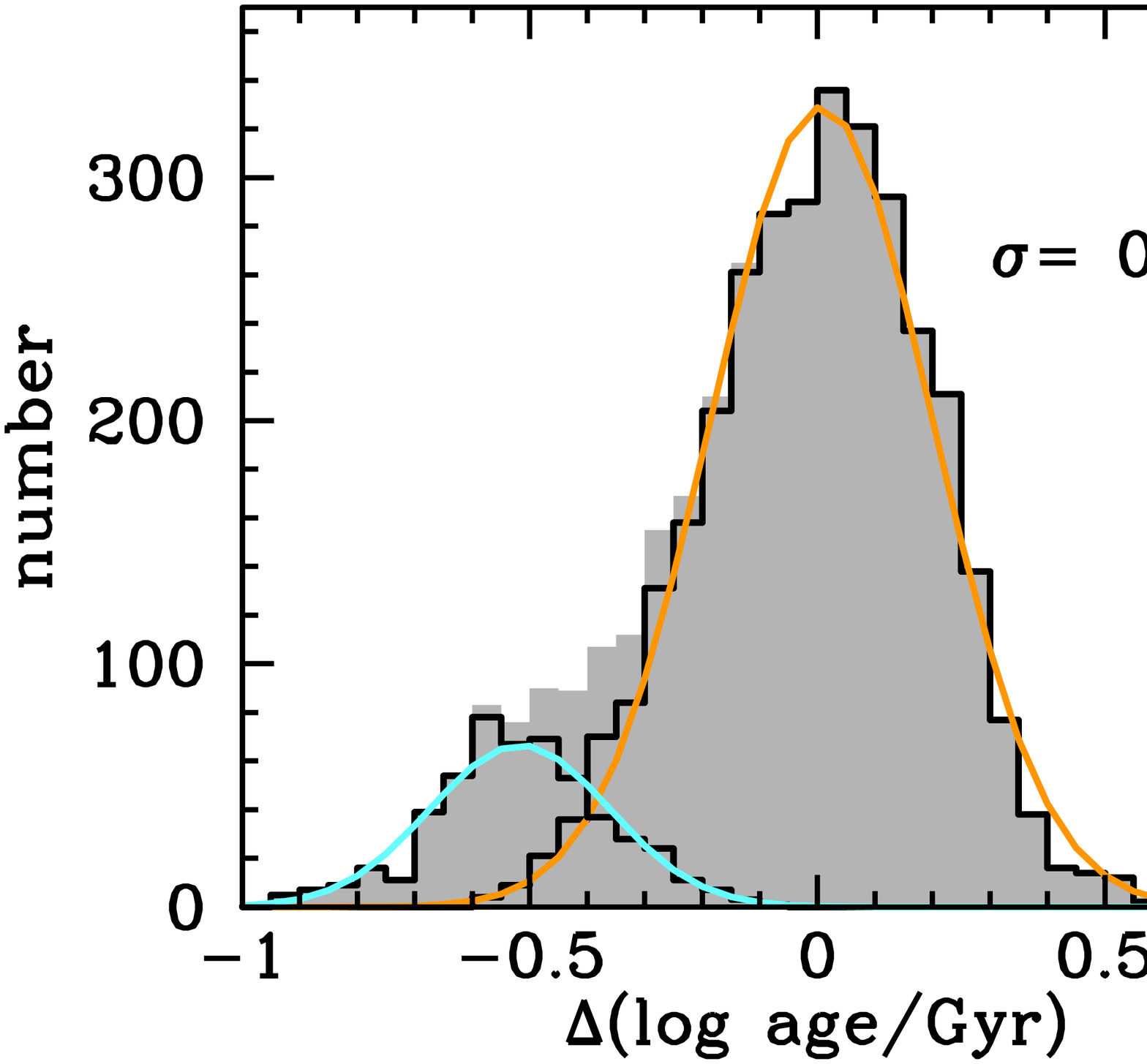}
\caption{Left hand panel: shows the relationship between the derived light-average ages and velocity dispersion. The dashed orange line indicate the separation of an old red sequence population (orange contours) from a rejuvenated population (cyan contours) with light-average ages smaller than 2.5 Gyr \citep{thomas10}. The fraction of rejuvenated galaxies is indicated below the separation line. The whole sample is shown as grey-scaled filled contours. The orange solid line is a least-square fit to the red sequence population and the parameters of the fit are given at the top of the panel. The dashed black line is the analogous fit from T10 for comparison.  Median 1$\sigma$-errors are shown in the lower right corner. Right hand panel: shows the distribution of deviation in age from the least-square fit to the red sequence population with the same colour coding as in the left hand panel. The standard deviation of the gaussian fitted to the distribution of the red sequence population (orange line) is indicated in the upper right corner.}
\label{age}
\end{figure*}

For [$\alpha$/Fe] TMJ25 produces overall higher element ratios by $\sim$0.1 dex compared to T10. In this case the choice of models matters as well as accounting for individual element ratios. In the right hand panel we have also included a comparison between the J11 [Mg/Fe] ratios and the [$\alpha$/Fe] ratios from T10 (least-square fit blue line). In fact the [Mg/Fe] ratios agree very well with the TMJ25 [$\alpha$/Fe] ratios as the blue and dash-dotted are very similar. Hence the [$\alpha$/Fe] ratios derived without taking individual element ratios into account reflect [Mg/Fe] rather than [O/Fe]. This is probably due to the fact that the Lick indices are more sensitive to variations in the abundance of Mg than O (see Section~\ref{signals}).

The TMJ25 setup consider all Lick indices and the J11 setup a selection of 18 indices best calibrated with galactic globular clusters (see Section~\ref{spp}). The difference seen between these two setups may therefore arise from the choice of indices instead of accounting for element ratios. To evaluate this we also derived ages, [$\alpha$/Fe] ratios and total metallicities using the TMJ models without element ratios, but for the 18 indices considered in this work. These result are very similar to that of TMJ25, such that the different choices of indices do not matter significantly.



\subsection{Ages}
\label{ageSec}

The relationship between luminosity-weighted age and velocity dispersion is presented in the left hand panel of Fig~\ref{age}. The full sample is shown with grey-scaled, filled contours. We reproduce the result from T10 of having a bimodal distribution of ages in analogy to a red sequence population (log(age)$>$0.4, orange contours) and a rejuvenated blue cloud population (log(age)$<$0.4, cyan contours) of low mass early-type galaxies, identified thanks to the purely visual classification of the MOSES sample (see Section~\ref{data}). 
The fraction of rejuvenated galaxies is 13.70$\%$, close the corresponding fraction of 10.15$\%$ found in T10. The rejuvenated population is further discussed in the following sections and in more detail in Section~\ref{rejuvenation}. Fig.~\ref{age} also shows a $\sigma$-clipped (3$\sigma$ limit) linear least-square fit to the red sequence population (orange solid line), along with the analogous from T10 for comparison (dashed black line). 
We find very similar trends, but with a slightly steeper slope, as in T10, i.e. increasing ages with increasing velocity dispersion. 
 


The distribution around the fit to the red sequence population is shown in the right hand panel of Fig.~\ref{age}, for the whole sample (grey histogram), red sequence population and rejuvenated population (black open histograms). Gaussians fitted to the rejuvenated and red sequence distributions are shown with the same colour coding as in the left hand panel, with the standard deviation of the latter given by the label. The rejuvenated population shows a distinct peak, offset from the red sequence population to younger ages by $\sim$0.55 dex.

\subsection{[Z/H]}
\label{ZSec}

\begin{figure*}
\includegraphics[angle=90,scale=0.35]{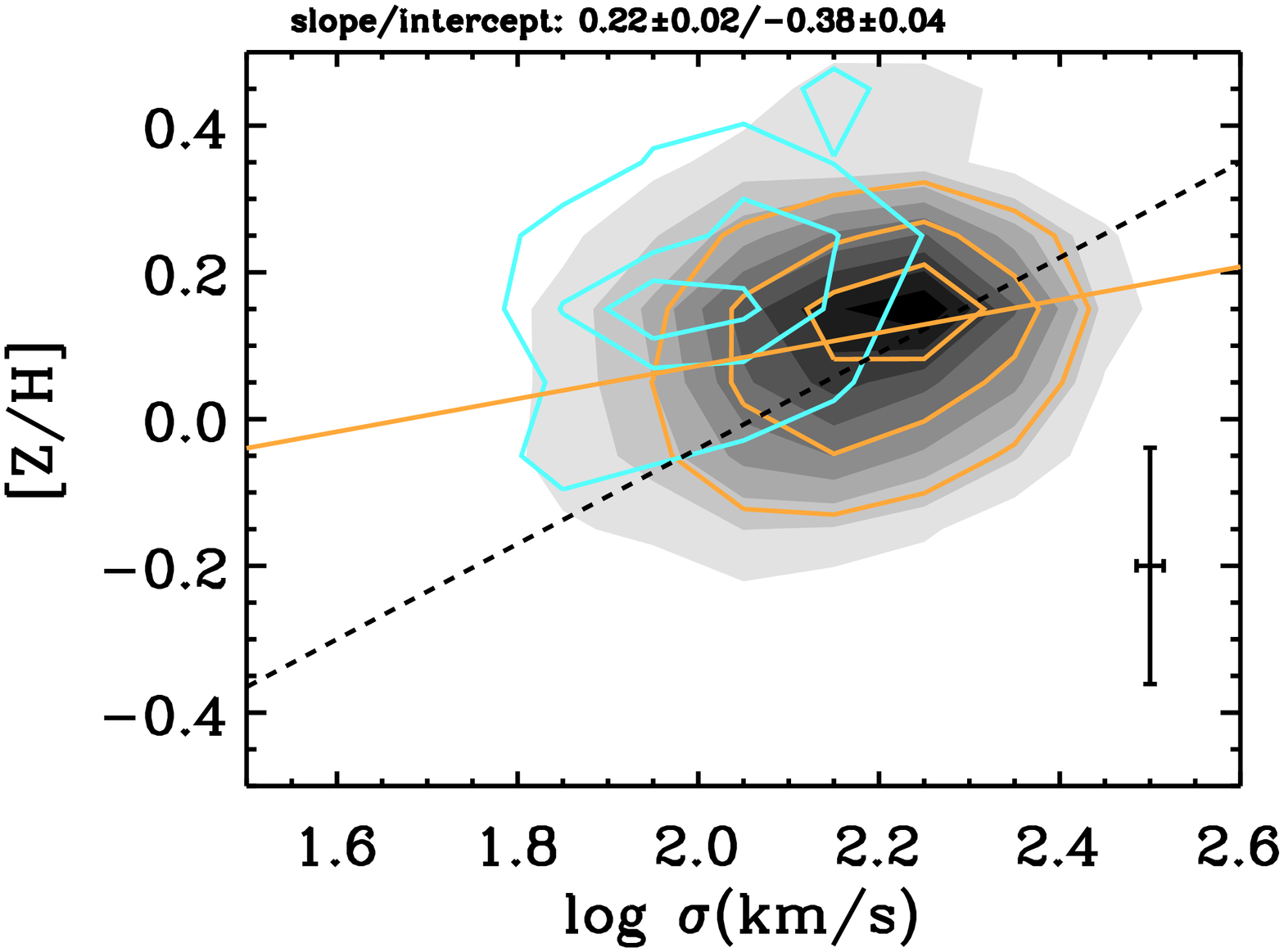}\includegraphics[scale=0.3]{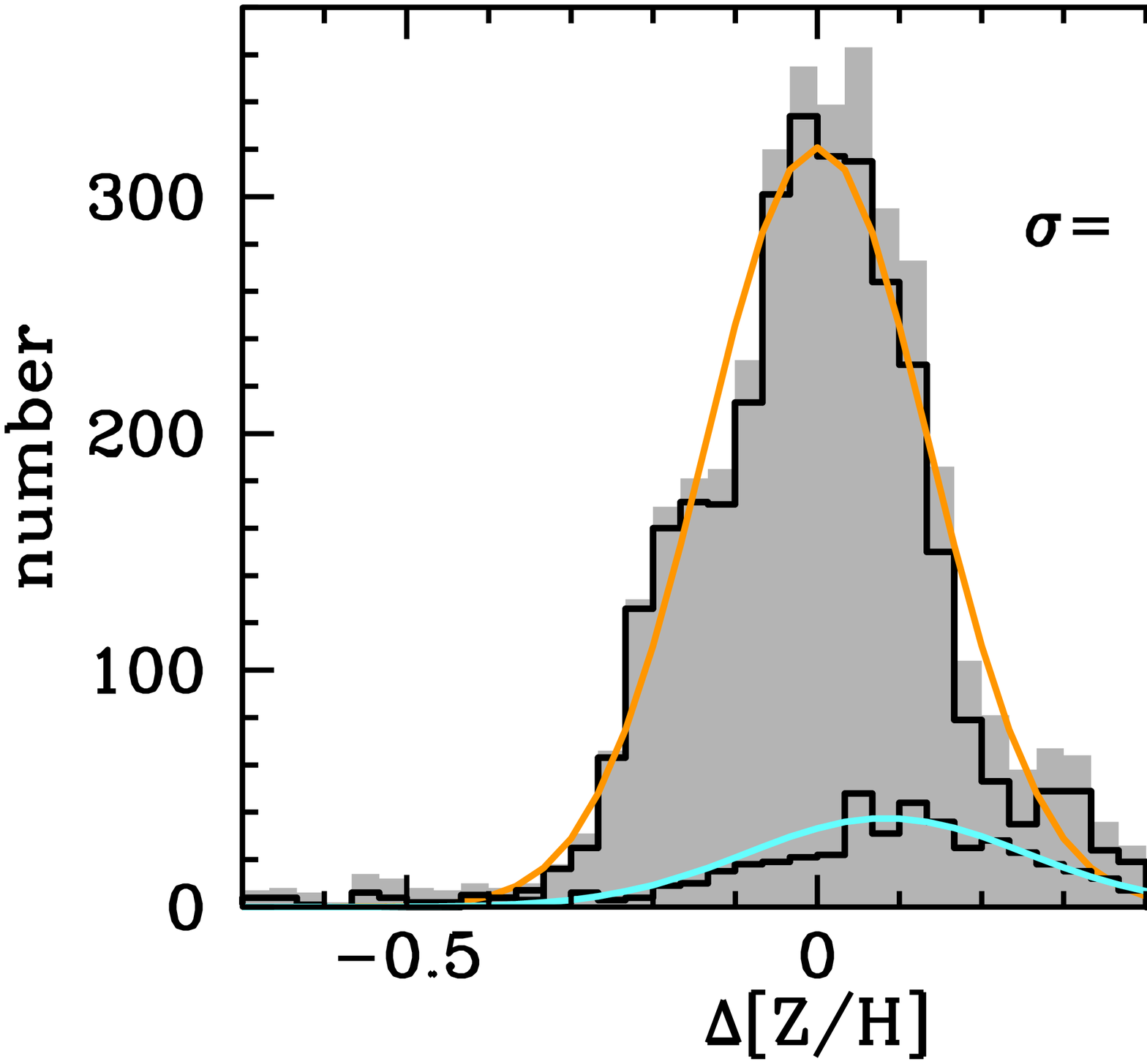}
\caption{Left hand panel: shows the relationship between the derived total metallicities and velocity dispersion. Rejuvenated objects with light-averaged ages smaller than 2.5 Gyr (see Fig.~\ref{age}) are presented with cyan contours and the orange contours show the old red sequence population. The whole sample is shown as grey-scaled filled contours. The orange solid line is a least-square fit to the red sequence population and the parameters of the fit are given at the top of the panel. The dashed black line is the analogous fit from T10 for comparison.  Median 1$\sigma$-errors are shown in the lower right corner. Right hand panel: shows the distribution of deviation in age from the least-square fit to the red sequence population with the same colour coding as in the left hand panel. The standard deviation of the gaussian fitted to the distribution of the red sequence population (orange line) is indicated in the upper right corner.}
\label{ZH}
\end{figure*}

The left hand panel of Fig.~\ref{ZH} shows the relationship between total metallicity and velocity dispersion, with the same colour coding as in Fig.~\ref{age}. The least-square fit to the red sequence population (orange line) indicate a strong correlation between total metallicity and velocity dispersion. This is in agreement with T10, but the slope of the fit derived here is smaller than the corresponding fit derived in T10 (black dashed line). In Section~\ref{T10Sec} we find that not accounting for element ratios produce the lower metallicities found in T10 compared to this work. This difference increases towards lower metallicities and produce the flatter slope of the [Z/H]-$\sigma$ relation.

\begin{figure}
\centering
\includegraphics[angle=90,scale=0.36]{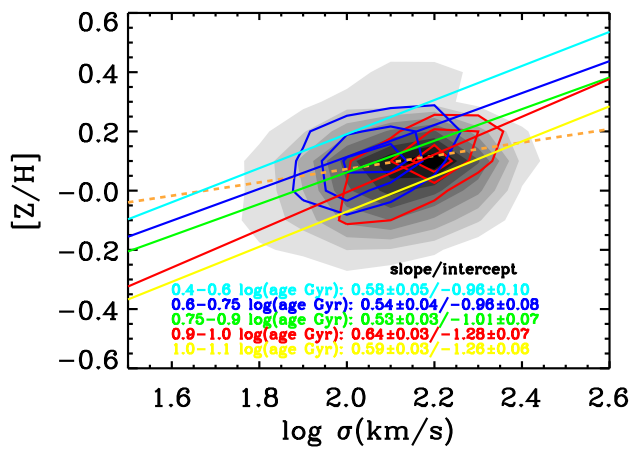}
\caption{Contour plots for the relationship between total metallicity and velocity dispersion in bins of narrow $\log$(age Gyr) intervals. Blue contours are for 0.6$<$$\log$(age Gyr)$<$0.75 and red contours for 0.9$<$$\log$(age Gyr)$<$1.0. Grey filled contours show the whole sample. Coloured lines are least-square fits to bins of varying age intervals, indicated by the correspondingly coloured labels that also give the parameters of the fits. The orange dashed line is the least-square fit to the whole sample (see Fig.~\ref{ZH})}
\label{ZH_age}
\end{figure}

The distribution around the fit to the red sequence population is shown in the right hand panel of Fig.~\ref{ZH}. 
The rejuvenated population (cyan gaussian) shows an offset towards higher metallicities by $\sim$0.1 dex compared to the red sequence population (orange gaussian), a behaviour also found by T10. 

In Fig.~\ref{ZH_age} we show least-square fits to the [Z/H]-$\sigma$ relationship in five narrow age-bins. 
These fits are indicated by the different colours with correspondingly coloured labels. The size of the bins were chosen to include a similar number of objects in each bin. Contours for two of the age bins are also shown, representing the fits with corresponding colours, and the grey-scaled filled contours represent the full sample. The fits to the age bins clearly show steeper slopes than for the entire red sequence population (dashed orange line). The fits to the age bins also show a parallel behaviour differing mainly in zero point offsets. This indicates a planar dependence on age and velocity dispersion for total metallicity, i.e. total metallicity correlate with velocity dispersion at fixed age and anti-correlate with age at fixed velocity dispersion. Such a behaviour has previously been found by \citet{trager00b}. 
The large scatter in [Z/H] found for the entire sample is partly an effect of this planar dependence. The shallower slope for the entire red sequence sample is due to the generally older ages found for galaxies with a higher velocity dispersion and vice versa (see Fig.~\ref{age}).  

\subsection{[Fe/H]}
\label{FeHSec}

\begin{figure*}
\includegraphics[angle=90,scale=0.35]{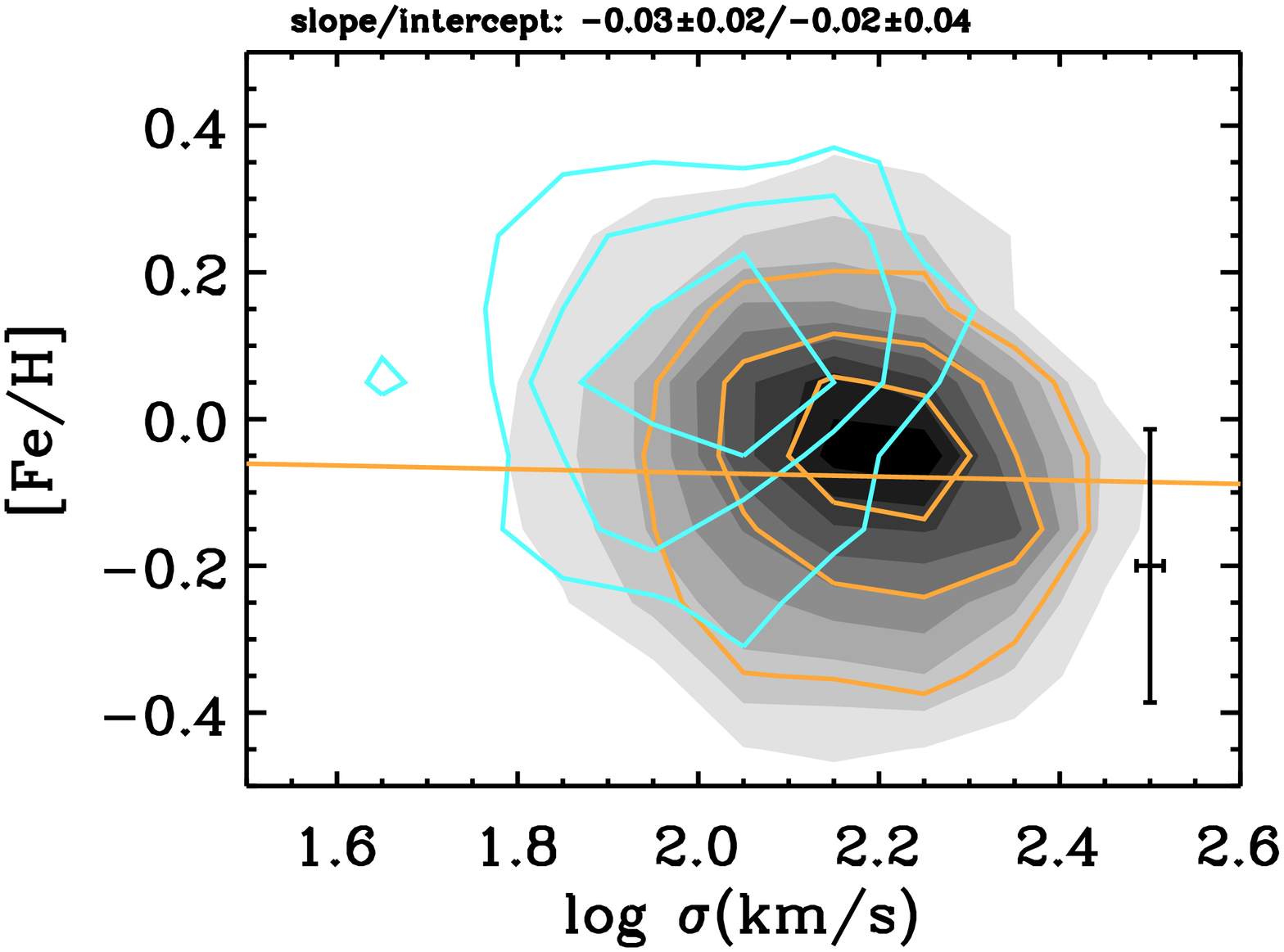}\includegraphics[scale=0.3]{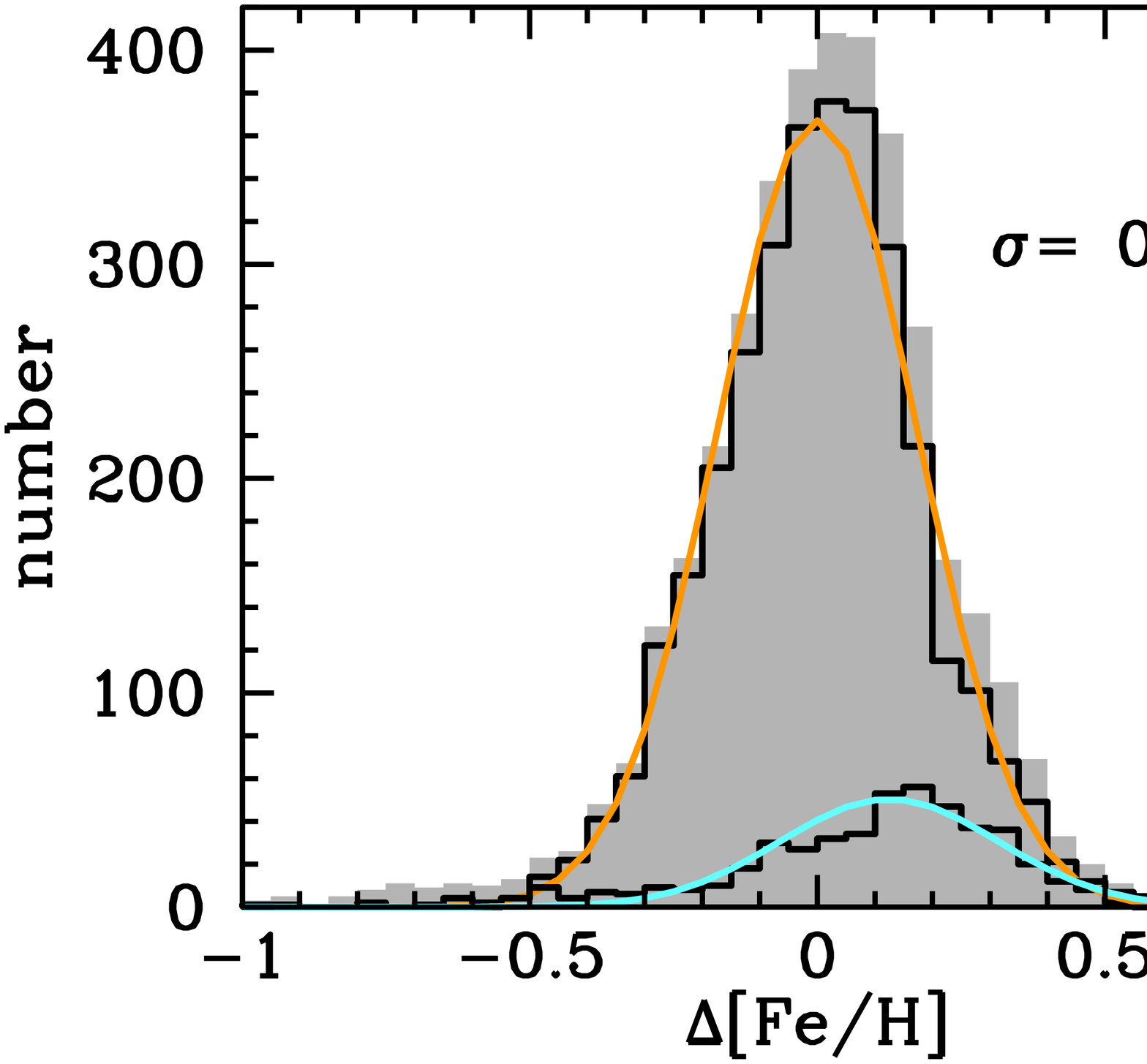}
\caption{Left hand panel: shows the relationship between iron abundance and velocity dispersion. Rejuvenated objects with light-averaged ages smaller than 2.5 Gyr (see Fig.~\ref{age}) are presented with cyan contours and the orange contours show the old red sequence population. The whole sample is shown as grey-scaled filled contours. The orange solid line is a least-square fit to the red sequence population and the parameters of the fit are given at the top of the panel.  Median 1$\sigma$-errors are shown in the lower right corner. Right hand panel: shows the distribution of deviation in age from the least-square fit to the red sequence population with the same colour coding as in the left hand panel. The standard deviation of the gaussian fitted to the distribution of the red sequence population (orange line) is indicated in the upper right corner.}
\label{FeH}
\end{figure*}

Iron abundances derived using Eq.~\ref{final} in Section~\ref{feh} are presented in the left hand panel of Fig.~\ref{FeH} as a function of velocity dispersion, with the same colour coding as in Fig.~\ref{age} for the results of this work. 
The least-square fit (orange line) shows no correlation between iron abundance and velocity dispersion for the full red sequence population. 
Hence the [Fe/H]-$\sigma$ relation is significantly flatter than the [Z/H]-$\sigma$ relationship presented in the previous section. Similar patterns have also been found by \citet{trager00b} and \citet{graves07}, but the slopes of their [Fe/H]-$\sigma$ relations are different from zero. Similar to this work \citet{price11} find a relatively flat [Fe/H]-$\log \sigma$ slope for a sample of 222 passive galaxies.

\begin{figure}
\centering
\includegraphics[angle=90,scale=0.36]{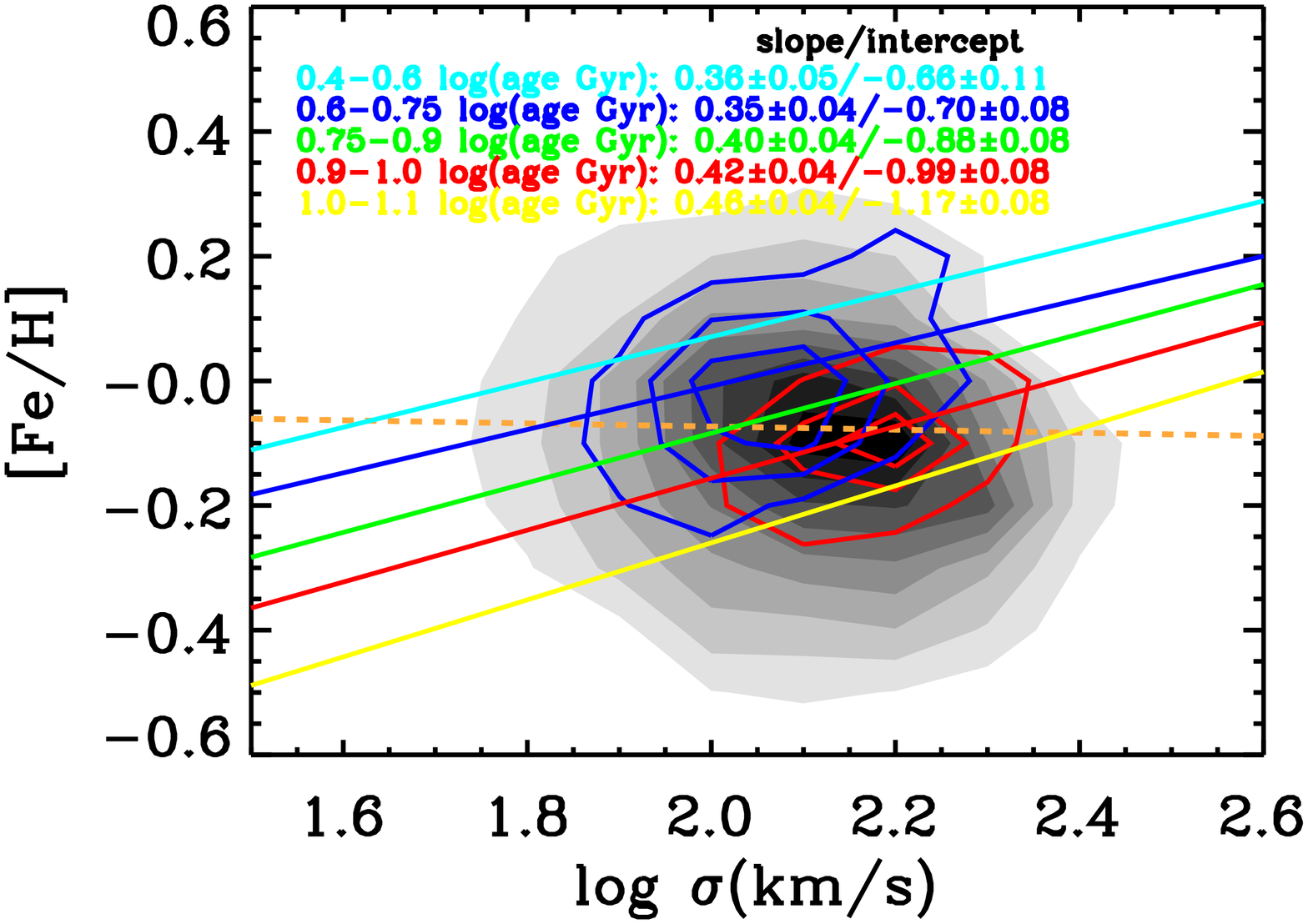}
\caption{Contour plots for the relationship between iron abundance and velocity dispersion in bins of narrow $\log$(age Gyr) intervals. Blue contours are for 0.6$<$$\log$(age Gyr)$<$0.75 and red contours for 0.9$<$$\log$(age Gyr)$<$1.0. Grey filled contours show the whole sample. Coloured lines are least-square fits to bins of varying age intervals, indicated by the correspondingly coloured labels that also give the parameters of the fits. The orange dashed line is the least-square fit to the whole sample (see Fig.~\ref{FeH})}
\label{FeH_age}
\end{figure}

\begin{figure*}
\includegraphics[angle=90,scale=0.35]{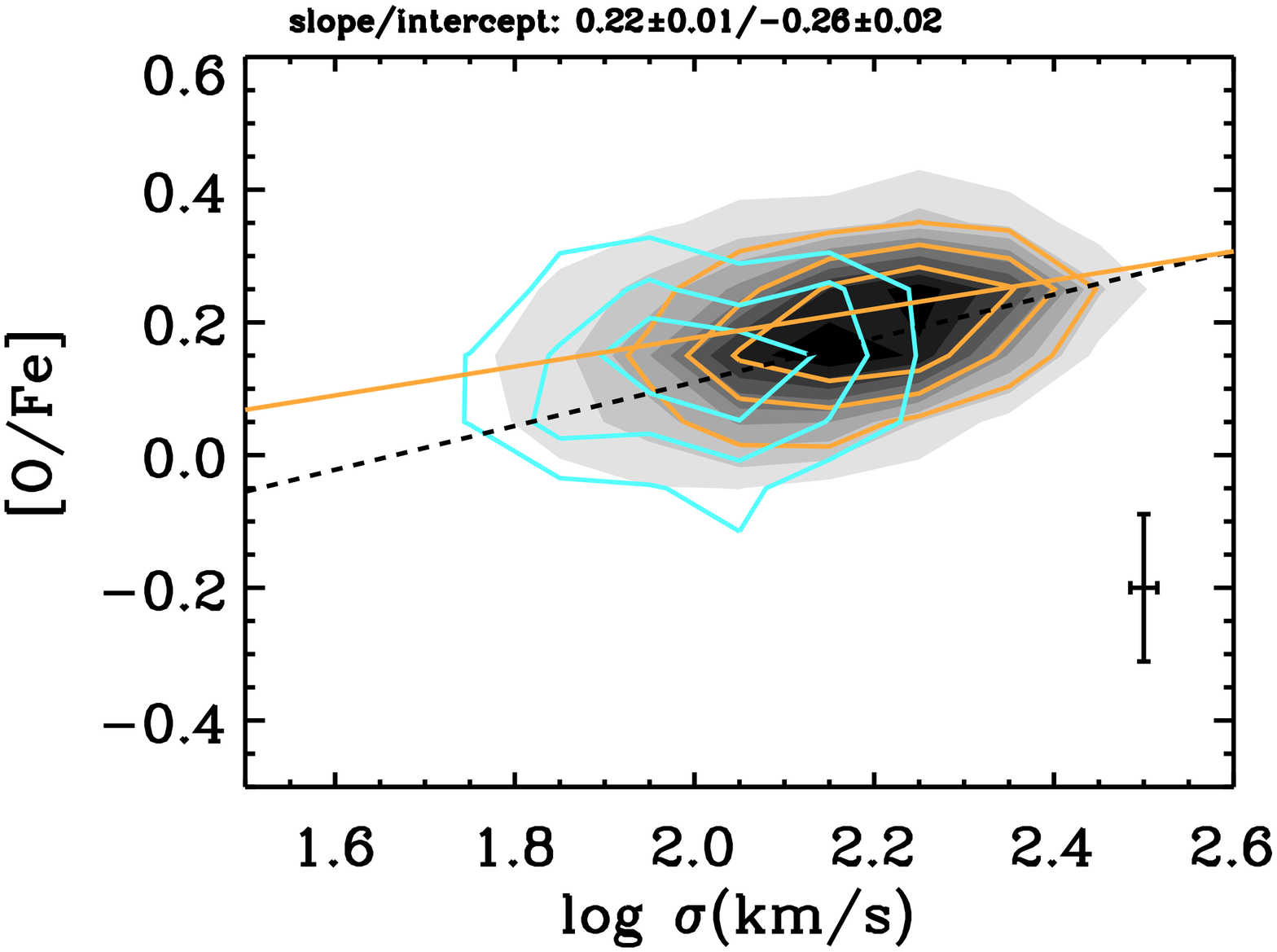}\includegraphics[scale=0.3]{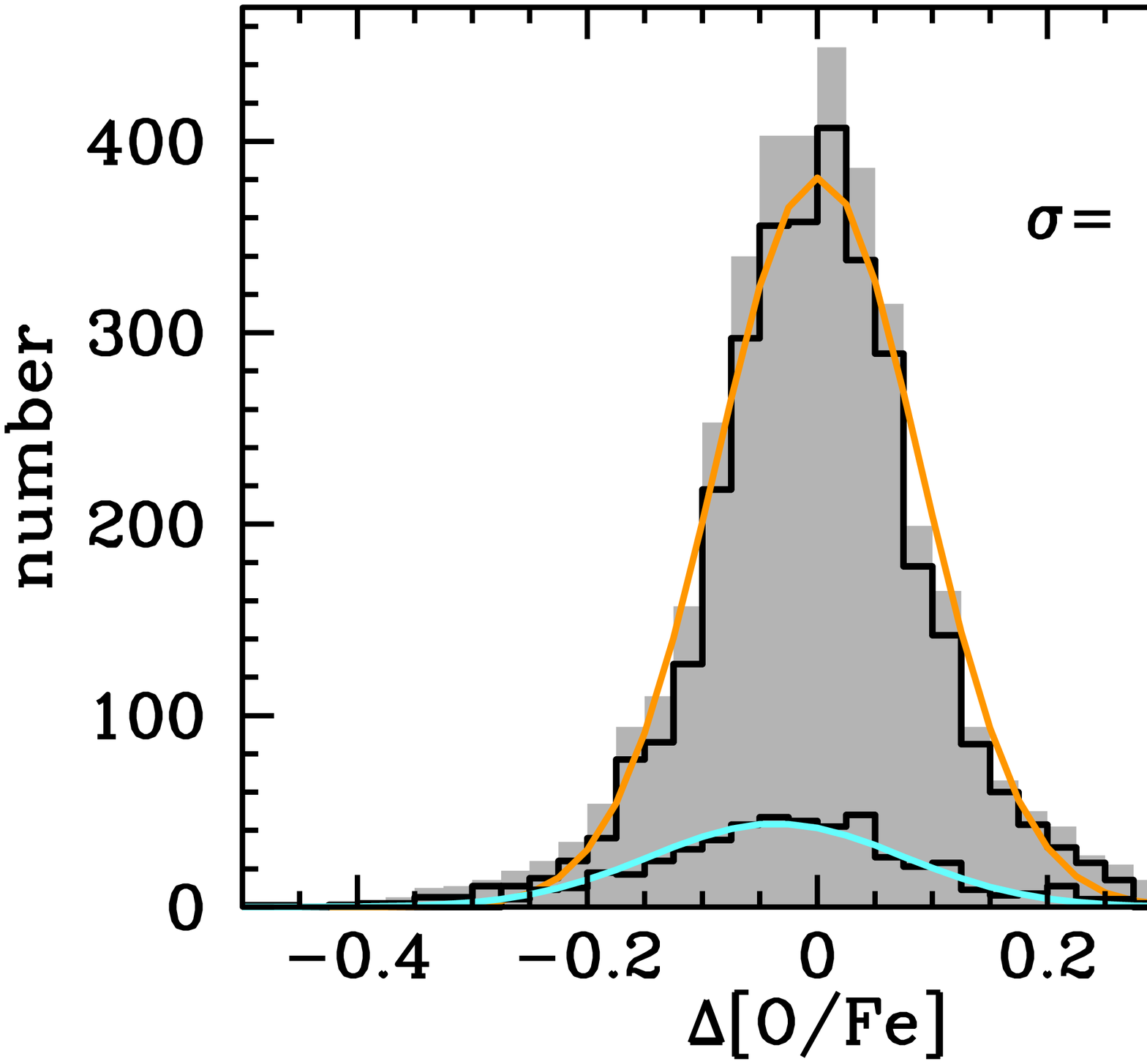}\\
\includegraphics[angle=90,scale=0.35]{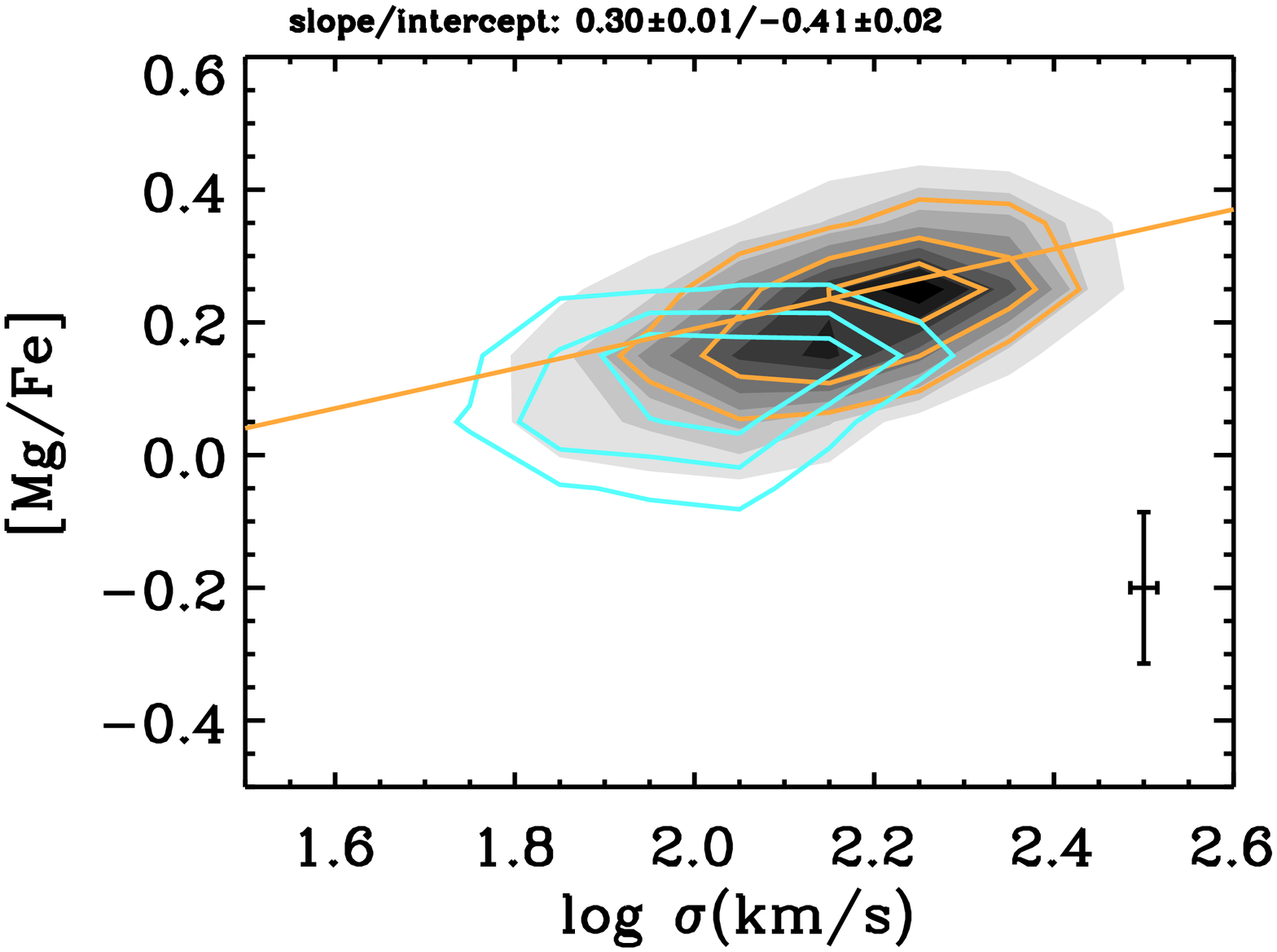}\includegraphics[scale=0.3]{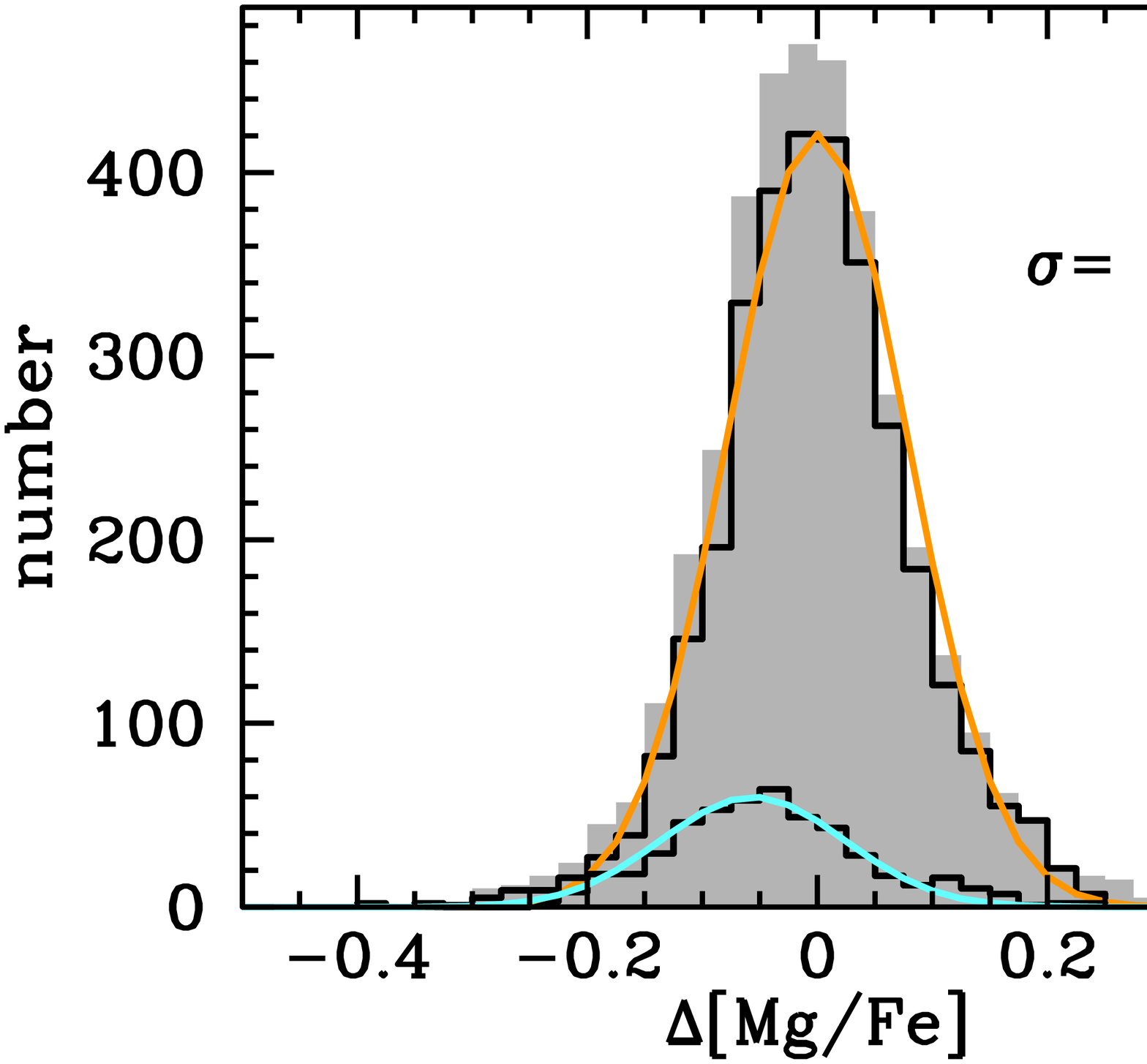}
\caption{Left hand panels: show the relationships for the derived O/Fe (upper) and Mg/Fe ratios (lower) with velocity dispersion. Rejuvenated objects with light-averaged ages smaller than 2.5 Gyr (see Fig.~\ref{age}) are presented with cyan contours and the orange contours show the old red sequence population. The whole sample is shown as grey-scaled filled contours. The orange solid lines are least-square fits to the red sequence population and the parameters of the fits are given at the top of the panels.  Median 1$\sigma$-errors are shown in the lower right corners. Right hand panels: show the distribution of deviation in age from the least-square fits to the red sequence population for O/Fe (upper) and Mg/Fe (lower), with the same colour coding as in the left hand panels. The standard deviation of the gaussians fitted to the distributions of the red sequence population (orange lines) are indicated in the upper right corners.}
\label{oxyMg}
\end{figure*}

The distribution around the fit to the red sequence population is shown in the right hand panel of Fig.~\ref{FeH}. The rejuvenated population (cyan gaussian) shows an offset ($\sim$0.15 dex) towards higher [Fe/H] compared to the red sequence population (orange gaussian). This offset is more pronounced than for [Z/H] (compare with the right hand panel of Fig.~\ref{ZH}). 


In Fig.~\ref{FeH_age} we show least-square fits to the [Fe/H]-$\sigma$ relationship in five narrow age-bins, with the same colour coding and bin sizes as in Fig.~\ref{ZH_age}. Similar to the total metallicity case, we find a planar dependence on velocity dispersion and age, i.e. iron abundance correlate with velocity dispersion at fixed age and anti-correlate with age at fixed velocity dispersion. 
Such a behaviour have again been found by \citet{trager00b} and also by \citet{smith09}. 
The planar dependence is partly responsible for the large scatter in iron abundance found for the full sample. 
Compared to total metallicity we find flatter slopes for the [Fe/H]-$\sigma$ relationships in the age-bins as well as for the full sample.
 The fits to the age bins in Fig.~\ref{FeH_age} clearly show steeper slopes than for the full red sequence population sample (dashed orange line), since older ages are in general found for galaxies with a higher velocity dispersion and vice versa (see Fig.~\ref{age} and Section~\ref{ZSec}). 
Hence, at fixed age the effective Fe yields are higher in more massive systems. However, Fe enrichment is suppressed in older galaxies because of time-scale dependent contribution from SN Ia. Therefore the relation becomes flat for the overall population and the [Fe/H]-age anti-correlation is steeper than the [Z/H]-age anti-correlation.


\begin{figure*}
\centering
\includegraphics[angle=90,scale=0.35]{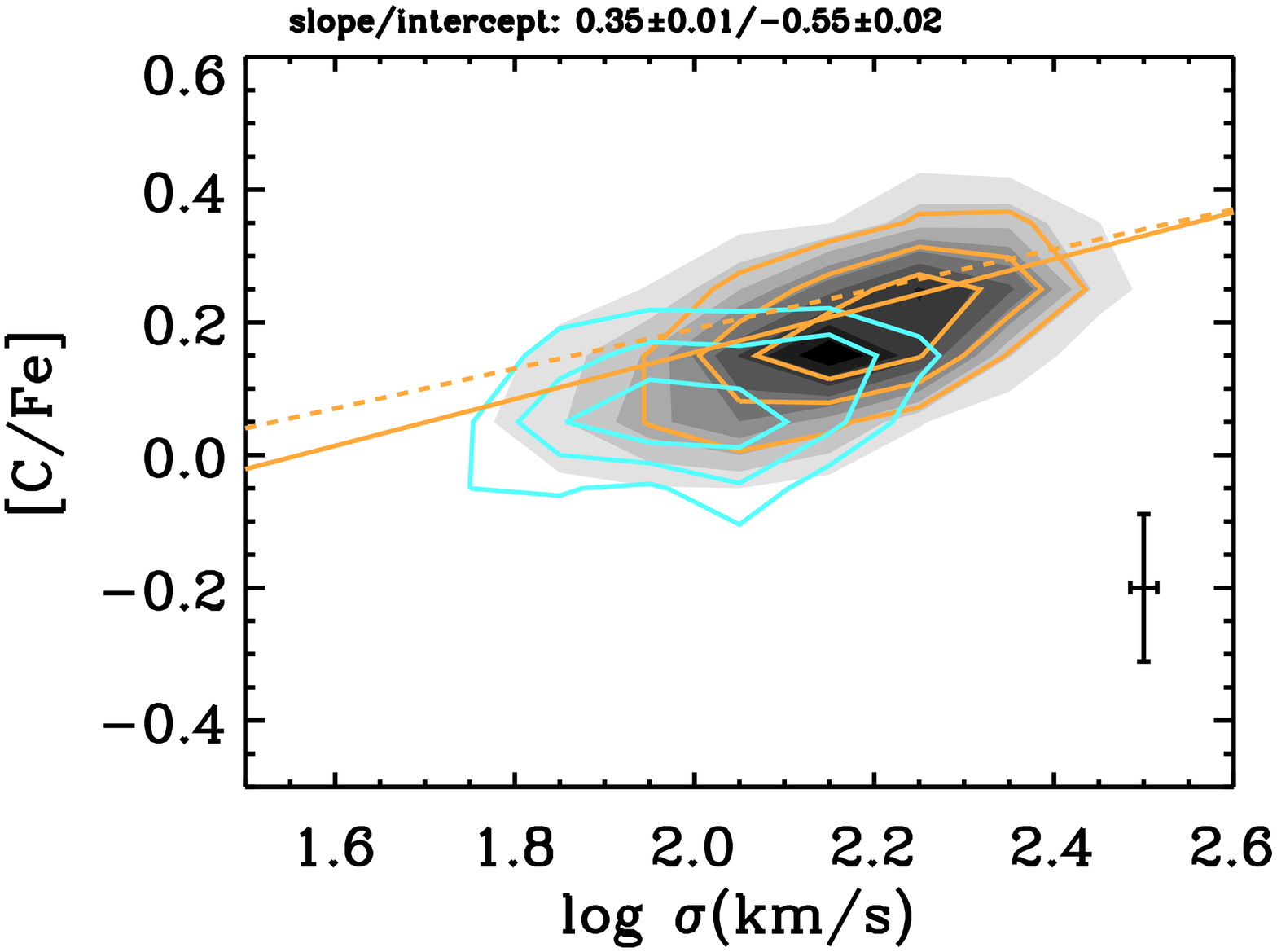}\includegraphics[scale=0.3]{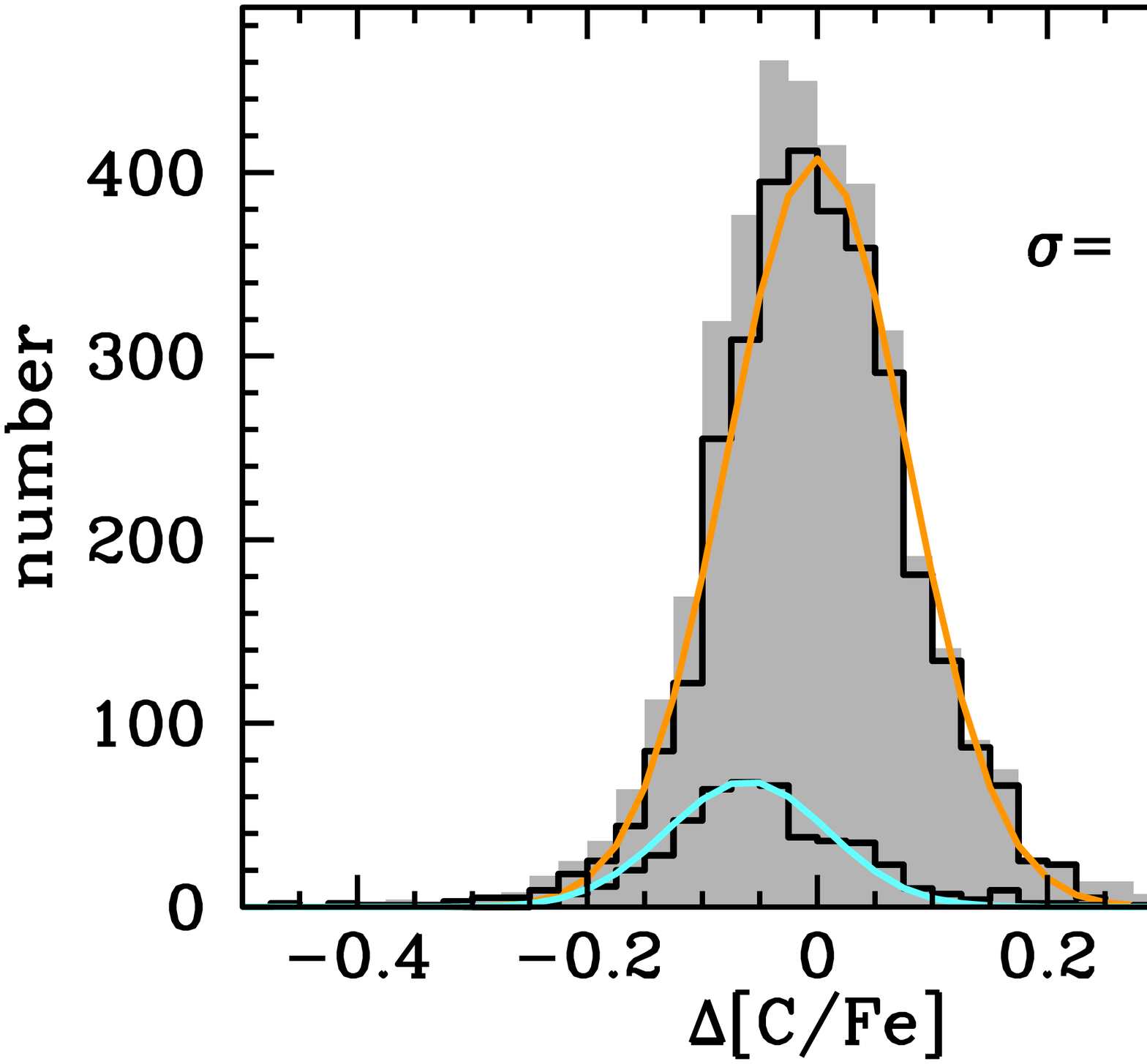}\\
\includegraphics[angle=90,scale=0.35]{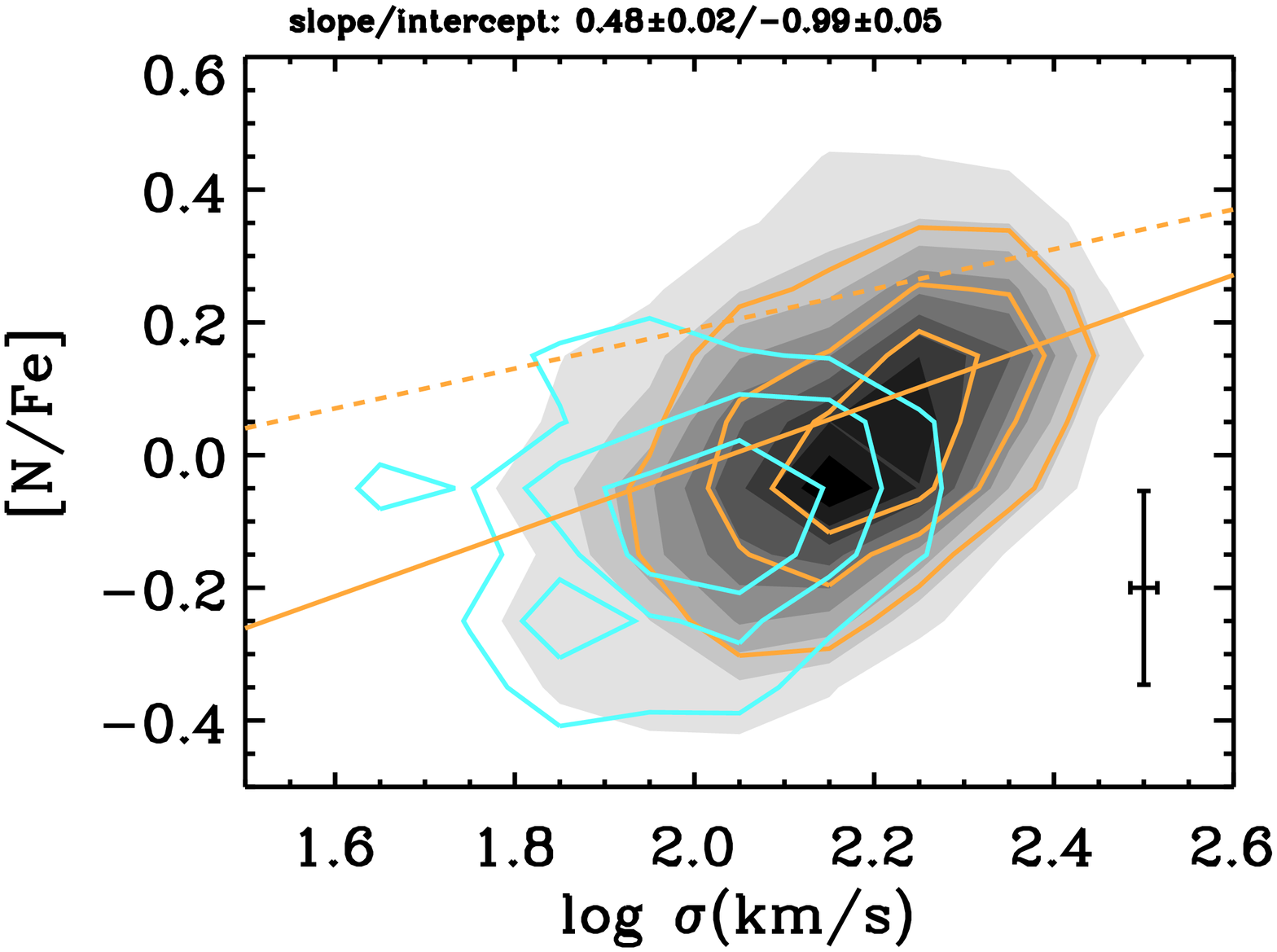}\includegraphics[scale=0.3]{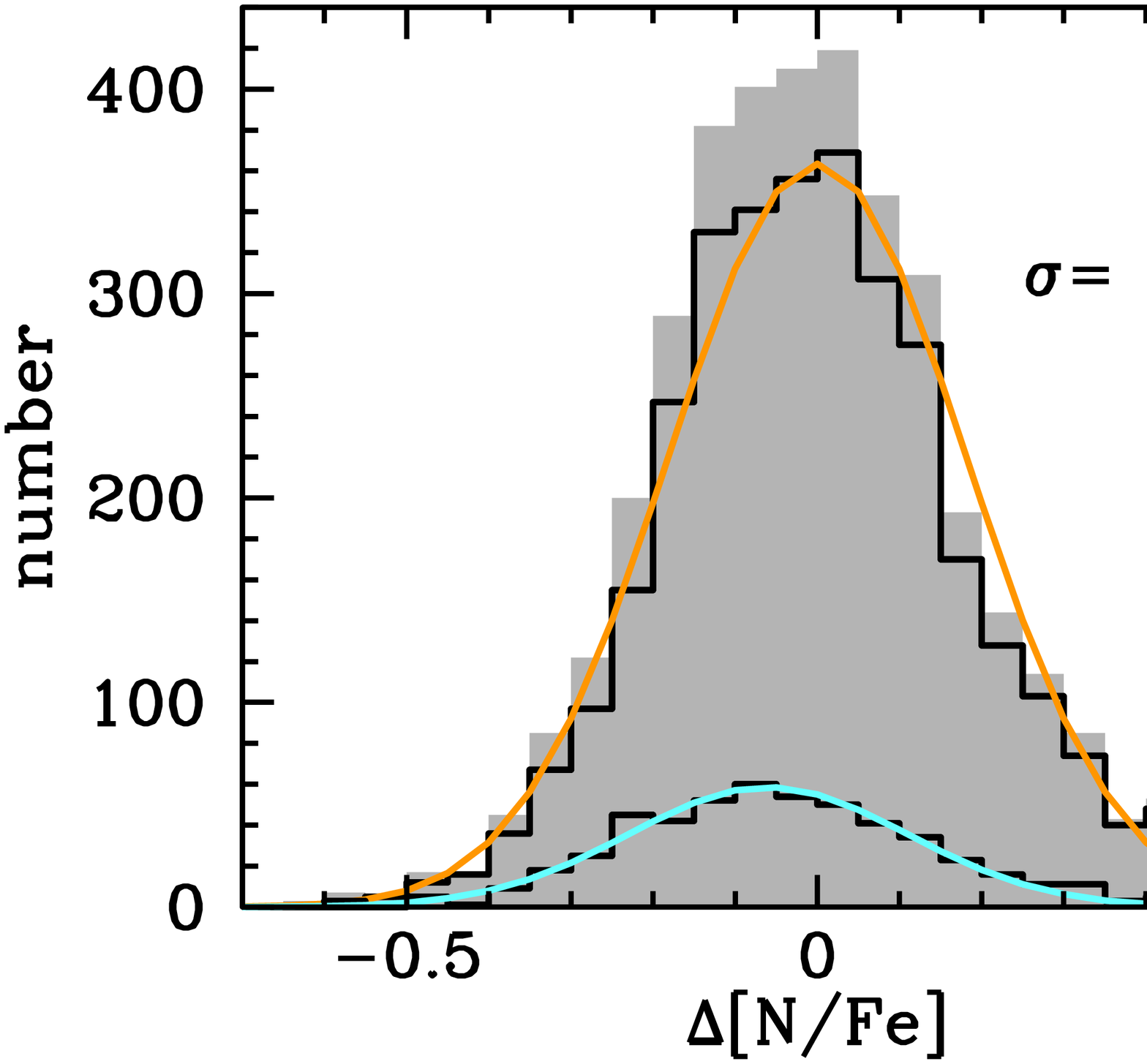}
\caption{Left hand panels: show the relationships for the derived C/Fe (upper) and N/Fe ratios (lower) with velocity dispersion. Rejuvenated objects with light-averaged ages smaller than 2.5 Gyr (see Fig.~\ref{age}) are presented with cyan contours and the orange contours show the old red sequence population. The whole sample is shown as grey-scaled filled contours. The orange solid lines are least-square fits to the red sequence population and the parameters of the fits are given at the top of the panels.  Median 1$\sigma$-errors are shown in the lower right corners. Right hand panels: show the distribution of deviation in age from the least-square fits to the red sequence population for C/Fe (upper) and N/Fe (lower), with the same colour coding as in the left hand panels. The standard deviation of the gaussians fitted to the distributions of the red sequence population (orange lines) are indicated in the upper right corners.}
\label{carbN}
\end{figure*}

\subsection{[O/Fe]}
\label{OSec}

The relationship between [O/Fe] ratio and velocity dispersion is presented in the upper left panel of Fig~\ref{oxyMg}, 
with the same colour coding as in Fig.~\ref{age}. 
For the red sequence population we find a somewhat shallower slope for the [O/Fe]-$\sigma$ relation (orange line) compared to the [$\alpha$/Fe]-$\sigma$ relation from T10 (dashed black line), but the general agreement is very good
, i.e. a tight correlation between [O/Fe] and velocity dispersion. The shallower slope can be explained by metallicity dependent C yields that depress the amount of O in more metal-rich systems (see discussion in Section~\ref{ICarb}). 

Delayed enrichment of Fe-like elements from SNIa explosion compared to the prompt enrichment of $\alpha$-elements from SNII explosion,
results in lower $\alpha$/Fe ratios for objects with more extended star formation histories. Hence
T10 interpreted the higher [$\alpha$/Fe] ratios found in more massive galaxies as shorter formation time-scales for such objects, and evidence for down-sizing of early-type galaxies. This interpretation gets support from the [O/Fe]-$\sigma$ relation found in this work, since O belongs to the $\alpha$-elements.

The upper right panel of Fig.~\ref{oxyMg} shows the distribution around the fit to the red sequence population. 
The rejuvenated population (cyan gaussian) shows an offset towards lower [O/Fe] ratios compared to the red sequence population (orange gaussian). This offset is less pronounced compared to what T10 found for [$\alpha$/Fe]. 

\subsection{[Mg/Fe]}
\label{MgSec}

The [Mg/Fe]-$\sigma$ relationship is presented in the lower left panel of Fig.~\ref{oxyMg}, with the same colour coding as in Fig.~\ref{age} for the contours and least-square fit. 
For the red sequence population [Mg/Fe] clearly increases with velocity dispersion, a pattern also found by \citet{sanchez06}, \citet{graves07}, \citet{smith09} and \citet{price11}. 
The [Mg/Fe]-$\sigma$ relationship shows a similar trend as [O/Fe]-$\sigma$ (compare with the upper left panel of Fig.~\ref{oxyMg}). A steeper slope is found for the former relationship, which is in better agreement with the slope of [$\alpha$/Fe]-$\sigma$ relation from T10. Thus the T10 [$\alpha$/Fe] ratios trace [Mg/Fe] rather than [O/Fe] as discussed in Section~\ref{T10Sec}. 

The distribution around the fit to the red sequence population is shown in the lower right panel of Fig.~\ref{oxyMg}, with the same colour coding as in Fig.~\ref{age}. The rejuvenated population (cyan gaussian) is offset towards lower [Mg/Fe] ratios compared to the red sequence population (orange gaussian). This offset is more pronounced than for [O/Fe] (see upper right panel of Fig.~\ref{oxyMg}, Section~\ref{OSec}). This is again in better agreement with the results for [$\alpha$/Fe] in T10, which suggests that the [$\alpha$/Fe] ratio derived in T10 is closest to the true [Mg/Fe] ratio. 

\subsection{[C/Fe]}
\label{CSec}		

The [C/Fe]-$\sigma$ relationship is shown in the upper left panel of Fig.~\ref{carbN}, with the same colour coding as in Fig.~\ref{age} for the contours and the least-square fit to the red sequence population. 
The [Mg/Fe]-$\sigma$ relationship is included for comparison (dashed orange line). 
Considering the red sequence population we find a very similar trend for the [C/Fe]-$\sigma$ relationship (solid orange line) as for [Mg/Fe]-$\sigma$ (dashed orange line), but the latter trend is slightly steeper. 
The stronger trend for [C/Fe] can be explained by metallicity dependent C yields (see discussion in Section~\ref{ICarb}).

Increasing [C/Fe] ratios with velocity dispersion has also been found by \citet{sanchez06}, \citet{graves07} and \citet{smith09}. \citet{sanchez06} report a stronger increase in [Mg/Fe] than [C/Fe] for a sample of 98 early-type galaxies. \citet{graves07} and \citet{smith09} study $\sim$6000 red sequence SDSS early-type galaxies and 147 red sequence galaxies, respectively, and find, in agreement with this work, a shallower trend for [Mg/Fe] than for [C/Fe]. A similar pattern was also found by \citet{price11}. 

The distribution around the fit to the red sequence population is shown in the upper right panel of Fig.~\ref{carbN}. The rejuvenated population (cyan gaussian) shows an offset of lower [C/Fe] ratios compared to the red sequence population (orange gaussian). Again very similar to [Mg/Fe] we find a clear offset towards lower [C/Fe] ratios for the rejuvenated population compared to the red sequence population. Hence we find that the [$\alpha$/Fe] ratio derived in T10 resembles both [Mg/Fe] and [C/Fe]. 
In fact the slope of the [$\alpha$/Fe]-$\sigma$ relationship from T10 (0.33) falls right in between the slope for [Mg/Fe]-$\sigma$ (0.30) and [C/Fe]-$\sigma$ (0.35). Following the discussion in Section~\ref{T10Sec} this is probably due to the fact that T10 use absorption line indices sensitive to various element abundance variations to derive [$\alpha$/Fe] without considering individual abundance ratios. These indices show the strongest signals to Mg and C, such that their derived [$\alpha$/Fe] ratios reflect [Mg/Fe] and [C/Fe] rather than [O/Fe]. 

In Section~\ref{ICarb} we discuss the controversy of the origin of C and that the similarity between the derived [Mg/Fe] and [C/Fe] ratios reflect a significant contribution of C from massive stars. 

\begin{figure*}
\centering
\includegraphics[angle=90,scale=0.35]{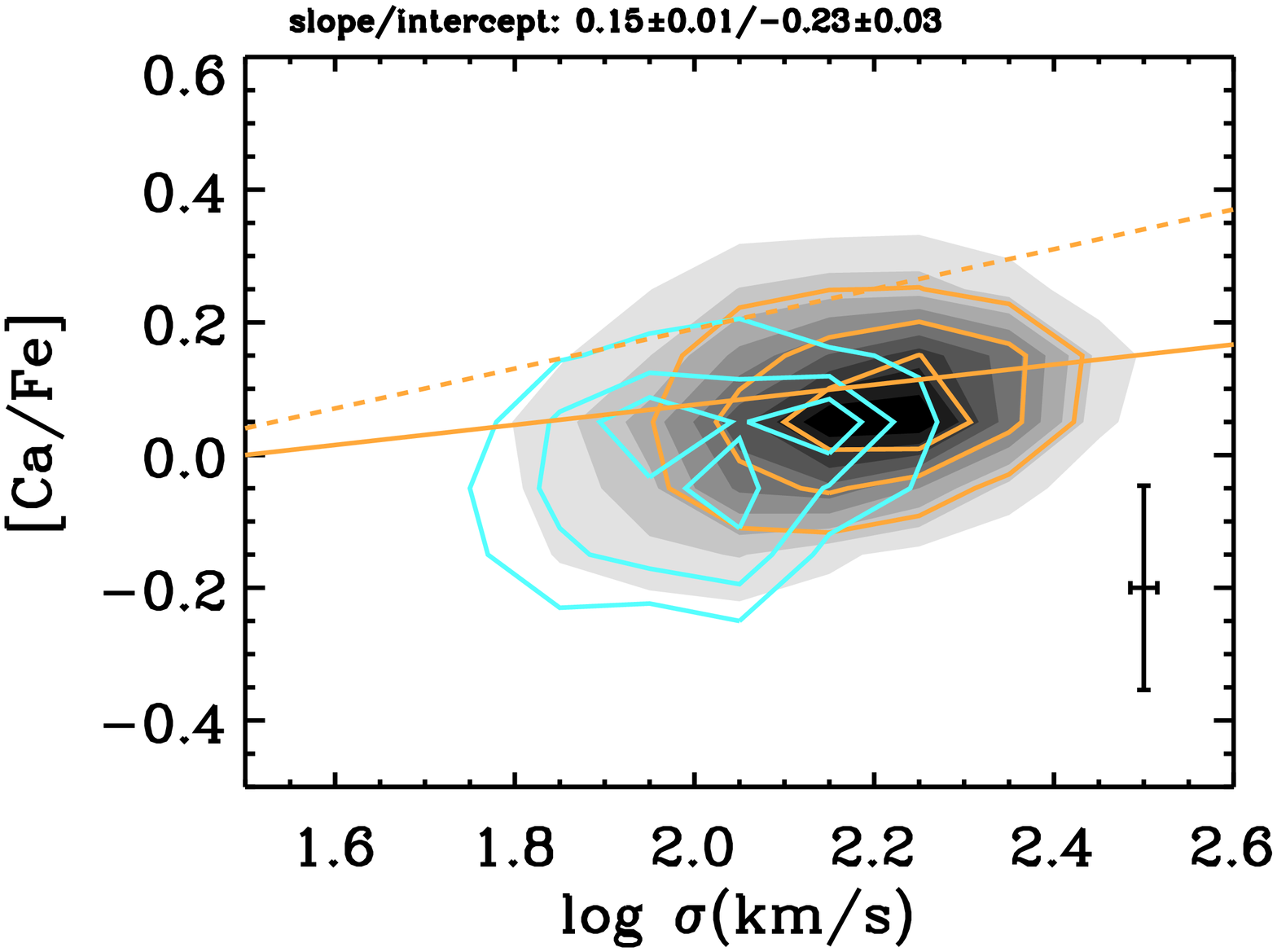}\includegraphics[scale=0.3]{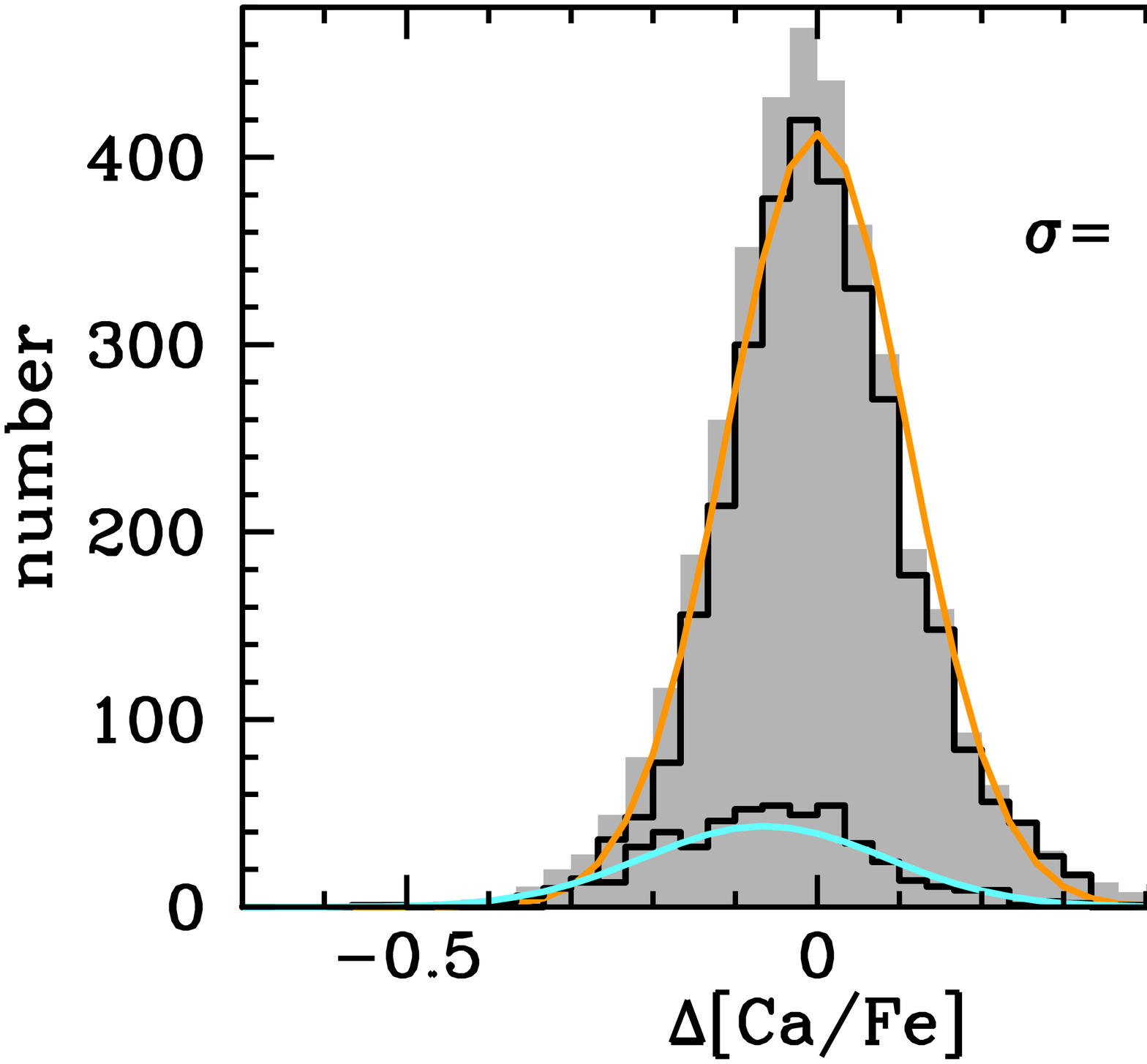}\\
\includegraphics[angle=90,scale=0.35]{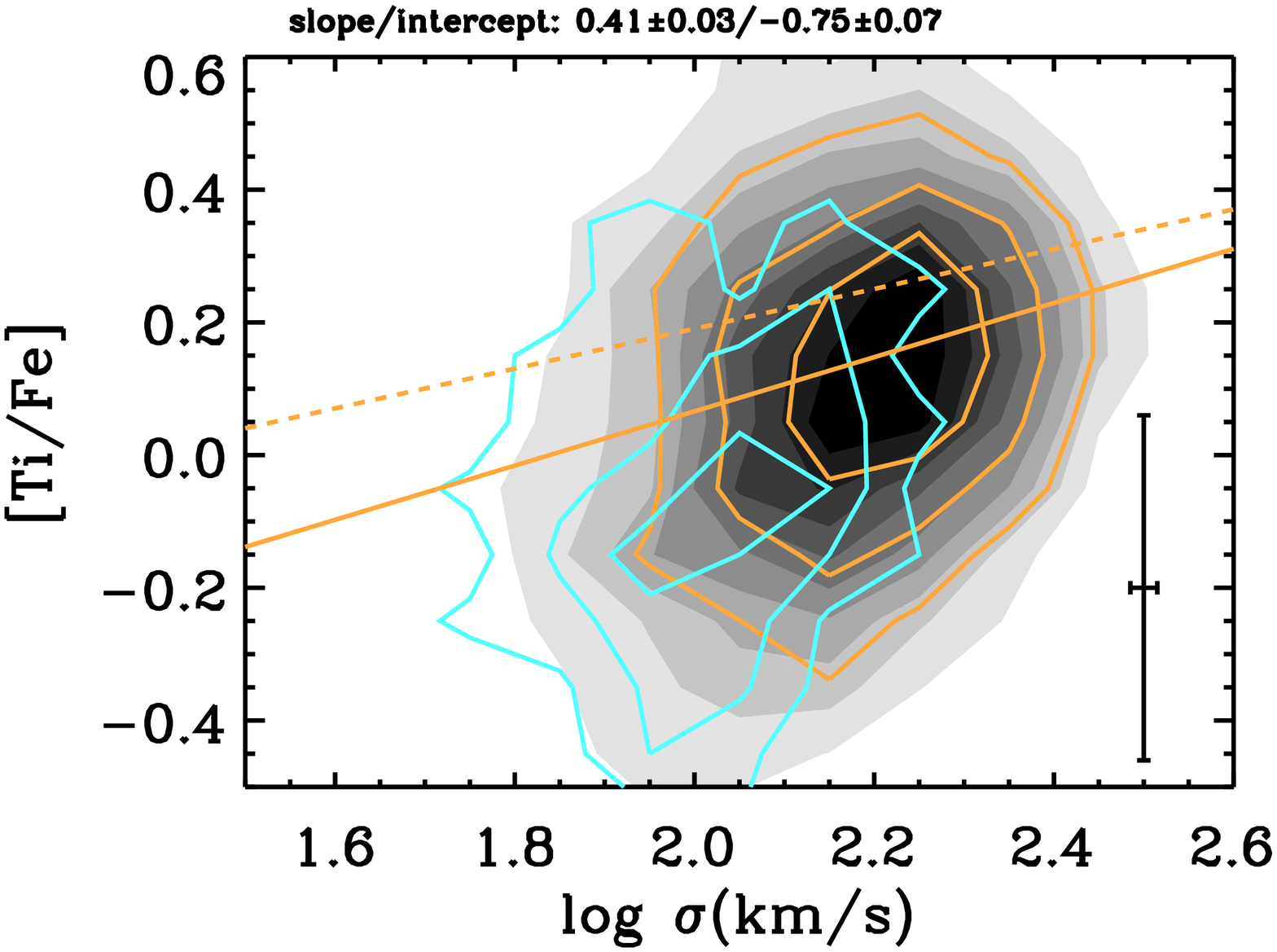}\includegraphics[scale=0.3]{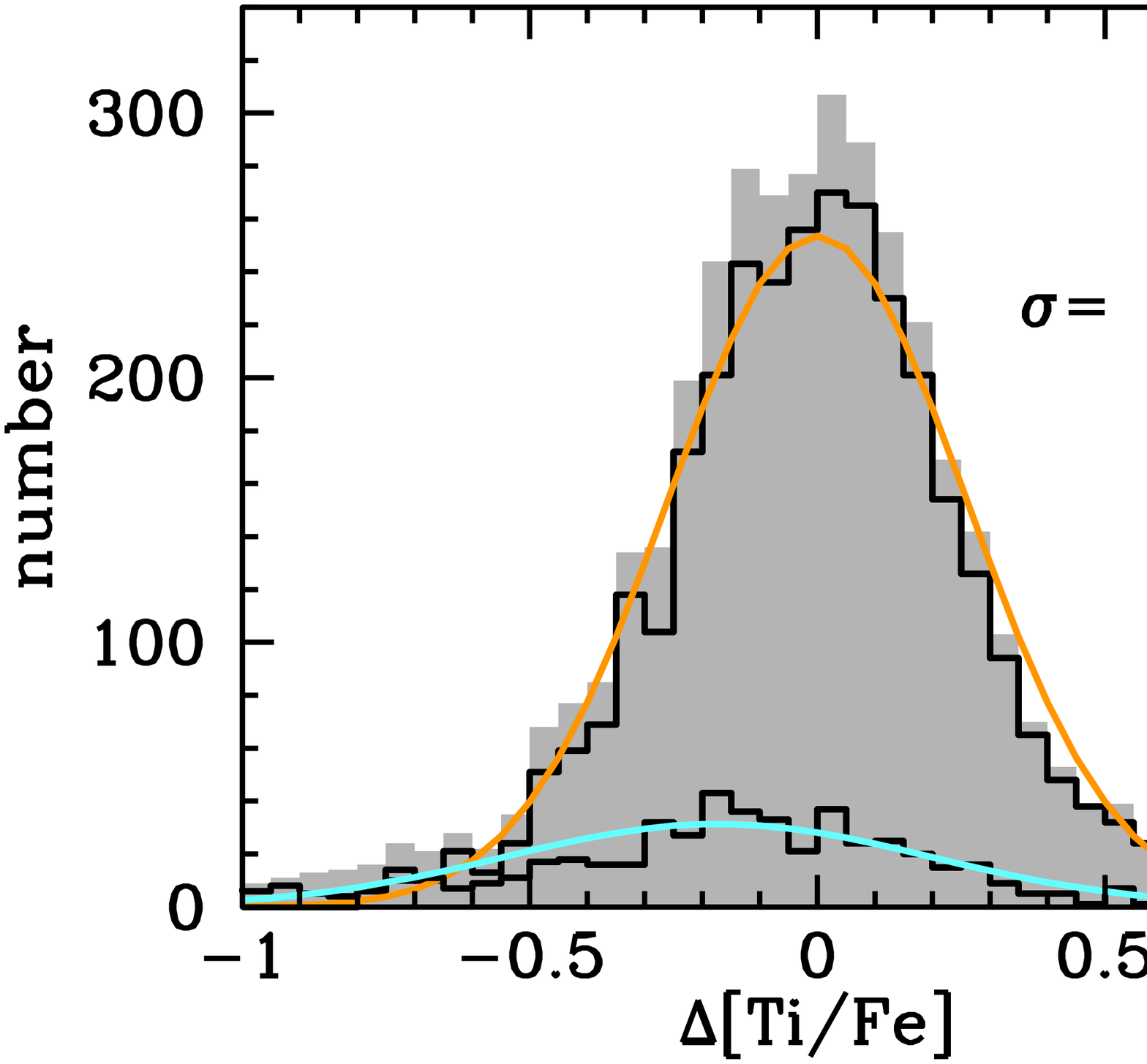}\\
\caption{Left hand panels: show the relationships for the derived Ca/Fe (upper) and Ti/Fe ratios (lower) with velocity dispersion. Rejuvenated objects with light-averaged ages smaller than 2.5 Gyr (see Fig.~\ref{age}) are presented with cyan contours and the orange contours show the old red sequence population. The whole sample is shown as grey-scaled filled contours. The orange solid lines are least-square fits to the red sequence population and the parameters of the fits are given at the top of the panels.  Median 1$\sigma$-errors are shown in the lower right corners. Right hand panels: show the distribution of deviation in age from the least-square fits to the red sequence population for Ca/Fe (upper) and Ti/Fe (lower), with the same colour coding as in the left hand panels. The standard deviation of the gaussians fitted to the distributions of the red sequence population (orange lines) are indicated in the upper right corners.}
\label{CaTiFig}
\end{figure*}

\subsection{[N/Fe]}
\label{NSec}

The [N/Fe]-$\sigma$ relationship is shown in the lower left panel of Fig.~\ref{carbN}, with the same colour coding as in Fig.~\ref{age} for the contours and least-square fit to the red sequence population. 
The [Mg/Fe]-$\sigma$ relationship is shown for comparison (dashed orange line). 
Despite a fairly large scatter we find a clear correlation between [N/Fe] and velocity dispersion for the red sequence population, i.e. [N/Fe] strongly increases with velocity dispersion.
This correlation shows a significantly steeper trend than the analogous found for [Mg/Fe] and [C/Fe]. The [N/Fe] ratios are also offset towards lower abundance ratios by $\sim$0.2 dex compared to [Mg/Fe] and [C/Fe]. 
An increase in [N/Fe] with velocity dispersion has also been found by \citet{sanchez06}, \citet{graves07} and \citet{smith09}. The result of \citet{sanchez06} show steeper trends with velocity dispersion for [Mg/Fe] than for [N/Fe], while \citet{smith09} report similar trends, steeper than [Mg/Fe], for [N/Fe] and [C/Fe]. \citet{graves07} and \citet{price11} find, again in agreement with this work, the steepest trends with velocity dispersion for [N/Fe]. Also in agreement with this work they find [N/Fe] to be offset towards lower abundance ratios compared to [Mg/Fe] and [C/Fe]. However, this offset is dependent on their adopted, fixed value for [O/Fe] (see Section~\ref{method_lit}).


The distributions around the fit to the red sequence population are shown in the lower right panel of Fig.~\ref{carbN}. The rejuvenated population (cyan gaussian) shows no clear offset compared to the red sequence population (orange gaussian), considering the fairly large scatter in the data.

The differences in the derived [C/Fe] and [N/Fe] ratios are interesting as both elements are expected to be enriched through the same processes. A thorough discussion on the origins of C and N are given in Section~\ref{ICarb} and~\ref{INitro}, respectively, and related to the derived element ratios.

\subsection{[Ca/Fe] and [Ti/Fe]}
\label{CaTiSec}

Trends with velocity for [Ca/Fe] and [Ti/Fe] are shown in the upper left hand and lower left hand panels of Fig.~\ref{CaTiFig}, respectively, with the same colour codings as in Fig.~\ref{age}. 
The derived [Ca/Fe] ratios show very weak trends with velocity dispersion and a fairly small scatter. The trend is significantly flatter than compared to the [Mg/Fe]-$\sigma$ relationship (dashed orange line) and [Ca/Fe] is generally lower than [Mg/Fe] by $\sim$0.1-0.2 dex. This implies an underabundance of Ca compared to Mg, such that Ca scales more closely with Fe rather than Mg. 
The [Ti/Fe] ratios show a very large scatter, which is due to just a weak response to Ti found for the Lick indices (see Fig.~\ref{response}), requiring very high S/N observations to constrain [Ti/Fe]. Thus, we can't draw any strong conclusions on the [Ti/Fe]-$\sigma$ derived trends. Still, when compared to the [Mg/Fe]-$\sigma$ relationship (dashed orange line) the data suggest an overall under-abundance of Ti compared to Mg with an offset similar to Ca. \cite*{milone00} instead find that [Ti/Fe]$\approx$[Mg/Fe] using the TiO bands for a sample of  12 galaxies. 

The under-abundance of Ca compared to Mg, C and O confirms previous findings \citep{thomas03b,graves07,smith09,price11}. 
This has been interpreted as a contribution of Ca from SNIa besides SNII \citep{TJM10}.  
Since Ti is a heavier element than Ca the under-abundance of both of these elements imply that type Ia Supernovae also contributes to Ti. 
These results hint that the contribution from type Ia Supernovae is  dependent on atomic number for the $\alpha$-elements. 
This pattern of lower Ca and Ti abundances compared to Mg, O, and C has also been found for the stellar populations of the Milky Way \citep[][and references therein]{TJM10}. 
Hence heavy $\alpha$-elements are universally produced in SNIa.

The right hand panels of Fig.~\ref{CaTiFig} show the distributions around the red sequence populations for Ca (upper panel) and Ti (lower panel), with the same colour coding as in Fig.~\ref{age}. The rejuvenated population (cyan gaussian) show weaker abundance ratios compared to the red sequence population. This is however less pronounced than for Mg and Ca because of a significant contribution from both SN II and SN Ia to Ca. 
Clearly the standard deviation of the rejuvenated population is fairly large indicating that the exact position of the rejuvenated [Ca/Fe] peak is not very well defined. For Ti the peak of the rejuvenated population is even less well defined and no conclusions regarding the this population can be drawn from Ti. 

\section{Discussion}
\label{disc}

We present a method for simultaneously deriving the element abundance ratios [C/Fe], [N/Fe], [Mg/Fe], [Ca/Fe] and [Ti/Fe] for unresolved stellar populations, together with the classical stellar population parameters age, [Z/H] and [O/Fe] (representing [$\alpha$/Fe], see Section~\ref{signals}). The method is based on the new flux-calibrated stellar population models of absorption line indices presented in \citet{TMJ10}. It has been calibrated on galactic globular clusters in \citet{TJM10}, showing a good agreement for the derived parameters with the corresponding of high resolution spectra of individual cluster stars. 
We apply the method to a sample of 3802 SDSS early-type galaxies. The same sample of galaxies was used in T10 who derived the stellar populations parameters age, [Z/H] and [$\alpha$/Fe], based on the TMB/K models. Hence the analysis presented here can be regarded as an extension to the work of T10. The results derived here are in good agreement with the results of T10, i.e. age, total metallicity and [$\alpha$/Fe] increase with increasing velocity dispersion. This is remarkable given that we use new models and a new method for measuring the stellar population parameters that includes individual
element abundance ratios.

\begin{table}
\center
\caption{Average offsets in abundance ratios, metallicity and iron abundance for the rejuvenated population compared to the red sequence population. First column states the parameters, second column the offsets and the third column gives the standard deviation of the distributions for the rejuvenated population.}
\label{offsets}
\begin{tabular}{ccc}
\hline
\bf Parameter & \bf offset & \bf $\sigma$ \\
\hline
$[$O/Fe$]$   & -0.0265 & 0.129 \\
$[$C/Fe$]$   & -0.0577 & 0.0931 \\
$[$N/Fe$]$    & -0.0526 & 0.183 \\
$[$Mg/Fe$]$ & -0.0539 & 0.0980 \\
$[$Ca/Fe$]$ & -0.0600 & 0.130 \\
$[$Ti/Fe$]$   & -0.143 & 0.360  \\
$[$Z/H$]$      & 0.106 & 0.248  \\
$[$Fe/H$]$    & 0.147 & 0.266 \\
\hline
\end{tabular}
\end{table}

\subsection{Rejuvenated population}
\label{rejuvenation}

T10 find the sample of visually identified early-type galaxies to have, in addition to a dominant red sequence galaxy population, a sub-population of rejuvenated galaxies, i.e. galaxies with mainly old stellar populations that have experienced minor recent star formation producing stars that overshine the dominant old stellar populations and thus mimicking overall young stellar populations. The rejuvenated galaxy population was identified by a secondary peak in the derived age distribution showing younger ages than the primary peak and consists mainly of low mass early-type galaxies. This was supported by higher total metallicities and most importantly by lower [$\alpha$/Fe] ratios and detection of residual star formation through presence of emission lines. 
Schawinski et al. (2007) and T10 present the emission line diagnostics and show that most of the galaxies in the the young subpopulation have emission lines caused by star formation activity.

The higher total metallicities imply that residual star formation in the rejuvenated galaxies occurred in an ISM containing a fraction of gas that have been chemically enriched compared to pristine metal-poor gas \citep{thomas10}. 
In this work we further show that the rejuvenated population shows a stronger excess in iron abundance compared to total metallicity. This is due to the delayed Fe enrichment from SN Ia, which consequently produces lower [$\alpha$/Fe] ratios. 
Hence the extended star formation histories of the rejuvenated galaxies allow enough time for the ISM to be highly enriched in Fe. 

Table~\ref{offsets} summarises the offsets between the rejuvenated and red sequence population derived in Section~\ref{ageSec}~-~\ref{CaTiSec}. The standard deviations ($\sigma$) for the distributions of the rejuvenated population are also included. 
The result of T10 is reproduced in this work, where we find younger ages together with higher [Z/H], higher [Fe/H] and lower [E/Fe] ratios for the rejuvenated population. In more detail we find [O/Fe], [Mg/Fe], [C/Fe], [N/Fe] and [Ca/Fe] to be offset towards lower abundance ratios by 0.03-0.06 dex. 
However, the precision of the offsets varies significantly and shows the highest accuracy for [Mg/Fe] and [C/Fe], while the offset for [N/Fe] is not well determined. [Ti/Fe] shows the largest offset, but to a very low precision. The accuracy of the offsets is due to the sensitivity of the indices to the variation of the different element abundances (see Section~\ref{signals}). [Z/H] and [Fe/H] show positive offsets for the rejuvenated population, with [Fe/H] being more offset by $\sim$0.04 dex, i.e. reflecting the offset found for [E/Fe]. 

\subsection{Environment}
\label{environment}

T10 find environmental dependencies for the fraction of rejuvenated galaxies, while the red sequence galaxy population is unaffected by the environment for the classic parameters age, total metallicity and $\alpha$/Fe. 
In brief the fraction of rejuvenated galaxies increases with decreasing velocity dispersion and environmental density. 
We reproduce these results and do not further discuss environmental dependencies for the classic parameters, the reader is instead referred to T10 for more details. 
We have also looked at environmental dependencies for the various element abundance ratios and no such can be seen. 

Hence we do not confirm earlier findings of \citet{sanchez03} and \citet{sanchez06},  who report over-abundances of carbon and nitrogen
in environments of lower density for 98 early-type galaxies. The work of \citet{clemens06} studies a large sample of SDSS 
\citep[Sloan Digital Sky Survey,][]{york00} early-type galaxies and find, in agreement with the results derived here, that the environment does not  influence the enhancement of carbon.

\begin{figure}
\centering
\includegraphics[scale=0.35]{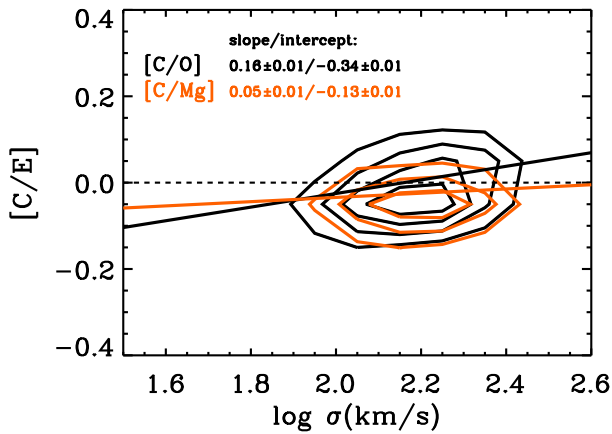}
\caption{[C/O] (black contours) and [C/Mg] (orange contours) as functions of velocity dispersion for the red sequence population. The solid lines are least-square fits to the relationships with corresponding colours and the dashed black line indicate solar abundance ratios. The slope and intercept of the least-square fits are given by the labels.}
\label{CO}
\end{figure}

\subsection{Lower limit on formation time-scales from [C/Mg]}
\label{ICarb}

Fig.~\ref{CO} shows contours and least-square fits for [C/O] (black contours and lines) and [C/Mg] (orange contours and lines) as functions of velocity dispersion for the red sequence population. Both of these abundance ratios are close to solar values. However, differences are noticeable. [C/Mg] is $\sim$0.0 for the most massive systems and $\sim$-0.05 for the less massive galaxies. [C/O] is instead $\sim$0.05 for the most massive systems and $\sim$-0.05 for the less massive galaxies.
The sources of C production have been the subject of a long standing debate. 
\citet{henry00} compared abundances of galactic and extra-galactic HII regions to models of chemical evolution. To match the data they needed stellar yields where massive stars dominate the production of C. 
\citet{cescutti09} found that stellar yields with metallicity dependent C produced in massive stars are needed to match models of chemical evolution with element abundance ratios derived from individual stellar spectra of the galactic stellar populations. Stellar yields take mass-loss and rotation into account. The latter cause newly synthesised C to be brought to the surface layers and ejected into the ISM \citep{MM02} by mass-loss. The mass-loss rate is dependent on metallicity such that more metal-rich massive stars experience higher mass-loss rates and consequently contribute higher abundances of C.

Hence recent results indicate that massive metal-rich stars are a major C source. 
Still, intermediate mass stars 
do also contribute significant amounts of C \citep{renzini81,vdH97}. Dredge-up on the asymptotic giant branch (AGB) phase of stellar evolution bring C up to the surface layers. This occurs following He-shell flashes when the convective envelope reach down to the inner layers where He-burning has taken place. The dredged-up C is then ejected into the ISM by mass-loss. \citet{vdH97} show that C yields from 3M$_{\odot}$ stars are significant. To reach solar C/Mg values star formation must continue over long enough time-scales to allow for the contribution of C from both massive and intermediate mass stars. This sets a lower limit for the star formation time-scales of $\sim$0.4 Gyr, which is the lifetime of a 3M$_{\odot}$ stars \citep{Castellani92,Bertelli09}. This is thus the lower limit for the formation of the most massive early-type galaxies of our sample, since these have [C/Mg]$\sim$0 (see Fig.~\ref{CO}). In a scenario where self-enriched bursts of star-formation build up the stellar populations of early-type galaxies, the bursts must last for at least $\sim$0.4 Gyr.
However, the formation time-scales must be long enough to allow the enrichment of the ISM to produce the high metallicities observed in massive early-type galaxies. Indeed, a similar time-scale ($\sim$0.4 Gyr) for the onset of galactic winds in models of chemical evolution can produce massive early-type galaxies with metallicities in agreement with observed values \citep{pipino10}.

Although weak, the trend of increasing [C/Mg] with velocity dispersion (see Fig.~\ref{CO}) 
could be caused by a metallicity dependent production of C in massive stars. The more massive galaxies are also more metal-rich (see Section~\ref{ZSec}).  
Maeder (1992) predict metallicity dependent C yields from increasing mass-loss in more metal-rich, massive stars. The net result is an ejection of large amount of Carbon before this element is turned into heavier elements. The side effect is that less Carbon is available for producing Oxygen. Thus Maeder (1992) also predicts a depression of Oxygen along with the enhancement of Carbon at higher metallicities, while at low metallicities the opposite situation is apparent. This behaviour was confirmed by e.g. Cescutti et al. (2009) for the Milky Way. 
In Fig.~\ref{CO} we have also seen that [C/O] increases with velocity dispersion such that super-solar and sub-solar [C/O] ratios are found for the most massive and least massive galaxies, respectively. 
Hence the flatter slope found for the [O/Fe]-$\sigma$ relationship compared to [Mg/Fe]-$\sigma$ can be explained by the metallicity dependent balance between C and O yields from massive stars, since higher velocity dispersion galaxies are more metal-rich. The effects discussed above are expected to be weak considering the short metallicity range covered. 

\subsection{The Nitrogen puzzle}
\label{INitro}

The origin of N is probably even more debated than the origin of C. 
It remains controversial whether N is a secondary and/or a primary element. N is produced in the CNO-cycle from C and O. Primary N then comes from the CNO processing in stars, 
while secondary N is produced from C and O already present in the collapsing gas-clouds forming the stars \citep[e.g.][]{matteucci86}. The amount of primary N produced is proportional to the abundances of other primary elements (e.g. C and O). The N abundance from secondary production is instead proportional to the initial C and O abundances. 
Primary and secondary production in intermediate mass stars, with masses in the range 4M$_{\odot}$$<$M$<$8M$_{\odot}$, are believed to be the dominant sources of N, 
where dredge-up, hot bottom burning and mass-loss eject N into the ISM \citep{renzini81,vdH97}. \cite{MM02} show that N can also be ejected by massive stars through the inclusion of stellar rotation. 
Observations indicate that primary N in low metallicity massive stars is required to explain observed trends of abundance ratios. \citet{izotov99} found HII regions in low metallicity blue compact galaxies to show a flat $\log$(N/O) trend with 12+$\log$(O/H) for their lowest metallicity galaxies, while the $\log$(N/O) ratio starts to increase at a certain metallicity (7.6$<$12+$\log$(O/H)$<$8.2). The interpretation is that the primary N is produced by massive stars at low metallicities and the contribution from intermediate mass stars is delayed in time and kicks in at higher metallicities.

\begin{figure}
\centering
\includegraphics[angle=90,scale=0.35]{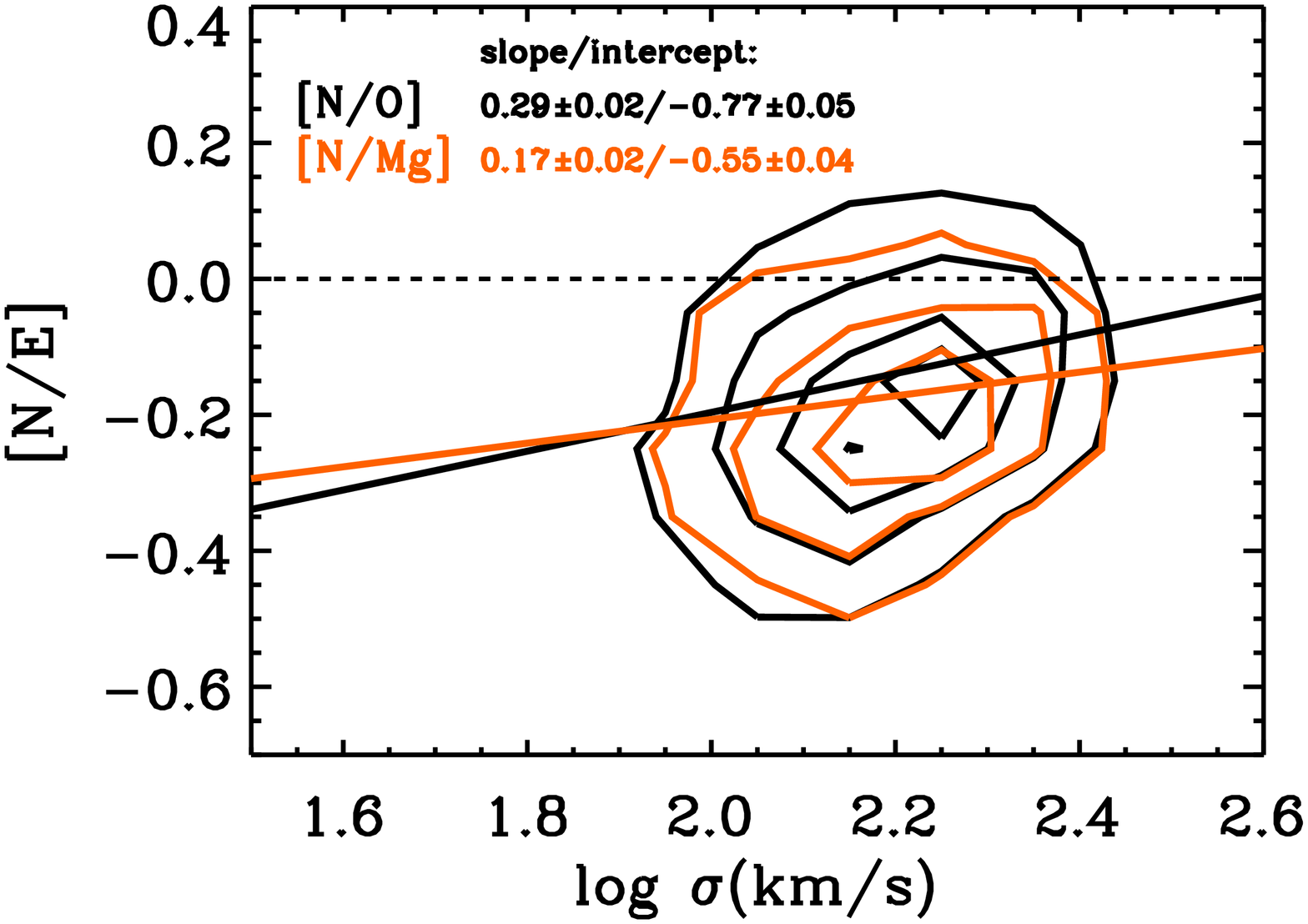}
\caption{[N/O] (black contours) and [N/Mg] (orange contours) as functions of velocity dispersion for the red sequence population. The solid lines are least-square fits to the relationships with corresponding colours and the dashed black line indicate solar abundance ratios. The slope and intercept of the least-square fits are given by the labels. 
}
\label{NO}
\end{figure}

Fig.~\ref{NO} shows the [N/O]-$\sigma$ (black contours) and [N/Mg]-$\sigma$ relationships (orange contours) together with least-square fits (solid lines) with corresponding colours. Fit parameters are given by the labels. An under-abundance of N of $\sim$0.2 dex compared to O and Mg is accompanied by a significant slope that is more prominent for the [N/O]-$\sigma$ relation.
The different origins of N compared to Mg and O make the N/Mg and N/O ratios useful formation time-scale indicators. 
Following our discussion of C/Mg as a lower limit time-scale indicator (see Section~\ref{CSec}), the low N/Mg ratios suggest that the formation time-scales of the red sequence sample are too short for the full N production to be reached. This could be the case if 
low mass stars with long life-times contribute significantly to the production of N. Such indications have been reported by \citet*{thuan10} from emission lines of SDSS star-forming galaxies, who find that significant amounts of N are produced in 1.5-2M$_{\odot}$ stars with lifetimes of 2-3 Gyr. 

In the low velocity dispersion regime (of our sample) we find [N/Fe]$<$0.0 (see lower left panel of Fig.~\ref{carbN}). This would require N to be produced over longer time-scales than Fe and could constrain the upper formation time-scale limit for the low mass systems, if the production sites of N were well constrained. Considering the stellar mass range suggested by \citet*{thuan10} for N contribution, an upper limit of $\sim$2.5 Gyr \citep[life-time of a 1.5 M$_{\odot}$ stars,][]{Castellani92,Bertelli08} may be implied for the galaxies in our sample with 1.8$<\sigma<$2.0 that have [N/Fe]$<$0.0. Future models with better constrained N yields are needed to quantify this to a higher degree.

However, besides the under-abundance of N the slopes of the N/Fe-$\sigma$, N/Mg-$\sigma$ and N/O-$\sigma$ relationships suggest that  abundance ratios are higher in more massive galaxies. Following the discussion above longer formation time-scales would be required for such systems compared to the lower mass galaxies. This is obviously in contradiction to the shorter formation time-scales of more massive systems implied by the better constrained, in terms of the sources of Mg production, Mg/Fe ratios. 
Hence in the scenario discussed above the N/E ratio does not hold as formation time-scale indicator for the most massive systems

Instead the Mg/Fe ratio would set an upper limit on the formation time-scales of the most massive systems and N/Fe provides an additional upper limit on lower mass galaxies. 
On the other hand if N is mainly produced in stars with masses above 4M$_{\odot}$, the delayed enrichment of Fe from SNIa will result in a correlation with velocity dispersion also for [N/Fe] (see discussion in Section~\ref{ICarb}). Still this correlation is stronger than for [O/Fe], [Mg/Fe] and [C/Fe]. In either case, the higher N/E ratios in the more massive systems still need to be explained.
Independent of the lower stellar mass limit for significant N contribution, at least three different scenarios can steepen the N/Fe-$\sigma$, N/O-$\sigma$ and N/Mg-$\sigma$ trends 

i) \textbf{Metallicity dependent secondary N production.} As mentioned secondary production of N will cause a strong increase in N abundances with increasing metallicity. With the higher velocity dispersion galaxies being overall more metal-rich the steep [N/Fe]-$\sigma$ trends could intuitively be explained by the secondary production of N in intermediate mass stars. 
The metallicity range covered is relatively small, however, and it seems contrived if metallicity effects in secondary N production can explain the steep [N/Fe]-$\sigma$ relationship alone. 

ii) \textbf{Primordial gas inflow.} The ratios between secondary and primary elements are useful indicators of primordial gas accretion, while the ratios between two primary elements are unaffected by such events \citep{koppen99,henry00}. 
Hence N/Mg and N/O ratios are 
tracers of primordial gas inflow \citep[e.g.][]{henry00} that results in lower N abundances compared to Mg and O. Fig.~\ref{NO} shows the [N/O]-$\sigma$ relationship. An under-abundance of N of $\sim$0.2 dex is accompanied by a significant slope. This trend could be caused by a stronger dilution of the ISM in low mass galaxies. If the high velocity dispersion galaxies experienced very intensive gas accretion over time-scales shorter than the star formation time-scale, high N abundances are achieved since dilution of the ISM does not take place over the entire star formation history. If low velocity dispersion galaxies instead experience gas accretion along with star formation the ISM instead gets diluted with primordial gas over the entire star formation history and the N abundances become lower. This scenario fits with the results of \citet{dekel06} who show that the shutdown of gas supply from cold streams is dependent on halo mass, such that it continues over longer time-scales in less massive halos. 
Such a scenario can steepen the slope of the [N/O,Mg]-$\sigma$ relationships and consequently [N/Fe]-$\sigma$, but also for this case it is not clear if it can produce the observed trends alone. 


iii) \textbf{Varying initial mass function (IMF).} IMF variations have been considered throughout the literature to explain variations of element ratios in early-type galaxies, especially the trend between [Mg/Fe] and velocity dispersion \citep[e.g.][]{thomas99b,trager00b,maraston03,matteucci94,sanchez06,smith09}. However, the variations considered also affect absorption line indices \citep{sanchez06} and galaxy scaling relations \citep[][and references therein]{Renzini93,maraston03} in negative ways questioning the plausibility of IMF dependencies on galaxy mass. 
Since intermediate mass stars are believed to be the dominant sources of N, IMF variations with galaxy mass dependent weights in the (approximate) stellar mass range 5M$_{\odot}$$<$M$<$8M$_{\odot}$ \citep[particularly high N yields,][]{vdH97} are needed to vary the N abundances. 
IMF variations with emphasis on such a specific mass range have not been observed and are unlikely to occur. 
This would affect C abundances as well, hence the more or less constant C/Mg ratios (over the velocity dispersion range covered) does not favour the IMF variations discussed.

The first two mechanisms (i-ii) could work together to cause the observed trends with velocity dispersion. To evaluate if this is plausible simulations of chemical evolution are needed taking these mechanisms into account. However, in \citet{pipino10} (P10) up-to-date models of chemical evolution fail at reproducing observed [N/Fe]-mass trends of early-type galaxies. 
P10 adopt the models of \citet{pipino04} (PM04) that reproduce the observed pattern of increasing Mg/Fe ratios with increasing stellar mass in early-type galaxies, by assuming star formation histories compatible with the down-sizing scenario and Mg contributed by type II Supernovae.  The models also reproduce the observed abundances of Ca by adopting yields where contributions to this element come from both type II and type Ia Supernovae.

Implementing different recipes of stellar yields P10 compare modified versions of the PM04 models to the observed element abundance ratios of \citet{graves07}. 
The modified versions mainly differ in the prescription of stellar mass-loss and rotation, which impacts on the abundances of C and N. Stellar rotation causes a mixing of elements in different stellar layers and mass-loss the ejection of newly synthesised element into the ISM. With the new prescriptions P10 are able to match the observed [C/Fe]-$\sigma$ trend, both in zero-point and slope. The results indicate that there must be a substantial contribution to the production of C from massive stars and metallicity dependent yields possible due to metallicity dependent mass-loss rates. For [N/Fe] they instead find a very large scatter between the different models such that the observed steep trend of the [N/Fe]-$\sigma$ relationship can not be simultaneously matched with the overall high [N/Fe] ratios. Also, to reach these high abundance ratios they must adopt a prescription that is not physically justified. 

\section{Conclusions}
\label{conc}

We present light-averaged ages, metallicities and element abundance ratios for 3802 SDSS early-type galaxies drawn from the MOSES catalogue \citep{schawinski07} with visual morphology classifications. 
Using the flux-calibrated TMJ models of absorption line indices, which are based on the MILES stellar library, we have developed a method for simultaneously deriving the element abundance ratios [C/Fe], [O/Fe] (inferred from [$\alpha$/Fe]), [N/Fe], [Mg/Fe], [Ca/Fe] and [Ti/Fe]. The models are well calibrated with galactic globular clusters with independent measurements of stellar population parameters and element ratios \citep[TMJ,][]{TJM10}.


We study the relationships between the stellar population parameters and galaxy stellar velocity dispersion. In agreement with the literature stellar population age and total metallicity correlate with velocity dispersion. [Fe/H] instead does not show such a correlation over the entire parameter range covered, but for a fixed age a steep trend is found for the [Fe/H]-$\sigma$ relation. This trend is shallower than the analogous for [Z/H] due to suppressed Fe enrichment in more massive galaxies because of time-scale dependent contribution from SN Ia

Similar trends are found for [O/Fe], [Mg/Fe] and [C/Fe], i.e. strong correlations with velocity dispersion in agreement with the literature. The first two are expected to be similar, since both O and Mg belong to the group of $\alpha$-elements produced in massive stars through type II Supernovae. This is also in favour of the down-sizing scenario of early-type galaxies that set an upper limit on the star formation time-scales and where more massive systems experience shorter time-scales \citep[e.g.][]{thomas10}. 

The C/Mg ratios are close to solar values, which instead sets a lower limit for the formation time-scales of early-type galaxies. Stars with masses down to $\sim$3 M$_{\odot}$ contribute significantly to the production of C. To reach solar C/Mg ratios formation time-scales need to be long enough for such stars to eject C into the ISM. The inferred lower formation time-scale limit is then $\sim$0.4 Gyr, which is the life-time of a 3 M$_{\odot}$ star. 

The [N/Fe] ratios are overall lower by $\sim$0.2 dex compared to [O/Fe] and [Mg/Fe] and the trend with velocity dispersion is very steep, i.e. more massive galaxies have significantly higher [N/Fe] ratios.
The observed [N/Fe]-$\sigma$ trends are difficult to interpret due to uncertainties in the origin of N. 
The zero-point and slope of this relationship can not be simultaneously matched by up-to-date models of chemical evolution \citep{pipino10}. Either the theoretical stellar yields have to be increased by a significant factor or other prescriptions have to be incorporated into the models that affect the N yields. Such prescriptions could be: 1. N yields with a stronger dependence on metallicity, 
since more massive early-type galaxies are more metal-rich. 2. A dependence on galaxy mass for the ratio between the time-scale of star formation and the time-scale of primordial gas inflow, which affects the N/O, N/Mg and N/Fe ratios due to the secondary nature of N. 

We do not find any dependence on environmental density for the element ratios studied. This is in contradiction to previous studies that have reported environmental dependencies for C and N abundances. Hence difference formation scenarios for field and cluster early-type galaxies can not be inferred from the element ratios studied in this work.

The [Ca/Fe] ratios do not correlate significantly with velocity dispersion and are close to solar values over the entire velocity dispersion range covered. Although tentative, due to large errors, Ti shows a behaviour similar to Ca. 
This indicates an atomic number dependent contribution from type Ia Supernovae to the production of $\alpha$-elements, i.e. the yields from type Ia Supernovae are higher for heavier $\alpha$-elements. This is now universally found since similar patterns have been found in the stellar populations of the Milky Way \citep[][and references therein]{TJM10} and puts strong constraints on supernova nucleosynthesis.



\section*{ACKNOWLEDGMENTS}

We would like to thank the referee Brigitte Rocca-Volmerange for very valuable comments that helped improve the manuscript.


{}

\label{lastpage}

\end{document}